\newif\ifpagetitre            \pagetitretrue
\newtoks\hautpagetitre        \hautpagetitre={\hfil}
\newtoks\baspagetitre         \baspagetitre={\hfil}
\newtoks\auteurcourant        \auteurcourant={\hfil}
\newtoks\titrecourant         \titrecourant={\hfil}

\newtoks\hautpagegauche       \newtoks\hautpagedroite
\hautpagegauche={\hfil\the\auteurcourant\hfil}
\hautpagedroite={\hfil\the\titrecourant\hfil}

\newtoks\baspagegauche \baspagegauche={\hfil\tenrm\folio\hfil}
\newtoks\baspagedroite \baspagedroite={\hfil\tenrm\folio\hfil}

\headline={\ifpagetitre\the\hautpagetitre
\else\ifodd\pageno\the\hautpagedroite
\else\the\hautpagegauche\fi\fi}

\footline={\ifpagetitre\the\baspagetitre
\global\pagetitrefalse
\else\ifodd\pageno\the\baspagedroite
\else\the\baspagegauche\fi\fi}

\vsize=9.0in\voffset=1cm
\looseness=2


\message{fonts,}

\font\tenrm=cmr10
\font\ninerm=cmr9
\font\eightrm=cmr8
\font\teni=cmmi10
\font\ninei=cmmi9
\font\eighti=cmmi8
\font\ninesy=cmsy9
\font\tensy=cmsy10
\font\eightsy=cmsy8
\font\tenbf=cmbx10
\font\ninebf=cmbx9
\font\tentt=cmtt10
\font\ninett=cmtt9

\font\ninesl=cmsl9
\font\eightsl=cmsl8

\font\nineit=cmti9
\font\eightit=cmti8

\skewchar\ninei='177 \skewchar\eighti='177
\skewchar\ninesy='60 \skewchar\eightsy='60

\def\eightpoint{\def\rm{\fam0\eightrm} 
\normalbaselineskip=9pt
\normallineskiplimit=-1pt
\normallineskip=0pt

\textfont0=\eightrm \scriptfont0=\sevenrm \scriptscriptfont0=\fiverm
\textfont1=\ninei \scriptfont1=\seveni \scriptscriptfont1=\fivei
\textfont2=\ninesy \scriptfont2=\sevensy \scriptscriptfont2=\fivesy
\textfont3=\tenex \scriptfont3=\tenex \scriptscriptfont3=\tenex
\textfont\itfam=\eightit  \def\it{\fam\itfam\eightit} 
\textfont\slfam=\eightsl \def\sl{\fam\slfam\eightsl} 

\setbox\strutbox=\hbox{\vrule height6pt depth2pt width0pt}%
\normalbaselines \rm}

\def\ninepoint{\def\rm{\fam0\ninerm} 
\textfont0=\ninerm \scriptfont0=\sevenrm \scriptscriptfont0=\fiverm
\textfont1=\ninei \scriptfont1=\seveni \scriptscriptfont1=\fivei
\textfont2=\ninesy \scriptfont2=\sevensy \scriptscriptfont2=\fivesy
\textfont3=\tenex \scriptfont3=\tenex \scriptscriptfont3=\tenex
\textfont\itfam=\nineit  \def\it{\fam\itfam\nineit} 
\textfont\slfam=\ninesl \def\sl{\fam\slfam\ninesl} 
\textfont\bffam=\ninebf \scriptfont\bffam=\sevenbf
\scriptscriptfont\bffam=\fivebf \def\bf{\fam\bffam\ninebf} 
\textfont\ttfam=\ninett \def\tt{\fam\ttfam\ninett} 

\normalbaselineskip=11pt
\setbox\strutbox=\hbox{\vrule height8pt depth3pt width0pt}%
\let \smc=\sevenrm \let\big=\ninebig \normalbaselines
\parindent=1em
\rm}

\def\tenpoint{\def\rm{\fam0\tenrm} 
\textfont0=\tenrm \scriptfont0=\ninerm \scriptscriptfont0=\fiverm
\textfont1=\teni \scriptfont1=\seveni \scriptscriptfont1=\fivei
\textfont2=\tensy \scriptfont2=\sevensy \scriptscriptfont2=\fivesy
\textfont3=\tenex \scriptfont3=\tenex \scriptscriptfont3=\tenex
\textfont\itfam=\nineit  \def\it{\fam\itfam\nineit} 
\textfont\slfam=\ninesl \def\sl{\fam\slfam\ninesl} 
\textfont\bffam=\ninebf \scriptfont\bffam=\sevenbf
\scriptscriptfont\bffam=\fivebf \def\bf{\fam\bffam\tenbf} 
\textfont\ttfam=\tentt \def\tt{\fam\ttfam\tentt} 

\normalbaselineskip=11pt
\setbox\strutbox=\hbox{\vrule height8pt depth3pt width0pt}%
\let \smc=\sevenrm \let\big=\ninebig \normalbaselines
\parindent=1em
\rm}

\message{fin format jgr}

\auteurcourant={R.G.Novikov}
\titrecourant={Approximate inverse scattering at fixed energy in
three dimensions}

\magnification=1200
\font\Bbb=msbm10
\def\C{\hbox{\Bbb C}}
\def\R{\hbox{\Bbb R}}
\def\N{\hbox{\Bbb N}}
\def\S{\hbox{\Bbb S}}
\def\pa{\partial}
\def\ep{\varepsilon}
\def\b{\backslash}
\def\v{\varphi}

\vskip 2 mm
\centerline{\bf The $\bar\pa$-approach to
approximate inverse scattering}
\centerline{\bf at fixed energy in three dimensions}
\vskip 2 mm
\centerline{\bf R. G. Novikov}

\vskip 4 mm

\noindent
{\ninerm CNRS, Labotaire de Math\'ematiques Jean Leray (UMR 6629),
Universit\'e de Nantes, BP 92208,}

\noindent
{\ninerm F-44322, Nantes cedex 03, France}

\noindent
{\ninerm e-mail: novikov@math.univ-nantes.fr}

\vskip 2 mm
{\bf Abstract}

We develop the $\bar\pa$-approach to inverse scattering at fixed energy
in dimensions $d\ge 3$ of [Beals, Coifman 1985] and [Henkin, Novikov 1987].
As a result we propose a stable method for nonlinear approximate
finding a potential $v$ from its scattering amplitude $f$ at fixed
energy $E>0$ in dimension $d=3$. In particular, in three dimensions  we
stably reconstruct $n$-times smooth potential $v$ with sufficient decay
at infinity, $n>3$, from its scattering amplitude $f$ at fixed energy
$E$ up to $O(E^{-(n-3-\ep)/2})$ in the uniform norm as $E\to +\infty$ for
any fixed arbitrary small
$\ep>0$ (that is with almost the same decay rate of the error
for $E\to +\infty$ as in the linearized case near zero potential).

\vskip 2 mm
{\bf 1. Introduction}
\vskip 2 mm
Consider the Schr\"odinger equation
$$-\Delta\psi + v(x)\psi=E \psi,\ \ x\in\R^d,\ \ d\ge 2,\ \ E>0,\eqno(1.1)$$
where
$$v\in W_s^{n,1}(\R^d)\ \ {\rm for\ some}\ \ n\in\N,\ \ n>d-2,\ \
{\rm and\ some}\ \ s>0,\eqno(1.2)$$
where
$$W_s^{n,1}(\R^d)=\{u:\ \Lambda^s\pa^Ju\in L^1(\R^d)\ \ {\rm for}\ \
|J|\le n\},\eqno(1.3)$$
where
$$\eqalign{
&J\in (\N\cup 0)^d,\ \ |J|=\sum_{i=1}^dJ_i,\ \ \pa^Ju(x)=
{\pa^{|J|}u(x)\over \pa x_1^{J_1}\ldots\pa x_d^{J_d}},\cr
&\Lambda^sw(x)=(1+|x|^2)^{s/2}w(x),\ \ x\in\R^d.\cr}$$
For equation (1.1) we consider the scattering amplitude $f(k,l)$, where
$(k,l)\in {\cal M}_E$,
$${\cal M}_E=\{k,l\in\R^d:\ k^2=l^2=E\},\ \ E>0.\eqno(1.4)$$
For definitions of the scattering amplitude see, for example, [F3] and
[FM].
Given $v$, to determine $f$ one can use, in particular, the integral
equation (2.5) of Section 2.

In the present work we consider, in particular, the following inverse
scattering problem for equation (1.1):

\vskip 2 mm
{\bf Problem 1.}
Given $f$ on ${\cal M}_E$ at fixed energy $E>0$, find $v$ on
$\R^d$ (at least approximately, but sufficiently stably for numerical
implementations).

Note that the Schr\"odinger equation (1.1) at fixed energy $E$ can be
considered also as the acoustic equation at fixed frequency $\omega$,
where $E=\omega^2$ (see, for example, Section 5.2 of [HN]). Therefore,
Problem 1 is also a basic problem of the monochromatic ultrasonic tomography.
Actually, the  creation of effective reconstruction methods for inverse
scattering in multidimensions (and especially in three dimensions)
was formulated as a very important problem many times in the mathematical
literature, see, for example, [Gel], [F3], [Gro]. In particular, as it is
mentioned in [Gro]: "For example, an efficient inverse scattering
algorithm would revolutionize medical diagnostics, making ultrasonic
devices at least as efficient as current X-ray analysis". The results
of the present work can be considered as a step to this objective.

Suppose, first, that
$$\|v\|_s^{n,1}=\max_{\scriptstyle |J|\le n}\|\Lambda^s
\pa^Jv\|_{L^1(\R^d)} \eqno(1.5)$$
is so small for fixed $n$, $s$, $d$ of (1.1), (1.2) and some fixed
$E_0>0$ that the following well-known Born approximation
$$\eqalignno{
&f(k,l)\approx {\hat v}(k-l),\ \ (k,l)\in {\cal M}_E,\ \ E\ge E_0,&(1.6)\cr
&{\hat v}(p)=\bigl({1\over 2\pi}\bigr)^d\int\limits_{\R^d}e^{ipx}v(x)dx,\ \
p\in\R^d,&(1.7)\cr}$$
is completely satisfactory. Then Problem 1 (for fixed $E\ge E_0$) is
reduced to finding $v$ from $\hat v$ on ${\cal B}_{2\sqrt{E}}$, where
$${\cal B}_r=\{p\in\R^d:\ |p|<r\},\ \ r>0.\eqno(1.8)$$
This linearized inverse scattering problem can be solved by the formula
$$v(x)=v_{appr}(x,E) + v_{err}(x,E),\eqno(1.9)$$
where
$$v_{appr}(x,E)=\int\limits_{{\cal B}_{2\sqrt{E}}}e^{-ipx}{\hat v}(p)dp,\ \
v_{err}(x,E)=\int\limits_{\R^d\b {\cal B}_{2\sqrt{E}}}e^{-ipx}{\hat v}(p)dp,
\eqno(1.10)$$
$x\in\R^d$, $E\ge E_0$. If $v$ satisfies (1.2), $n>d$, and
$\|v\|_0^{n,1}\le C$ then
$$\eqalignno{
&|{\hat v}(p)|\le C_1(n,d)\, C\,(1+|p|)^{-n},\ \ p\in\R^d,&(1.11a)\cr
&|v_{err}(x,E)|\le C_2(n,d)\,C\,E^{-(n-d)/2},\ \ x\in\R^d,\ \ E\ge E_0,
&(1.11b)\cr}$$
where $C_1(n,d)$, $C_2(n,d)$ are some positive constants and
$\|v\|_0^{n,1}$ is defined by (1.5).

If $v$ satisfies (1.2) and, in addition, is compactly supported or
exponentially decaying at infinity, then $\hat v$ on ${\cal B}_{2\sqrt{E}}$
uniquely determines $\hat v$ on $\R^d\b  {\cal B}_{2\sqrt{E}}$
(at fixed $E>0$) by an analytic continuation and, therefore, in the Born
approximation (1.6) $f$ on ${\cal M}_E$ (at fixed $E\ge E_0$) uniquely
determines $v$ on $\R^d$. However, this determination is not
sufficiently stable for direct numerical implementation.

In [No3], [No5] it was shown, in particular, that if $v$ is a bounded,
measurable, real function on $\R^d$ and, in addition, is compactly
supported or exponentially decaying at infinity, then $f$ on ${\cal M}_E$
at fixed $E>0$ uniquely determines $v$ almost everywhere on $\R^d$ for
$d\ge 3$. In [No2], [No3], [No4] similar results were given also for $d=2$
under the condition that $v$ is sufficiently small in comparison with
fixed $E$. However, these determinations of [No2], [No3], [No4], [No5] are not
sufficiently stable for direct numerical implementation (because of the
nature of Problem 1 explained already for the linearized case (1.6)).
Actually, any precise reconstruction of $v$, where
$v\in C^n(\R^d)$ and $supp\,v\in {\cal B}_r$ for fixed $n\in\N$ and
$r>0$, from $f$ on ${\cal M}_E$ at fixed $E>0$ is exponentially unstable
(see [Ma]). In the present work we will not discuss such
reconstructions in detail.

In [HN] for $d\ge 2$ it was shown, in particular, that if $v$ is
measurable, real function on $\R^d$ and $|v(x)|<C(1+|x|)^{-d-\ep}$
for some positive $\ep$ and $C$, then for any fixed $E$ and $\delta$,
where $0<\delta<E$, $f$ on $\cup_{\lambda\in
[E-\delta,E+\delta]}{\cal M}_{\lambda}$ uniquely determines $\hat v$ on
${\cal B}_{2\sqrt{E}}$.
However, unfortunately, this determination of [HN] involves an analytical
continuation and, therefore, is not sufficiently stable for direct
numerical implementation.

Note also that in [Ch] an efficient numerical algorithm for the
reconstruction from multi-frequency scattering data was proposed in two
dimensions, but  [Ch] gives no rigorous mathematical
theorem.

On the other hand
in [No6], [No7] we succeeded to give stable
approximate solutions of nonlinearized
Problem 1 for $d=2$ and $v$ satisfying (1.2), $n>d=2$, with the same
decay rate of the error terms for $E\to +\infty$ as in the linearized case
(1.6), (1.9), (1.10), (1.11b) (or, more precisely, with the error terms
decaying
as $O(E^{-(n-2)/2})$ in the uniform norm as $E\to +\infty$). Note that
in [No6], [No7] we proceed from the fixed-energy inverse scattering
reconstruction procedure developed in [No1], [GM], [No2], [No4] for $d=2$.
(In turn, the works [No1], [GM], [No2], [No4] are based, in particular, on the
nonlocal Riemann-Hilbert problem approach of [M], the $\bar\pa$-approach of
[ABF] and some results of [F2], [F3], [GN].) Note also that the
reconstruction procedure of [No1], [GM], [No2], [No4] was implemented
numerically
in [BBMRS], [BMR].

In the present work we succeeded, in particular, to give a stable
 approximate
solution of nonlinearized Problem 1 for $d=3$ and $v$ satisfying (1.2),
$n>d=3$, with the error term decaying as $O(E^{-(n-3-\ep)/2})$ in the
uniform norm as $E\to +\infty$ for any fixed arbitrary small $\ep>0$
(that is with almost the same decay rate of the error term for
$E\to +\infty$ as in the linearized case (1.6), (1.9), (1.10), (1.11b)).

Note that before the works [No6], [No7] in dimension $d=2$ and the
present work in dimension $d=3$, even for real $v$ of the Schwartz classs
on $\R^d$, $d\ge 2$, no result was given, in general, in the literature on
finding $v$ on $\R^d$ from $f$ on ${\cal M}_E$ with the error
decaying more rapidly than $O(E^{-1/2})$ in the uniform norm as
$E\to +\infty$ (see related discussion given in [No6]).

The aforementioned
result of the present work
is a corollary (see Corollary 3 of Section 8)
of the method developed in Sections 3,4,5,6,7 (for $d=3$) for
approximate finding $\hat v$ on ${\cal B}_{2\tau\sqrt{E}}$ from $f$ on
${\cal M}_E$ at fixed $E>0$, where $\tau\in ]0,1[$ is a parameter of our
approximate reconstruction. More precisely, in the present work for $v$
satisfying (1.2), $n>2$, $d=3$, and $\|v\|_s^{n,1}\le C$ we give a
stable method for approximate finding $\hat v$ on ${\cal B}_{2\tau\sqrt{E}}$
from
$f$ on ${\cal M}_E$ (at fixed $E\ge E(s,n,\mu_0,C)\to +\infty$ as
$C\to +\infty$) with the error  decaying as $O(E^{-(n-\mu_0)/2})$ in the
norm $\|\cdot\|_{E,\tau,\mu_0}$ as $E\to +\infty$ for fixed $C$,
$\mu_0$ and
$\tau$, where $\mu_0\ge 2$, $0<\tau<\tau(s,n,\mu_0,C)\to 0$ as
$C\to +\infty$,
$$\|w\|_{E,\tau,\mu_0}=\sup_{\scriptstyle p\in {\cal B}_{2\tau\sqrt{E}}}
(1+|p|)^{\mu_0}|w(p)|,\eqno(1.12)$$
see Theorems 1 and 2 of Section 8. Our reconstruction procedure can be
summarized as follows (where $d=3$):

1. From $f$ on ${\cal M}_E$ via the Faddeev equation (2.20) we find the
Faddeev generalized scattering amplitude $h_{\gamma}(k,l)$ for
$(k,l)\in {\cal M}_E$, $\gamma\in\S^{d-1}$, $\gamma k=0$;

2. From $h_{\gamma}(k,l)$, $(k,l)\in {\cal M}_E$, $\gamma\in\S^{d-1}$,
$\gamma k=0$, $\gamma l=0$, via formulas (2.8), (5.1), (5.9a), (5.11) and
nonlinear integral
equation (5.34) derived in Section 5 we find an approximation
${\tilde H}_{E,\tau}$ to the Faddeev generalized "scattering" amplitude
$H$ on $\Omega_E^{\tau}\b Re\,\Omega_E^{\tau}$ for some $\tau\in ]0,1[$,
where
$$\eqalign{
&\Omega_E^{\tau}=\{k\in\C^d,\ p\in {\cal B}_{2\tau\sqrt{E}}:\ k^2=E,\ p^2=
2kp\},\cr
&Re\,\Omega_E^{\tau}=\{k\in\R^d,\ p\in {\cal B}_{2\tau\sqrt{E}}:\ k^2=E,\ p^2=
2kp\};\cr}\eqno(1.13)$$

3. From ${\tilde H}_{E,\tau}$ on $\Omega_E^{\tau}\b Re\,\Omega_E^{\tau}$ via
formulas (7.2) we find approximations ${\hat v}_{\pm}(\cdot, E,\tau)$ to
$\hat v$ on ${\cal B}_{2\tau\sqrt{E}}$.

This reconstruction procedure (with estimates for the difference
${\hat v}-{\hat v}(\cdot,E,\tau)$ on ${\cal B}_{2\tau\sqrt{E}}$) is
presented in detail in Theorem 1 of Section 8.
A stability estimate for this
procedure with respect to errors in $f$ is given in Theorem 2 of Section 8.
For the case when nonlinear integral equation (5.34) in our
reconstruction procedure is approximately solved just by the very first
approximation, this reconstruction is, actually, reduced to the approximate
reconstruction proposed and implemented numerically in [ABR]
proceeding from  [HN]. A numerical realization of the strict reconstruction
procedure of the present work is now in preparation by the authors of [ABR].

Note also that in Corollary 2 of Section 8 we give a generalization of
Theorem 1 to the case of approximate finding $\hat v$ on
${\cal B}_{2\tau\sqrt{E}}$ from $f$ given only on
$${\cal M}_{E,\tau}=\{(k,l)\in {\cal M}_E:\ |k-l|<2\tau\sqrt{E}\},\eqno(1.14)
$$
where $E>0$, $0<\tau<1$.

In the present work, besides some results of [F3], we proceed from the
$\bar\pa$-approach to inverse scattering at fixed energy in dimension
$d\ge 3$ of [BC] and [HN] and some estimates of [ER1] and [No7].

The present paper is composed as follows. In Section 2 we give some
preliminaries concerning the scattering amplitude $f$ and its Faddeev's
extension $h$. Our method for approximate nonlinear inverse scattering
at fixed energy in dimension $d=3$ is developed in Sections 3,4,5,6,7 and
is summarized in Section 8. The main technical proofs are given in
Sections 9,10,11,12.

In the present work we consider, mainly, the case of the most
important dimension $d=3$. However, only restrictions in time prevent us
from generalizing all main results of the present work to the case $d>3$.

\vskip 2 mm

{\bf 2. Scattering data and some preliminaires}
\vskip 2 mm
Note that in this paper we always suppose that real degrees of positive
real values denote positive real values.
Note also that through $c_j$ we shall denote some positive constants
(which can be given explicitly).

Consider
$$C^{\alpha,\mu}(\R^d)=\{u\in C(\R^d):\ \|u\|_{\alpha,\mu}< +\infty\},\ \
\alpha\in ]0,1], \ \mu\in\R,\eqno(2.1)$$
where
$$\eqalignno{
&\|u\|_{\alpha,\mu}=\|\Lambda^{\mu}u\|_{\alpha},&(2.2a)\cr
&\Lambda^{\mu}u(p)=(1+|p|^2)^{\mu/2}u(p),\ \ p\in\R^d,&(2.2b)\cr
&\|w\|_{\alpha}=\sup_{p,\xi\in\R^d,\ |\xi|\le 1}(|w(p)|+|\xi|^{-\alpha}
|w(p+\xi)-w(p)|),&(2.2c)\cr}$$
Consider also ${\cal H}_{\alpha,\mu}$ defined as the  closure of
$C_0^{\infty}(\R^d)$ (the  space of infinitely smooth functions with
compact support) in $\|\cdot\|_{\alpha,\mu}$.

Let $\hat v$ be defined by (1.7). If $v$ satisfies (1.2), then
$$\hat v\in {\cal H}_{\alpha,n}(\R^d),\ \ {\rm where}\ \ \alpha=\min(1,s).
\eqno(2.3)$$

For equation (1.1), where
$$\hat v\in {\cal H}_{\alpha,\mu}(\R^d)\ \ {\rm for\ some}\ \
\alpha\in ]0,1[\ \
 {\rm and\ some\ real}\ \ \mu>d-2,\eqno(2.4)$$
we consider the function $f(k,l)$, where $k,l\in\R^d$, $k^2=E$, of the
classical scattering theory. Given $v$, to determinate $f$ one can use the
following integral equation
$$f(k,l)={\hat v}(k-l)-\int\limits_{\R^d}
{{\hat v}(m-l)f(k,m)dm\over {m^2-k^2-i0}},\eqno(2.5)$$
where $k,l\in\R^d$, $k^2>0$, and where at fixed $k$ the function $f$ is
sought in $C^{\alpha,\mu}(\R^d)$. In addition, $f$ on ${\cal M}_E$
defined by (1.4)
is the scattering amplitude for equation (1.1).

Note that
$$\eqalignno{
&{\cal M}_E=\S^{d-1}_{\sqrt{E}}\times\S^{d-1}_{\sqrt{E}},\ \ {\rm where}
&(2.6)\cr
&\S^{d-1}_r=\{m\in\R^d:\ |m|=r\},\ \ r>0. &(2.7)\cr}$$

For equation (1.1), where $\hat v$ satisfies (2.4), we consider also the
Faddeev functions
$$\eqalign{
&h_{\gamma}(k,l)=H_{\gamma}(k,k-l),\ \ {\rm where}\ \ k,l\in\R^d,\ k^2=E,\
\gamma\in\S^{d-1},\ \ {\rm and}\cr
&h(k,l)=H(k,k-l),\ \ {\rm where}\ \ k,l\in\C^d\b\R^d,\ k^2=E,\ \ Im\,k=Im\,l
\cr
&{\rm (see\ [F3],\ [HN],\ [No7])}:\cr}\eqno(2.8)$$
$$H_{\gamma}(k,p)={\hat v}(p)-\int\limits_{\R^d}
{{\hat v}(p+\xi)H_{\gamma}(k,-\xi)d\xi\over {\xi^2+2(k+i0\gamma)\xi}},\ \
k\in\R^d\b 0,\ \gamma\in\S^{d-1},\ p\in\R^d,\eqno(2.9)$$
where at fixed $\gamma\in\S^{d-1}$ and $k\in\R^d\b 0$ we consider (2.9)
as an equation for $H_{\gamma}(k,\cdot)\in C^{\alpha,\mu}(\R^d)$,
$$H(k,p)={\hat v}(p)-\int\limits_{\R^d}
{{\hat v}(p+\xi)H(k,-\xi)d\xi\over {\xi^2+2k\xi}},\ \
k\in\C^d\b\R^d,\  p\in\R^d,\eqno(2.10)$$
where at fixed $k\in\C^d\b\R^d$ we consider (2.10) as an equation for
$H(k,\cdot)\in C^{\alpha,\mu}(\R^d)$. In addition, $h$ on
$\{k,l\in\C^d\b\R^d:\ Im\,k=Im\,l,\ k^2=l^2=E\}$ or that is the same $H$ on
$\Omega_E\b Re\,\Omega_E$, where
$$\eqalign{
&\Omega_E=\{k\in\C^d,\ \ p\in\R^d:\ p^2=2kp,\ k^2=E\},\cr
&Re\,\Omega_E=\{k\in\R^d,\ \ p\in\R^d:\ p^2=2kp,\ k^2=E\},\cr}\eqno(2.11)$$
can be considered as the scattering amplitude in the complex domain for
equation (1.1).

Consider the operator ${\tilde A}^+(k)$ from (2.5) and the operators
$A_{\gamma}(k)$, $A(k)$ from (2.9), (2.10):
$$\eqalignno{
&({\tilde A}^+(k)U)(l)=\int\limits_{\R^d}
{{\hat v}(m-l)U(m)dm\over {m^2-k^2-i0}},\ \ l\in\R^d,\ k\in\R^d\b 0,&(2.12)
\cr
&(A_{\gamma}(k)U)(p)=\int\limits_{\R^d}
{{\hat v}(p+\xi)U(-\xi)d\xi\over {\xi^2+2(k+i0\gamma)\xi}},\ \ p\in\R^d,\
 \gamma\in\S^{d-1},\ k\in\R^d\b 0,&(2.13)\cr
&(A(k)U)(p)=\int\limits_{\R^d}
{{\hat v}(p+\xi)U(-\xi)d\xi\over {\xi^2+2k\xi}},\ \ p\in\R^d,\
 \ k\in\C^d\b\R^d.&(2.14)\cr}$$

If $\hat v$ satisfies (2.4), then
$$\eqalign{
&\|\Lambda_k^{\mu}{\tilde A}^+(k)\Lambda_k^{-\mu}u\|_{\alpha}\le
(1/2)|k|^{-\sigma}c_1(\alpha,\mu,\sigma,d)\|\hat v\|_{\alpha,\mu}
\|u\|_{\alpha},\cr
&{\rm where}\ \ k\in\R^d,\ k^2\ge 1,\ \ {\rm and}\ \
(\Lambda_ku)(l)=(1+|k-l|^2)^{1/2}u(l),\ \ l\in\R^d,\cr}\eqno(2.15)$$
$$\eqalign{
&\|\Lambda^{\mu}A_{\gamma}(k)\Lambda^{-\mu}u\|_{\alpha}\le
(1/2)|k|^{-\sigma}c_1(\alpha,\mu,\sigma,d)\|\hat v\|_{\alpha,\mu}
\|u\|_{\alpha},\cr
&{\rm for}\ \ \gamma\in\S^{d-1},\ \ k\in\R^d,\ \ k^2\ge 1,\cr}\eqno(2.16)$$
$$\eqalign{
&\|\Lambda^{\mu}A(k)\Lambda^{-\mu}u\|_{\alpha}\le
|Re\,k|^{-\sigma}c_1(\alpha,\mu,\sigma,d)\|\hat v\|_{\alpha,\mu}
\|u\|_{\alpha},\cr
&{\rm for}\ \  k\in\C^d\b\R^d,\ \ \R\ni k^2\ge 1,\cr}\eqno(2.17)$$
where $u\in C^{\alpha,0}(\R^d)$, $0\le\sigma< min(1,\mu-d+2)$.
Estimate (2.15) follows from Theorem 2.1 of [ER1] for $d=3$ and Theorem 1.1
of [ER2] for $d=2$. (Note also that (2.15) is a development  of a
related estimate from [F1] for $d=3$.) Estimates (2.16), (2.17) are
given in Proposition 1.1 of [No7].

If   $v$ satisfies (2.4) and
$$
\|\hat v\|_{\alpha,\mu}\le N<
{E^{\sigma/2}\over c_1(\alpha,\mu,\sigma,d)}\ \ {\rm for\ some}\ \
\sigma\in ]0, min(1,\mu-d+2)[\ \ {\rm and\ some}\ \ E\ge 1,\eqno(2.18)$$
then the following results are valid:

I. For $k^2=E$ equations (2.5), (2.9) and (2.10) considered as mentioned
above (for fixed $k$ or for fixed $k$ and $\gamma$) are uniquely solvable
(by the method of successive approximations).

II. The following formulas hold (see [F3], [HN]):
$$\eqalign{
&h_{\gamma}(k,l)=h(k+i0\gamma,l+i0\gamma)\ \ {\rm for}\ \ k,l\in\R^d,\
k^2=E,\ \gamma\in\S^{d-1},\cr
&f(k,l)=h_{k/|k|}(k,l)\ \ {\rm for}\ \ k,l\in\R^d,\ k^2=E;\cr}\eqno(2.19)$$
$$
h_{\gamma}(k,l)=f(k,l)+{\pi i\over \sqrt{E}}
\int\limits_{\S^{d-1}_{\sqrt{E}}}
h_{\gamma}(k,m)\chi((m-k)\gamma)f(m,l)dm,\eqno(2.20)$$
where
$$
\chi(s)=0\ \ {\rm for}\ \ s\le 0,\ \chi(s)=1\ \ {\rm for}\ \ s>0,\eqno(2.21)
$$
$k,l\in\R^d$,\ $k^2=E$, $\gamma\in\S^{d-1}$, $dm$ is the standard measure
on $\S_{\sqrt{E}}^{d-1}$.

III. The following $\bar\pa$- equation holds (see [BC], [HN]):
$$\eqalign{
&{\pa\over \pa\bar k_j}H(k,p)=-2\pi\int\limits_{\R^d}
\xi_jH(k,-\xi)H(k+\xi,p+\xi)\delta(\xi^2+2k\xi)d\xi,\cr
&j=1,\ldots,d,\ \ k\in\C^d\b\R^d,\ \  k^2=E,\ \ p\in\R^d,\cr}\eqno(2.22)$$
 where
$\delta$ is the Dirac function; in addition, for $d\ge 3$
$$\int\limits_{\R^d}u(\xi)\delta(\xi^2+2k\xi)d\xi=
\int_{\{\xi\in\R^d:\ \xi^2+2k\xi=0\}}
{u(\xi)\over |J(k,\xi)|}|d\xi_3\wedge\ldots d\xi_d|,\eqno(2.23)$$
where $J(k,\xi)=4[(\xi_1+Re\,k_1)Im\,k_2-(\xi_2+Re\,k_2)Im\,k_1]$
is the Jacobian of the map
$(\xi_1,\ldots,\xi_d)\to (\xi^2+2Re\,k\xi, 2Im\,k\xi,\xi_3,\ldots,\xi_d)$
and $u$ is a test function.

IV. The following estimates are valid:
$$|H(k,p)-{\hat v}(p)|\le {\eta\over {1-\eta}}N
(1+p^2)^{-\mu/2},
\eqno(2.24)$$
where
$$\eta=|Re\,k|^{-\sigma} c_1(\alpha,\mu,\sigma,d)
N,\eqno(2.25)$$
$k\in\C^d\b\R^d$, $k^2=E$, $p\in\R^d$ (and where $\eta<1$ due to (2.18) and
the inequality $|Re\,k|\ge E^{1/2}$), and, as a corollary,
$${\hat v}(p)=H(k,p)+O\bigl({1\over (1+p^2)^{\mu/2}|k|^{\sigma}}\bigr)\ \
{\rm as}\ \ |k|=(|Re\,k|^2+|Im\,k|^2)^{1/2}\to\infty,\eqno(2.26)$$
where $k\in\C^d\b\R^d$, $k^2=E$, $p\in\R^d$;
$$\eqalignno{
&|f(k,l)|\le {N\over {1-\eta}}(1+|k-l|^2)^{-\mu/2},\ \ k,l\in\R^d,\ \
k^2=E,&(2.27a)\cr
&|f(k,l)-{\hat v}(k-l)|\le {\eta\over {1-\eta}}N
(1+|k-l|^2)^{-\mu/2},\ \
k,l\in\R^d,\ k^2=E,&(2.27b)\cr}$$
$$\eqalign{
&|f(k,l)-f(k^{\prime},l^{\prime})|
\le\cr
&c_2(\mu)
{N\over {1-\eta}}
(1+|k-l|^2)^{-\mu/2}
(|l-l^{\prime}|^{\alpha}+|k-k^{\prime}|^{\alpha}),\cr
&k,k^{\prime},l,
l^{\prime}\in\R^d,\ k^2={k^{\prime}}^2=l^2={l^{\prime}}^2=E,\ \
|k-k^{\prime}|\le 1,\ \ |l-l^{\prime}|\le 1,\cr}\eqno(2.28a)$$

$$\eqalign{
&|f(k,l)-{\hat v}(k-l)-(f(k^{\prime},l^{\prime})-
{\hat v}(k^{\prime}-l^{\prime}))|
\le\cr
&c_2(\mu)
{\eta\over {1-\eta}}N
(1+|k-l|^2)^{-\mu/2}
(|l-l^{\prime}|^{\alpha}+|k-k^{\prime}|^{\alpha}),\cr
&k,k^{\prime},l,
l^{\prime}\in\R^d,\ k^2={k^{\prime}}^2=l^2={l^{\prime}}^2=E,\ \
|k-k^{\prime}|\le 1,\ \ |l-l^{\prime}|\le 1,\cr}\eqno(2.28b)$$
where
$$\eqalignno{
&c_2(\mu)=c_2^{\prime}(\mu)c_2^{\prime\prime}(\mu),&(2.28c)\cr
&c_2^{\prime}(\mu)=\sup_{\scriptstyle p,p^{\prime}\in\R^d,\atop\scriptstyle
|p-p^{\prime}|\le 1}{(1+|p|^2)^{\mu/2}\over (1+|p^{\prime}|^2)^{\mu/2}},
&(2.28d)\cr
&c_2^{\prime\prime}(\mu)=1+
\sup_{\scriptstyle p,p^{\prime}\in\R^d,\atop\scriptstyle
|p-p^{\prime}|\le 1}
{|(1+|p|^2)^{\mu/2}-(1+|p^{\prime}|^2)^{\mu/2}|
\over (1+|p^{\prime}|^2)^{\mu/2}|p-p^{\prime}|},&(2.28c)\cr}$$

$$|H_{\gamma}(k,p)-{\hat v}(p)|\le {\eta\over {1-\eta}}
N(1+p^2)^{-\mu/2},\ \
k,p\in\R^d,\ k^2=E,\ \gamma\in\S^{d-1},\eqno(2.29a)$$
$$\eqalign{
&|H_{\gamma}(k,p)-{\hat v}(p)-(H_{\gamma}(k,p^{\prime})-{\hat v}(p^{\prime}))|
\le {c_2^{\prime\prime}(\mu)\eta\over {1-\eta}}
N(1+p^2)^{-\mu/2}|p-p^{\prime}|^{\alpha},\cr
&k,p,p^{\prime}\in\R^d,\ k^2=E,\  |p-p^{\prime}|\le 1,\
\gamma\in\S^{d-1},\cr}\eqno(2.29b)$$
$$
|H(k,p)|\le {1\over {1-\eta}}N(1+p^2)^{-\mu/2},\ \
k\in\C^d\b\R^d,\ k^2=E,\ p\in\R^d,\eqno(2.30)$$
where
$$\eta=E^{-\sigma/2}c_1(\alpha,\mu,\sigma,d)N
\eqno(2.31)$$
(and where $\eta<1$ due to (2.18)).

Estimates (2.27), (2.29)-(2.31) follow from (2.5), (2.9), (2.10),
(2.15)-(2.17). Estimate (2.28) follows from (2.5), (2.15) and the
symmetry
$$f(k,l)=f(-k,-l),\ \ k,l\in\R^d,\ \ k^2=l^2=E>0.\eqno(2.32)$$

Consider
$$C^{\alpha}(\S_r^{d-1})=\{u\in C(\S_r^{d-1}):\ \|u\|_{C^{\alpha}(\S_r^{d-1})}
< +\infty\},\ \alpha\in [0,1[,\ r>0,\eqno(2.33)$$
where
$$\eqalign{
&\|u\|_{C^{\alpha}(\S_r^{d-1})}= \|u\|_{C(\S_r^{d-1})}=\sup_{m\in\S_r^{d-1}}
|u(m)|\ \ {\rm for}\ \ \alpha=0,\cr
&\|u\|_{C^{\alpha}(\S_r^{d-1})}=\max(\|u\|_{C(\S_r^{d-1})},
\|u\|^{\prime}_{C^{\alpha}(\S_r^{d-1})}),\cr
&\|u\|^{\prime}_{C^{\alpha}(\S_r^{d-1})}=\sup_{\scriptstyle m,m^{\prime}\in
\S_r^{d-1},\atop \scriptstyle |m-m^{\prime}|\le 1}
|m-m^{\prime}|^{-\alpha}|u(m)-u(m^{\prime})|\ \ {\rm for}\ \ \alpha\in ]0,1[.
\cr}\eqno(2.34)$$

Consider
$$C^{\alpha}({\cal M}_E)=\{u\in C({\cal M}_E):\
\|u\|_{C^{\alpha}({\cal M}_E),0}< +\infty\},\ \ \alpha\in [0,1[,\ E>0,
\eqno(2.35)$$
where
$$\eqalign{
&\|u\|_{C^{\alpha}({\cal M}_E),\mu}=\|u\|_{C({\cal M}_E),\mu}=\cr
&\sup_{(k,l)\in {\cal M}_E}(1+|k-l|^2)^{\mu/2}|u(k,l)|\ \
{\rm for}\ \ \alpha=0,\ \mu\ge 0,\cr
&\|u\|_{C^{\alpha}({\cal M}_E),\mu}=\max\,(\|u\|_{C({\cal M}_E),\mu},
\|u\|^{\prime}_{C^{\alpha}({\cal M}_E),\mu}),\cr
&\|u\|^{\prime}_{C^{\alpha}({\cal M}_E),\mu}=
\sup_{\scriptstyle (k,l),(k^{\prime},
l^{\prime})\in {\cal M}_E,\atop\scriptstyle |k-k^{\prime}|\le 1,
|l-l^{\prime}|\le 1}
(1+|k-l|^2)^{\mu/2}(|k-k^{\prime}|^{\alpha}+|l-l^{\prime}|^{\alpha})^{-1}
\times\cr
&|u(k,l)-u(k^{\prime},l^{\prime})|\ \
{\rm for}\ \ \alpha\in ]0,1[,\ \mu\ge 0.
\cr}\eqno(2.36)$$

If assumptions (2.4), (2.18) are fulfilled, then, as a corollary of
(2.27), (2.28),
$$\eqalign{
&f\in C^{\alpha}({\cal M}_E),\ \
\|f-{\hat v}\|_{C^{\alpha}({\cal M}_E),\mu}\le c_2(\mu){\eta\over {1-\eta}}
N,\cr
&\|f\|_{C^{\alpha}({\cal M}_E),\mu}\le {c_2(\mu)\over {1-\eta}}
N,\cr}\eqno(2.37)$$
where $\eta$ is given by (2.31), $\hat v={\hat v}(k-l)$, $(k,l)\in {\cal M}_E$
.

Consider (2.20) for $k^2=l^2=E$ as a family of equations parametrized by
$\gamma$ and $k$ for finding  $h_{\gamma}(k,\cdot)$ on $\S^{d-1}_{\sqrt{E}}$
from the scattering amplitude $f$ on ${\cal M}_E$. Consider the operator
$B_{\gamma}(k)$ from (2.20):
$$\eqalign{
&(B_{\gamma}(k)U)(l)={\pi i\over \sqrt{E}}\int\limits_{\S^{d-1}_{\sqrt{E}}}
U(m)\chi((m-k)\gamma)f(m,l)dm,\cr
&l\in\S^{d-1},\ \ \gamma\in\S^{d-1},\ \ k\in\R^d,\ \ k^2=E>0.\cr}\eqno(2.38)
$$
As a variation of a related result of [F3], we have that if $\hat v$
satisfies (2.4) and if at fixed $\gamma\in\S^{d-1}$ and $k\in\R^d$,
$k^2=E>0$, equations (2.5), (2.9) (considered as mentioned above) are
uniquely solvable, then (2.20) is uniquely solvable in
$C^{\beta}(\S^{d-1}_{\sqrt{E}})$ for any fixed $f\in [0,\alpha]$.
Besides, the following estimate holds:
$$\eqalign{
&\|\Lambda_k^{\mu}B_{\gamma}(k)\Lambda_k^{-\mu}
u\|_{C^{\beta}(\S^{d-1}_{\sqrt{E}})}
\le c_3(\beta,\mu,\sigma,d)E^{-\sigma/2}\|f\|_{C^{\beta}({\cal M}_E),\mu}
\times\cr
&\|u\|_{C(\S^{d-1}_{\sqrt{E}})}\ \ {\rm for}\ \ f\in C^{\beta}({\cal M}_E),\
u\in C(\S^{d-1}_{\sqrt{E}}),\cr}\eqno(2.39)$$
where $\beta\in [0,1[$, $\mu>d-2$, $0\le\sigma<\min\,(1,\mu-d+2)$,
$\gamma\in\S^{d-1}$, $k\in\R^d$, $k^2=E\ge 1$, and
$$(\Lambda_ku)(l)=(1+|k-l|^2)^{1/2}u(l),\ \ l\in\S^{d-1}_{\sqrt{E}}.
\eqno(2.40)$$
In (2.39) we do not assume that $f$ is related with $\hat v$. Actually,
(2.39) is simpler than (2.15), (2.16), because (2.39) contains no
singular integrals.

\vskip 2 mm
{\bf 3. Coordinates on $\Omega_E^{\tau}$ for $E>0$, $\tau\in ]0,1]$ and
$d=3$}
\vskip 2 mm

Consider $\Omega_E^{\tau}$ and $Re\,\Omega_E^{\tau}$ defined by (1.13).
 Note that $\Omega_E^{\tau_1}\subset\Omega_E^{\tau_2}$,
$Re\,\Omega_E^{\tau_1}\subset Re\,\Omega_E^{\tau_2}$, $\tau_1<\tau_2$,
and $\Omega_E^{+\infty}=\Omega_E$, $Re\,\Omega_E^{+\infty}=Re\,\Omega_E$,
where $\Omega_E$ and $Re\,\Omega_E$ are defined by (2.11).

For our considerations for $d=3$ we introduce some convenient coordinates
on $\Omega_E^{\tau}$,  $E>0$,
$0<\tau\le 1$. Let
$$\eqalign{
&\Omega_{E,\nu}^{\tau}=\{k\in\C^3,\ p\in {\cal B}_{2\tau\sqrt{E}}\b
{\cal L}_{\nu}:\ k^2=E,\ p^2=2kp\},\cr
&Re\,\Omega_{E,\nu}^{\tau}=\{k\in\R^3,\ p\in {\cal B}_{2\tau\sqrt{E}}\b
{\cal L}_{\nu}:\ k^2=E,\ p^2=2kp\},\cr}\eqno(3.1)$$
where $E>0$, $0<\tau\le 1$,
$$\eqalignno{
&{\cal B}_r=\{p\in\R^3:\ |p|<r\},\ r>0,&(3.2)\cr
&{\cal L}_{\nu}=\{x\in\R^3:\ t\nu,\ t\in\R\},\ \ \nu\in\S^2.&(3.3)\cr}$$
Note that $\Omega_{E,\nu}^{\tau}$ is an open and dense subset of
$\Omega_E^{\tau}$ for $d=3$, $E>0$, $\tau\in ]0,1]$.

For $p\in\R^3\b {\cal L}_{\nu}$ consider $\theta(p)$ and $\omega(p)$ such
that
$$\eqalign{
&\theta(p),\ \omega(p)\ \ {\rm smoothly\ depend\ on}\ \ p\in\R^3\b
{\cal L}_{\nu},\cr
&{\rm take\ their\ values\ in}\ \ \S^2\ \ {\rm and}\cr
&\theta(p)p=0,\ \omega(p)p=0,\ \theta(p)\omega(p)=0.\cr}\eqno(3.4)$$
Note that (3.4) implies that
$$\omega(p)={p\times\theta(p)\over |p|}\ \ {\rm for}\ \ p\in\R^3\b
{\cal L}_{\nu} \eqno(3.5a)$$
or
$$\omega(p)=-{p\times\theta(p)\over |p|}\ \ {\rm for}\ \ p\in\R^3\b
{\cal L}_{\nu}, \eqno(3.5b)$$
where $\times$ denotes vector product.

To satisfy (3.4), (3.5a) we can take
$$\theta(p)={\nu\times p\over |\nu\times p|},\ \omega(p)={p\times\theta(p)
\over |p|},\ p\in\R^3\b {\cal L}_{\nu}.\eqno(3.6)$$

\vskip 2 mm
{\bf Lemma 1.}
{\it Let} $E>0$, $\nu\in\S^2$. {\it Let} $\theta$, $\omega$ {\it satisfy}
(3.4). {\it Then the following formulas give a diffeomorphism between}
$\Omega_{E,\nu}^{\tau}$ {\it and} $(\C\b 0)\times ({\cal B}_{2\tau\sqrt{E}}\b
{\cal L}_{\nu})$ {\it for} $\tau\in ]0,1]$:
$$(k,p)\to (\lambda,p),\ \ {\it where}\ \ \lambda=\lambda(k,p)=
{k(\theta(p)+i\omega(p))\over (E-p^2/4)^{1/2}},\eqno(3.7)$$
$$\eqalign{
&(\lambda,p)\to (k,p),\ \ {\it where}\ \ k=k(\lambda,p,E)=
\kappa_1(\lambda,p,E)\theta(p)+\kappa_2(\lambda,p,E)\omega(p)+p/2, \cr
&\kappa_1(\lambda,p,E)=(\lambda+1/\lambda){(E-p^2/4)^{1/2}\over 2},\
\kappa_2(\lambda,p,E)=(1/\lambda-\lambda){i(E-p^2/4)^{1/2}\over 2},\cr}
\eqno(3.8)$$
{\it where} $(k,p)\in \Omega_{E,\nu}^{\tau}$,
$(\lambda,p)\in (\C\b 0)\times ({\cal B}_{2\tau\sqrt{E}}\b {\cal L}_{\nu})$.
{\it In addition, formulas} (3.7), (3.8) {\it give also diffeomorphisms
between} $Re\,\Omega_{E,\nu}^{\tau}$ {\it and}
${\cal T}\times ({\cal B}_{2\tau\sqrt{E}}\b {\cal L}_{\nu})$
{\it and between}
$\Omega_{E,\nu}^{\tau}\b Re\,\Omega_{E,\nu}^{\tau}$ {\it and}
$(\C\b (0\cup {\cal T}))\times ({\cal B}_{2\tau\sqrt{E}}\b {\cal L}_{\nu})$
{\it for} $\tau\in ]0,1]$, {\it where}
$${\cal T}=\{\lambda\in\C:\ |\lambda|=1\}.\eqno(3.9)$$

Actually, Lemma 1 follows from   properties (3.4) and the result that
formulas (3.7), (3.8) for $\lambda(k)$ and $k(\lambda)$ at fixed
$p\in {\cal B}_{2\sqrt{E}}\b {\cal L}_{\nu}$ give a diffeormorphism between
$\{k\in\C^3:\ k^2=E,\ p^2=2kp\}$ and $\C\b 0$. The latter result
follows from the fact (see [No4]) that the following formulas
$$\lambda={(k_1+ik_2)\over E^{1/2}},\ \ k_1=(\lambda+1/\lambda){E^{1/2}\over
2},\ \  k_2=(1/\lambda-\lambda){iE^{1/2}\over 2} $$
give a diffeomorphism between $\{k\in\C^2:\ k^2=E\}$, $E>0$, and $\C\b 0$.

Note that for $k$ and $\lambda$ of (3.7), (3.8) the following formulas
hold:
$$\eqalign{
&|Im\,k|={(E-p^2/4)^{1/2}\over 2}||\lambda|-1/|\lambda||,\cr
&|Re\,k|=\biggl({{E-p^2/4}\over 4}(|\lambda|+1/|\lambda|)^2+
p^2/4\biggr)^{1/2},\cr}\eqno(3.10)$$
where $(k,p)\in\Omega_{E,\nu}^{\tau}$, $(\lambda,p)\in (\C\b 0)\times
({\cal B}_{2\sqrt{E}}\b {\cal L}_{\nu})$, $E>0$, $\nu\in\S^2$,
$\tau\in ]0,1]$.

We consider $\lambda$, $p$ of Lemma 1 as coordinates on
$\Omega_{E,\nu}^{\tau}$ and on $\Omega_E^{\tau}$ for $E>0$, $\tau\in ]0,1]$,
 $d=3$.

\vskip 2 mm
{\bf 4. $\bar\pa$-equation for $H$ on
$\Omega_E^{\tau}\b Re\,\Omega_E^{\tau}$ for $E>0$, $\tau\in ]0,1[$ and
$d=3$}

\vskip 2 mm
{\bf Lemma 2.}
{\it Let} $\nu\in\S^2$ {\it and let} $\theta$, $\omega$ {\it satisfy}
(3.4), (3.5a). {\it Let  assumptions} (2.4), (2.18) {\it be fulfilled. Then}
$$\eqalign{
&{\pa\over \pa\bar\lambda}H(k(\lambda,p,E),p)=\cr
&-{\pi\over 4}\int\limits_{-\pi}^{\pi}\biggl((E-p^2/4)^{1/2}
{sgn\,(|\lambda|^2-1)(|\lambda|^2+1)\over \bar\lambda |\lambda|}
(\cos\,\v-1)-|p|{1\over \bar\lambda}\sin\,\v\biggr)\times\cr
&H(k(\lambda,p,E),-\xi(\lambda,p,E,\v))
H(k(\lambda,p,E)+\xi(\lambda,p,E,\v),p+\xi(\lambda,p,E,\v))d\v \cr}
\eqno(4.1)$$
{\it for} $\lambda\in\C\b ({\cal T}\cup 0)$, $p\in {\cal B}_{2\tau\sqrt{E}}\b
{\cal L}_{\nu}$, $\tau\in ]0,1]$, {\it where} $\lambda$ {\it and} $p$
{\it are coordinates of Lemma 1}, $k(\lambda,p,E)$ {\it is defined in}
(3.8) ({\it and also depends on} $\nu$, $\theta$, $\omega$),
$$\eqalignno{
&\xi(\lambda,p,E,\v)=Re\,k(\lambda,p,E)(\cos\v-1)+k^{\perp}(\lambda,p,E)
\sin\v,&(4.2)\cr
&k^{\perp}(\lambda,p,E)={Im\,k(\lambda,p,E)\times Re\,k(\lambda,p,E)\over
|Im\,k(\lambda,p,E)|}.&(4.3)\cr}$$

Proof of Lemma 2 is given in Section 9.
In this proof we deduce (4.1) from (2.22).

Note that on the left-hand side of (4.1)
$$(k(\lambda,p,E),p)\in\Omega^{\tau}_{E,\nu}\b Re\,\Omega^{\tau}_{E,\nu},$$
whereas on the right-hand side of (4.1)
$$\eqalign{
&(k(\lambda,p,E),-\xi(\lambda,p,E,\v))\in\Omega_E\b Re\,\Omega_E,\cr
&(k(\lambda,p,E)+
\xi(\lambda,p,E,\v),p+\xi(\lambda,p,E,\v))\in
\Omega_E\b Re\,\Omega_E,\cr}$$
 but not necessarily
$$\eqalign{
&(k(\lambda,p,E),-\xi(\lambda,p,E,\v))\in\Omega_E^{\tau}\b
Re\,\Omega_E^{\tau},\cr
&(k(\lambda,p,E)+
\xi(\lambda,p,E,\v),p+\xi(\lambda,p,E,\v))\in
\Omega_E^{\tau}\b Re\,\Omega_E^{\tau}\cr}$$
for $\lambda\in\C\b ({\cal T}\cup 0)$, $p\in {\cal B}_{2\tau\sqrt{E}}\b
{\cal L}_{\nu}$, $\tau\in ]0,1]$, $\v\in ]-\pi,\pi [$.

Therefore, consider $\chi_rH$, where $\chi_r$ denotes the multiplication
operator by the function $\chi_r(p)$, where
$$\chi_r(p)=1\ \ {\rm for}\ \ |p|<r,\ \ \chi_r(p)=0\ \ {\rm for}\ \
|p|\ge r,\ \ {\rm where}\ \ p\in\R^3,\ r>0.\eqno(4.4)$$
Note that
$$\eqalign{
&\chi_{2\tau\sqrt{E}}H(k,p)=H(k,p)\ \ {\rm for}\ \ (k,p)\in\Omega_E^{\tau}\b
Re\,\Omega_E^{\tau},\cr
&\chi_{2\tau\sqrt{E}}H(k,p)=0\ \ {\rm for}\ \ (k,p)\in(\Omega_E\b
Re\,\Omega_E)\b\Omega_E^{\tau},\cr}\eqno(4.5)$$
where $E>0$, $\tau\in ]0,1]$.

Further, (under the assumptions of Lemma 1) consider
$$\eqalign{
&\{U_1,U_2\}(\lambda,p,E)=\cr
&-{\pi\over 4}\int\limits_{-\pi}^{\pi}\biggl((E-p^2/4)^{1/2}
{sgn\,(|\lambda|^2-1)(|\lambda|^2+1)\over \bar\lambda |\lambda|}
(\cos\,\v-1)-|p|{1\over \bar\lambda}\sin\,\v\biggr)\times\cr
&U_1(k(\lambda,p,E),-\xi(\lambda,p,E,\v))
U_2(k(\lambda,p,E)+\xi(\lambda,p,E,\v),p+\xi(\lambda,p,E,\v))d\v,\cr}
\eqno(4.6)$$
where $U_1$ and $U_2$ are test functions on $\Omega_E\b Re\,\Omega_E$,
$k(\lambda,p,E)$ and $\xi(\lambda,p,E,\v)$ are defined by (3.8), (4.2),
$\lambda\in\C\b ({\cal T}\cup 0)$, $p\in {\cal B}_{2\tau\sqrt{E}}\b
{\cal L}_{\nu}$, $\tau\in ]0,1]$. Now we can write (4.1) as
$$
{\pa\over \pa\bar\lambda}\chi_{2\tau\sqrt{E}}H(k(\lambda,p,E),p)=
\{H,H\}(\lambda,p,E)=
\{\chi_{2\tau\sqrt{E}}H,\chi_{2\tau\sqrt{E}}H\}(\lambda,p,E)+
R_{E,\tau}(\lambda,p),\eqno(4.7)$$
$$\eqalign{
&R_{E,\tau}(\lambda,p)=\cr
&\{(1-\chi_{2\tau\sqrt{E}})H,\chi_{2\tau\sqrt{E}}H\}(\lambda,p,E)+
\{\chi_{2\tau\sqrt{E}}H,(1-\chi_{2\tau\sqrt{E}})H\}(\lambda,p,E)+ \cr
&\{(1-\chi_{2\tau\sqrt{E}})H,(1-\chi_{2\tau\sqrt{E}})H\}(\lambda,p,E),
\cr}\eqno(4.8)$$
where
$\lambda\in\C\b ({\cal T}\cup 0)$, $p\in {\cal B}_{2\tau\sqrt{E}}\b
{\cal L}_{\nu}$, $\tau\in ]0,1]$. Thus, (4.7) is a $\bar\pa$-equation
for $H$ on $\Omega_E^{\tau}\b Re\,\Omega_E^{\tau}$ up to the remainder
$R_{E,\tau}$. Let us estimate $R_{E,\tau}$  under  assumptions (2.4), (2.18)
for $d=3$, $\mu\ge 2$ (see estimate (4.14) given below).

Let
$$|||U|||_{E,\mu}=\sup_{(k,p)\in\Omega_E\b Re\,\Omega_E}
(1+|p|)^{\mu}\,|U(k,p)| \eqno(4.9)$$
for $U\in L^{\infty}(\Omega_E\b Re\,\Omega_E)$, $E>0$, $\mu\ge 0$.

\vskip 2 mm
{\bf Lemma 3.}
{\it Let the assumptions of Lemma 1 be fulfilled. Let}
$U_1,U_2\in L^{\infty}(\Omega_E\b Re\,\Omega_E)$,
$|||U_1|||_{E,\mu}<\infty$, $|||U_2|||_{E,\mu}<\infty$
 {\it for some} $\mu\ge 2$.
{\it Then}:
$$\{U_1,U_2\}(\cdot,\cdot,E)\in L^{\infty}((\C\b ({\cal T}\cup 0))\times
({\cal B}_{2\tau\sqrt{E}}\b {\cal L}_{\nu})),\ \ \tau\in ]0,1[,
\eqno(4.10)$$
{\it and}
$$|\{U_1,U_2\}(\lambda,p,E)|\le {c_4(\mu,\tau,E)|||U_1|||_{E,\mu}
|||U_2|||_{E,\mu}\over (1+|p|)^{\mu}(1+|\lambda|^2)},\eqno(4.11)$$
$\lambda\in\C\b ({\cal T}\cup 0)$,
$p\in {\cal B}_{2\tau\sqrt{E}}\b {\cal L}_{\nu}$, $\tau\in ]0,1[$,
{\it where} $c_4$ {\it is some positive constant such that}
$$c_4(\mu,\tau,E)\le {c_4^{\prime}(\mu)\over \sqrt{z}}
\bigl(1+{1\over \sqrt{z}}\bigr)+
{c_4^{\prime\prime}(\mu)\over \sqrt{E}\min\,(1,2\sqrt{z})},\ \
z={{1-\tau^2}\over 4\tau^2},\eqno(4.12)$$
{\it for some positive constants} $c_4^{\prime}$ {\it and}
$c_4^{\prime\prime}$.

Proof of Lemma 3 is given in Section 10.

\vskip 2 mm
{\bf Remark 1.}
Using the proof (given in Section 10) of (4.10) one can see also that
the variations of $U_1$, $U_2$ on the sets of zero measure in
$\Omega_E\b Re\,\Omega_E$ imply variations of $\{U_1,U_2\}(\cdot,\cdot,E)$
on sets of zero measure, only, in
$(\C\b ({\cal T}\cup 0))\times
({\cal B}_{2\tau\sqrt{E}}\b {\cal L}_{\nu})$.

Under assumptions (2.4), (2.18) for $d=3$, due to (2.30), (2.31), (4.4)
we have that
$$\eqalignno{
&|||H|||_{E,\mu}\le 2^{\mu/2}(1-\eta)^{-1}
N,\
|||\chi_{2\tau\sqrt{E}}H|||_{E,\mu}\le   2^{\mu/2}(1-\eta)^{-1}
N,&(4.13a)\cr
&|||(1-\chi_{2\tau\sqrt{E}})H|||_{E,\mu_0}\le 2^{\mu/2}(1-\eta)^{-1}
(1+2\tau\sqrt{E})^{\mu_0-\mu}
N,&(4.13b)\cr}$$
where $\eta$ is given by (2.31),  $\tau\in ]0,1]$,
$0\le\mu_0\le\mu$.

Under  assumptions (2.4), (2.18), $d=3$, $\mu\ge 2$, using estimates (4.13a),
(4.13b) and Lemma 3 we obtain the following estimate for
$R_{E,\tau}$:
$$|R_{E,\tau}(\lambda,p)|\le {c_4(\mu_0,\tau,E)2^{\mu}
N^2\over  (1-\eta)^2
(1+2\tau\sqrt{E})^{\mu-\mu_0}(1+|p|)^{\mu_0}(1+|\lambda|^2)}
\biggl(2+{1\over (1+2\tau\sqrt{E})^{\mu-\mu_0}}\biggr),\eqno(4.14)$$
where $\lambda\in\C\b ({\cal T}\cup 0)$,
$p\in {\cal B}_{2\tau\sqrt{E}}\b {\cal L}_{\nu}$, $\tau\in ]0,1[$,
$2\le\mu_0\le\mu$.

Estimate (4.14) for sufficiently great $\mu-\mu_0$ and formulas (4.12),
(2.31)
show, in particular, that $R_{E,\tau}$ rapidly vanishes when $\tau\sqrt{E}$
increases for $0<\tau\le\tau_1$ and fixed $\tau_1<1$.

\vskip 2 mm
{\bf 5. Approximate finding $H$ on $\Omega_E^{\tau}\b Re\,\Omega_E^{\tau}$
from $H_{\pm}$ on  $Re\,\Omega_E^{\tau}$  for}

{\bf $E>0$, $\tau\in ]0,1[$ and $d=3$}

\vskip 2 mm
Our next purpose is to obtain an integral equation for approximate finding
$H$ on $\Omega_E^{\tau}\b Re\,\Omega_E^{\tau}$ from $H_{\pm}$ on
$Re\,\Omega_E^{\tau}$ for $E>0$, $\tau\in ]0,1[$, $d=3$, where
$$H_{\pm}(k,p)=H_{\gamma^{\pm}(k,p)}(k,p),\ \
\gamma^{\pm}(k,p)=\pm\,{p\times (k-p/2)\over |p|\,|k-p/2|},\eqno(5.1)$$
$(k,p)\in Re\,\Omega_E^1$, $|p|\ne 0$, $E>0$, $d=3$ (where $H_{\gamma}$
is defined by means of (2.9)). Note that
$$(k-p/2)p=0,\ \ |k-p/2|=(E-p^2/4)^{1/2}\ \ {\rm for}\ \
(k,p)\in Re\,\Omega_E^1,\ \ E>0.\eqno(5.2)$$
We will give the aforementioned integral equation in the coordinates
$\lambda$, $p$ of Lemma 1 under  assumptions (3.5a). Note that in the
coordinates $\lambda$, $p$ of Lemma 1 under  assumption (3.5a) the
following formulas hold:
$$\eqalignno{
&H_{\pm}(k(\lambda,p,E),p)=H(k(\lambda(1\mp 0),p,E),p),&(5.3)\cr
&\gamma^{\pm}(k(\lambda,p,E),p)=\pm\,\biggl({-i\over 2}
\biggl({1\over \lambda}-\lambda\biggr)\theta(p)+{1\over 2}
\biggl(\lambda+{1\over \lambda}\biggr)\omega(p)\biggr),&(5.4)\cr}$$
where $\lambda\in {\cal T}$, $p\in {\cal B}_{2\sqrt{E}}\b {\cal L}_{\nu}$.

Consider
$${\cal D}_+=\{\lambda\in\C:\ |\lambda|<1\},\ \
{\cal D}_-=\{\lambda\in\C:\ |\lambda|>1\}.\eqno(5.5)$$
We will use the following formulas:
$$\eqalignno{
&u_+(\lambda)={1\over 2\pi i}\int\limits_{\cal T}u_+(\zeta){d\zeta\over
{\zeta-\lambda}}-{1\over \pi}\int\!\!\!\int_{{\cal D}_+}
{\pa u_+(\zeta)\over \pa\bar\zeta}{d\,Re\,\zeta\,d\,Im\,\zeta\over
{\zeta-\lambda}},\ \ \lambda\in {\cal D}_+\b 0,&(5.6a)\cr
&u_-(\lambda)=-{1\over 2\pi i}\int\limits_{\cal T}
u_-(\zeta){\lambda d\zeta\over
\zeta(\zeta-\lambda)}-{1\over \pi}\int\!\!\!\int_{{\cal D}_-}
{\pa u_-(\zeta)\over \pa\bar\zeta}{\lambda\,d\,Re\,\zeta\,d\,Im\,\zeta\over
\zeta(\zeta-\lambda)},\ \ \lambda\in {\cal D}_,&(5.6b)\cr}$$
where $u_+(\lambda)$ is continuous and bounded for $0<|\lambda|\le 1$,
$\pa u_+(\lambda)/\pa\bar\lambda$ is bounded for $0<|\lambda|<1$,
$u_-(\lambda)$ is continuous and bounded for $|\lambda|\ge 1$,
$\pa u_-(\lambda)/\pa\bar\lambda$ is bounded for $|\lambda|>1$ and
$\pa u_-(\lambda)/\pa\bar\lambda=O(|\lambda|^{-2})$ as $|\lambda|\to\infty$
(and where the integrals along the circle ${\cal T}$ are taken in the
counter-clockwise direction). Note that the aforementioned assumptions
on $u_{\pm}$ in (5.6) can be somewhat weakened. Formulas (5.6)
follow from the well-known Cauchy-Green formula
$$u(\lambda)={1\over 2\pi i}\int\limits_{\pa {\cal D}}u(\zeta){d\zeta\over
{\zeta-\lambda}}-{1\over \pi}\int\!\!\!\int_{\cal D}{\pa u(\zeta)\over
\pa\bar\zeta}{d\,Re\,\zeta\,d\,Im\,\zeta\over {\zeta-\lambda}},\ \
\lambda\in {\cal D},\eqno(5.7)$$
where ${\cal D}$ is a bounded open domain in $\C$ with sufficiently regular
boundary $\pa {\cal D}$ and $u$ is a sufficiently regular function in
${\bar {\cal D}}={\cal D}\cup\pa {\cal D}$. Assuming that $\lambda$, $p$
are the coordinates of Lemma 1 under  assumption (3.5a), consider
$$\eqalignno{
&H(\lambda,p,E)=H(k(\lambda,p,E),p),\ \ \lambda\in\C\b ({\cal T}\cup 0),
&(5.8)\cr
&H_{\pm}(\lambda,p,E)=H_{\pm}(k(\lambda,p,E),p),\ \ \lambda\in {\cal T},
&(5.9a)\cr
&H_+(\lambda,p,E)=H(k(\lambda,p,E),p),\ \ \lambda\in {\cal D}_+\b 0,
&(5.9b)\cr
&H_-(\lambda,p,E)=H(k(\lambda,p,E),p),\ \ \lambda\in {\cal D}_-,
&(5.9c)\cr}$$
where $p\in {\cal B}_{2\sqrt{E}}\b {\cal L}_{\nu}$ and
$H_{\pm}(k(\lambda,p,E),p)$ in (5.9a) are defined using (5.1) or (5.3).
One can show that, under  assumptions (2.4), (2.18), $d=3$, and for
$p\in {\cal B}_{2\sqrt{E}}\b {\cal L}_{\nu}$, $H_+(\lambda,p,E)$ is
continuous for $0<|\lambda|\le 1$ and $H_-(\lambda,p,E)$ is continuous for
$|\lambda|\ge 1$. Further, using (2.30), (4.7), (4.8),
(4.13) and Lemma 3 one
can see that, under  assumptions (2.4), (2.18), where $d=3$, $\mu\ge 2$, and
for
$p\in {\cal B}_{2\sqrt{E}}\b {\cal L}_{\nu}$, $H_{\pm}(\lambda,p,E)$
satisfy  the assumptions for $u_{\pm}(\lambda)$ mentioned in (5.6) and
(therefore)  formulas (5.6) hold for
$u_{\pm}(\lambda)=H_{\pm}(\lambda,p,E)$.

\vskip 2 mm
{\bf Proposition 1.}
{\it Let  assumptions} (2.4), (2.18), {\it where} $d=3$ $\mu\ge 2$, {\it be
fulfilled. Let} $\lambda$, $p$ {\it be the coordinates of Lemma 1 under
assumption} (3.5a). {\it Let}
$H(\lambda,p,E)$ {\it be defined by} (5.8). {\it Then}
$H_{E,\tau}=H(\lambda,p,E)$ {\it as a function of}
$\lambda\in\C\b ({\cal T}\cup 0)$ {\it and}
$p\in {\cal B}_{2\tau\sqrt{E}}\b {\cal L}_{\nu}$, {\it where} $\tau\in ]0,1[$,
{\it satisfies the following nonlinear integral equation}
$$H_{E,\tau}=H_{E,\tau}^0+M_{E,\tau}(H_{E,\tau})+Q_{E,\tau},\ \ \tau\in
]0,1[,\eqno(5.10)$$
where:
$$\eqalignno{
&H_{E,\tau}^0(\lambda,p)={1\over 2\pi i}\int\limits_{\cal T}H_+(\zeta,p,E)
{d\zeta\over {\zeta-\lambda}},\ \lambda\in {\cal D}_+\b 0,\
p\in {\cal B}_{2\tau\sqrt{E}}\b {\cal L}_{\nu},&(5.11a)\cr
&H_{E,\tau}^0(\lambda,p)=-{1\over 2\pi i}\int\limits_{\cal T}H_-(\zeta,p,E)
{\lambda d\zeta\over \zeta(\zeta-\lambda)},\ \lambda\in {\cal D}_-,\
p\in {\cal B}_{2\tau\sqrt{E}}\b {\cal L}_{\nu},&(5.11b)\cr}$$
{\it where} $H_{\pm}(\lambda,p,E)$ {\it are defined by} (5.9a);
$$\eqalign{
&M_{E,\tau}(U)(\lambda,p)=M_{E,\tau}^+(U)(\lambda,p)=\cr
&-{1\over \pi}\int\!\!\!\int_{{\cal D}_+}(U,U)_{E,\tau}(\zeta,p)
{d\,Re\,\zeta\,d\,Im\,\zeta\over {\zeta-\lambda}},\ \
\lambda\in {\cal D}_+\b 0,\
p\in {\cal B}_{2\tau\sqrt{E}}\b {\cal L}_{\nu},\cr}\eqno(5.12a)$$
$$\eqalign{
&M_{E,\tau}(U)(\lambda,p)=M_{E,\tau}^-(U)(\lambda,p)=\cr
&-{1\over \pi}\int\!\!\!\int_{{\cal D}_-}(U,U)_{E,\tau}(\zeta,p)
{\lambda d\,Re\,\zeta\,d\,Im\,\zeta\over \zeta(\zeta-\lambda)},\ \
\lambda\in {\cal D}_-,\
p\in {\cal B}_{2\tau\sqrt{E}}\b {\cal L}_{\nu},\cr}\eqno(5.12b)$$
$$\eqalign{
&(U_1,U_2)_{E,\tau}(\lambda,p)=\cr
&-{\pi\over 4}\int\limits_{-\pi}^{\pi}\biggl((E-p^2/4)^{1/2}
{sgn\,(|\lambda|^2-1)(|\lambda|^2+1)\over \bar\lambda |\lambda|}
(\cos\,\v-1)-|p|{1\over \bar\lambda}\sin\,\v\biggr)\times\cr
&U_1(z_1(\lambda,p,E,\v),-\xi(\lambda,p,E,\v))
U_2(z_2(\lambda,p,E,\v),p+\xi(\lambda,p,E,\v))\times\cr
&\chi_{2\tau\sqrt{E}}(\xi(\lambda,p,E,\v))
\chi_{2\tau\sqrt{E}}(p+\xi(\lambda,p,E,\v))d\v,\ \
\lambda\in\C\b ({\cal T}\cup 0),\ \
p\in {\cal B}_{2\tau\sqrt{E}}\b {\cal L}_{\nu},\cr}\eqno(5.13)$$
{\it where} $U$, $U_1$, $U_2$ {\it are test functions on}
$(\C\b ({\cal T}\cup 0))\times
({\cal B}_{2\tau\sqrt{E}}\b {\cal L}_{\nu})$,
$$\eqalign{
&z_1(\lambda,p,E,\v)={k(\lambda,p,E)
(\theta(-\xi(\lambda,p,E,\v))+i\omega(-\xi(\lambda,p,E,\v)))\over
(E-|\xi(\lambda,p,E,\v)|^2/4)^{1/2}},\cr
&z_2(\lambda,p,E,\v)={(k(\lambda,p,E)+\xi(\lambda,p,E,\v))
(\theta(p+\xi(\lambda,p,E,\v))+i\omega(p+\xi(\lambda,p,E,\v)))\over
(E-|p+\xi(\lambda,p,E,\v)|^2/4)^{1/2}}\cr}\eqno(5.14)$$
{\it for}  $\lambda\in\C\b ({\cal T}\cup 0)$,
$p\in {\cal B}_{2\sqrt{E}}\b {\cal L}_{\nu}$, $\v\in [-\pi,\pi]$, {\it and
where} $k(\lambda,p,E)$, $\xi(\lambda,p,E,\v)$ {\it are defined by} (3.8),
(4.2), $\theta$, $\omega$ {\it are the vector functions of} (3.4), (3.5a);
$$\eqalignno{
&Q_{E,\tau}(\lambda,p)=
-{1\over \pi}\int\!\!\!\int_{{\cal D}_+}R_{E,\tau}(\zeta,p)
{d\,Re\,\zeta\,d\,Im\,\zeta\over {\zeta-\lambda}},\ \
\lambda\in {\cal D}_+\b 0,\
p\in {\cal B}_{2\tau\sqrt{E}}\b {\cal L}_{\nu},&(5.15a)\cr
&Q_{E,\tau}(\lambda,p)=
-{1\over \pi}\int\!\!\!\int_{{\cal D}_-}R_{E,\tau}(\zeta,p)
{\lambda d\,Re\,\zeta\,d\,Im\,\zeta\over \zeta(\zeta-\lambda)},\ \
\lambda\in {\cal D}_-,\
p\in {\cal B}_{2\tau\sqrt{E}}\b {\cal L}_{\nu},&(5.15b)\cr}$$
{\it where} $R_{E,\tau}$ {\it is defined by} (4.8).

Proposition 1 follows from formulas (5.6) for $u_{\pm}(\lambda)=
H_{\pm}(\lambda,p,E)$ (defined by (5.9)), the $\bar\pa$- equation (4.7),
formula (4.6) and Lemma 1. We consider (5.10) as an integral equation
for finding $H_{E,\tau}$ from $H_{E,\tau}^0$ with unknown remainder
$Q_{E,\tau}$. Thus, actually, we consider (5.10) as an approximate
equation for finding $H_{E,\tau}$ from $H_{E,\tau}^0$.

Let
$$|||U|||_{E,\tau,\mu}=\sup_{\scriptstyle \lambda\in\C\b ({\cal T}\cup 0),
\atop\scriptstyle p\in {\cal B}_{2\tau\sqrt{E}}\b {\cal L}_{\nu}}
(1+|p|)^{\mu}|U(\lambda,p)| \eqno(5.16)$$
for $U\in L^{\infty}((\C\b ({\cal T}\cup 0))\times
({\cal B}_{2\tau\sqrt{E}}\b {\cal L}_{\nu}))$, where $E>0$, $\tau\in ]0,1[$,
$\nu\in\S^2$, $\mu>0$.

\vskip 2 mm
{\bf Lemma 4.}
{\it Let} $E>0$, $\nu\in\S^2$, $\tau\in ]0,1[$, $\mu\ge 2$. {\it Let}
$M_{E,\tau}$ {\it be defined by} (5.12) ({\it where} $\lambda$, $p$ {\it
are the coordinates of Lemma 1 under assumption} (3.5a)).
 {\it Let}
$U_1,U_2\in L^{\infty}((\C\b ({\cal T}\cup 0))\times
({\cal B}_{2\tau\sqrt{E}}\b {\cal L}_{\nu}))$,
$|||U_1|||_{E,\tau,\mu}<+\infty$,  $|||U_2|||_{E,\tau,\mu}<+\infty$.
{\it Then}
$$\eqalignno{
&M_{E,\tau}(U_j)\in L^{\infty}((\C\b ({\cal T}\cup 0))\times
({\cal B}_{2\tau\sqrt{E}}\b {\cal L}_{\nu})),\ j=1,2,&(5.17)\cr
&|||M_{E,\tau}(U_j)|||_{E,\tau,\mu}\le c_5c_4(\mu,\tau,E)
(|||U_j|||_{E,\tau,\mu})^2,\ j=1,2,&(5.18)\cr}$$
$$\eqalign{
&|||M_{E,\tau}(U_1)-M_{E,\tau}(U_2)|||_{E,\tau,\mu}\le  \cr
&\le c_5c_4(\mu,\tau,E)
(|||U_1|||_{E,\tau,\mu}+|||U_2|||_{E,\tau,\mu})
|||U_1-U_2|||_{E,\tau,\mu},\cr}\eqno(5.19)$$
{\it where} $c_4(\mu,\tau,E)$ {\it is the constant of} (4.11) {\it and}
$$c_5=\sup_{\lambda\in {\cal D}_+}{1\over \pi}\int\!\!\!\int_{{\cal D}_+}
{d\,Re\,\zeta\,d\,Im\,\zeta\over (1+|\zeta|^2)|\zeta-\lambda|}=
\sup_{\lambda\in {\cal D}_-}{1\over \pi}\int\!\!\!\int_{{\cal D}_-}
{|\lambda| d\,Re\,\zeta\,d\,Im\,\zeta\over
|\zeta|(1+|\zeta|^2)|\zeta-\lambda|}.\eqno(5.20)$$
Formulas (5.17), (5.18) follow from formulas (5.12) (5.13), Lemma 1,
formula (4.6) and Lemma 3. To obtain (5.19) we use also the formula
$$(U_1,U_1)_{E,\tau}-(U_2,U_2)_{E,\tau}=
(U_1-U_2,U_1)_{E,\tau}+(U_2,U_1-U_2)_{E,\tau}.\eqno(5.21)$$

\vskip 2 mm
{\bf Lemma 5.}
{\it Let  assumptions} (2.4), (2.18) {\it be fulfilled and} $\eta$
{\it be given by} (2.31), {\it where} $d=3$. {\it Let}
$$\delta={c_3(0,\mu,\sigma,3)N\over (1-\eta)E^{\sigma/2}}<1.
\eqno(5.22)$$
{\it Let}
$H_{\pm}(\lambda,p,E)$ {\it and}
$H_{E,\tau}^0(\lambda,p)$ {\it be defined by} (5.9a), (5.11),
$\tau\in ]0,1[$. {\it Then}
$$
|H_{\pm}(\lambda,p,E)-{\hat v}(p)|\le {\eta N\over (1-\eta)
(1+p^2)^{\mu/2}},\eqno(5.23)$$
$$
|H_{\pm}(\lambda,p,E)-H_{\pm}(\lambda^{\prime},p,E)|\le
{c_6(\alpha,\mu,\sigma,\beta)|\lambda-\lambda^{\prime}|^{\beta}N^2\over
(1-\eta)^2(1-\delta)(1+p^2)^{\mu/2}},\eqno(5.24)$$
{\it where} $\lambda,\lambda^{\prime}\in {\cal T}$,
$|\lambda-\lambda^{\prime}|\le (E-p^2/4)^{-1/2}$,
$0<\beta\le \min\,(\alpha,\sigma,1/2)$,
$p\in {\cal B}_{2\tau\sqrt{E}}\b {\cal L}_{\nu}$;
$$\eqalignno{
&H_{E,\tau}^0\in L^{\infty}((\C\b ({\cal T}\cup 0))\times
({\cal B}_{2\tau\sqrt{E}}\b {\cal L}_{\nu})),&(5.25a)\cr
&|||H_{E,\tau}^0|||_{E,\tau,\mu}\le 2^{\mu/2}N+
{c_7(\alpha,\mu,\sigma,\beta)N^2\over
(1-\eta)^2(1-\delta)E^{\beta/2}},\ \ 0<\beta<\min\,(\alpha,\sigma,1/2).
&(5.25b)\cr}$$

Lemma 5 is proved in Section 11.

\vskip 2 mm
{\bf Lemma 6.}
{\it Let assumptions} (2.4), (2.18) {\it be fulfilled and}
 $\eta$ {\it be given by} (2.31), {\it where} $d=3$, $\mu\ge 2$. {\it Let}
$Q_{E,\tau}$ {\it be defined by} (5.15), $\tau\in ]0,1[$. {\it Then}
$$\eqalignno{
&Q_{E,\tau}\in L^{\infty}((\C\b ({\cal T}\cup 0))\times
({\cal B}_{2\tau\sqrt{E}}\b {\cal L}_{\nu})),&(5.26)\cr
&|||Q_{E,\tau}|||_{E,\tau,\mu_0}\le
{3c_5c_4(\mu_0,\tau,E)2^{\mu}N^2\over
(1-\eta)^2(1+2\tau\sqrt{E})^{\mu-\mu_0}},\ 2\le\mu_0\le\mu.&(5.27)\cr}$$

Lemma 6 follows from (5.15), (4.14), (5.20).

Lemmas 4,5,6 show that, under the assumptions of Proposition 1, the
non-linear integral equation (5.10) for unknown $H_{E,\tau}$ can be
analyzed for $H_{E,\tau}^0$, $Q_{E,\tau}$, $H_{E,\tau}\in
L^{\infty}((\C\b ({\cal T}\cup 0))\times
({\cal B}_{2\tau\sqrt{E}}\b {\cal L}_{\nu}))$ using the norm
$|||\cdot|||_{E,\tau,\mu_0}$, where $2\le\mu_0\le\mu$.

Consider the equation
$$U=U^0+M_{E,\tau}(U),\ \ E>0,\ \ \tau\in ]0,1[, \eqno(5.28)$$
for unknown $U$. Actually, under the assumptions of Proposition 1, we
suppose that $U^0=H_{E,\tau}^0+Q_{E,\tau}$ or consider $U^0$ as an
approximation to $H_{E,\tau}^0+Q_{E,\tau}$.

\vskip 2 mm
{\bf Lemma 7.}
{\it Let} $E>0$, $\nu\in\S^2$, $\tau\in ]0,1[$, $\mu\ge 2$ {\it and}
$0<r< (2c_5c_4(\mu,\tau,E))^{-1}$. {\it Let} $M_{E,\tau}$ {\it be defined by}
(5.12) ({\it where} $\lambda$, $p$ {\it are the coordinates of Lemma 1
under assumption} (3.5a)). {\it Let}
$U^0\in L^{\infty}((\C\b ({\cal T}\cup 0))\times
({\cal B}_{2\tau\sqrt{E}}\b {\cal L}_{\nu}))$ {\it and}
$|||U^0|||_{E,\tau,\mu}\le r/2$. {\it Then equation} (5.28) {\it is
uniquely solvable for}
$U\in L^{\infty}((\C\b ({\cal T}\cup 0))\times
({\cal B}_{2\tau\sqrt{E}}\b {\cal L}_{\nu}))$, $|||U|||_{E,\tau,\mu}\le r$,
{\it and} $U$ {\it can be found by the method of successive approximations,
in addition}
$$|||U-M^n_{E,\tau,U^0}(0)|||_{E,\tau,\mu}\le
{(2c_5c_4(\mu,\tau,E)r)^n\over {1-2c_5c_4(\mu,\tau,E)r}}{r\over 2},\ \
n\in\N,\eqno(5.29)$$
{\it where} $M_{E,\tau,U^0}$ {\it denotes the map} $U\to U^0+M_{E,\tau}(U)$.

Lemma 7 is proved in Section 12.

\vskip 2 mm
{\bf Lemma 8.}
{\it Let the assumptions of Lemma 7 be fulfilled. Let also}
${\tilde U}^0\in L^{\infty}((\C\b ({\cal T}\cup 0))\times
({\cal B}_{2\tau\sqrt{E}}\b {\cal L}_{\nu}))$,
$|||{\tilde U}^0|||_{E,\tau,\mu}\le r/2$, {\it and} $\tilde U$ {\it denote
the solution of} (5.28) {\it with} $U^0$ {\it replaced by}
${\tilde U}^0$, {\it where}
${\tilde U}\in L^{\infty}((\C\b ({\cal T}\cup 0))\times
({\cal B}_{2\tau\sqrt{E}}\b {\cal L}_{\nu}))$,
$|||{\tilde U}|||_{E,\tau,\mu}\le r$. {\it Then}
$$|||U-\tilde U|||_{E,\tau,\mu}\le (1-2c_5c_4(\mu,\tau,E)r)^{-1}
|||U^0-{\tilde U}^0|||_{E,\tau,\mu}.\eqno(5.30)$$

Lemma 8 is proved in Section 12.

Proposition 1 and Lemmas 4,5,6,7,8 imply, in particular, the following
result.

\vskip 2 mm
{\bf Proposition 2.}
{\it Let assumptions} (2.4), (2.18), (5.22) {\it be fulfilled, where}
$d=3$, $\mu\ge 2$. {\it Let} $\nu\in\S^2$, $\tau\in ]0,1[$,
$2\le\mu_0\le\mu$, $0<\beta<\min\,(\alpha,\sigma,1/2)$ {\it and}
$$\max\,(r_1,r_2)< (2c_5c_4(\mu_0,\tau,E))^{-1},\eqno(5.31)$$
{\it where}
$$2^{\mu/2}N+
{c_7(\alpha,\mu,\sigma,\beta)N^2\over (1-\eta)^2(1-\delta)E^{\beta/2}}+
{3c_5c_4(\mu_0,\tau,E)2^{\mu}N^2\over
(1-\eta)^2(1+2\tau\sqrt{E})^{\mu-\mu_0}}= {r_1\over 2},\eqno(5.32a)$$
$${2^{\mu/2}N\over {1-\eta}}=r_2,\eqno(5.32b)$$
{\it and} $\eta$, $\delta$ {\it are given by} (2.31) (
$d=3$), (5.22).
{\it Let}
$$\max\,(r_1,r_2)\le r<(2c_5c_4(\mu_0,\tau,E))^{-1}.$$
{\it Then}
$$|||H_{E,\tau}-{\tilde H}_{E,\tau}|||_{E,\tau,\mu_0}\le
{3c_5c_4(\mu_0,\tau,E)2^{\mu}N^2\over
(1-2c_5c_4(\mu_0,\tau,E)r)
(1-\eta)(1+2\tau\sqrt{E})^{\mu-\mu_0}},\eqno(5.33)$$
{\it where} $H_{E,\tau}=H(\lambda,p,E)$ {\it as a function of}
$\lambda\in\C\b ({\cal T}\cup 0)$ {\it and}
$p\in {\cal B}_{2\tau\sqrt{E}}\b {\cal L}_{\nu}$ {\it (as in Proposition 1)
and} ${\tilde H}_{E,\tau}$ {\it is defined as the solution of}
$${\tilde H}_{E,\tau}=H^0_{E,\tau}+M_{E,\tau}({\tilde H}_{E,\tau}),\eqno(5.34)
$$
{\it where}
$|||{\tilde H}_{E,\tau}|||_{E,\tau,\mu_0}\le r$.

Using the definitions of $\eta$ and $\delta$ of (2.31), (5.22) we obtain that:
$$\eqalign{
&{\rm if}\ \ N\le c_8(\alpha,\mu,\mu_0,\sigma,E,\tau),\
\ {\rm then\ conditions}\cr
&(2.18), (5.22)\ \ {\rm and}\ \ (5.31)\ \ {\rm are\ fulfilled},
\cr}\eqno(5.35)$$
where $0<\alpha<1$, $2\le\mu_0\le\mu$, $0<\sigma<1$, $E\ge 1$,
$0<\tau<1$, $d=3$.

Due to (4.12), we have that:
$$c_4(\mu,\tau,E)\le\ep\ \ {\rm if}\  0<\tau\le\tau(\ep,\mu),\ \
E\ge E(\ep,\mu) \eqno(5.36)$$
for any arbitrary small $\ep>0$ and appropriate sufficiently small
$\tau(\ep,\mu)\in ]0,1[$ and sufficiently great $E(\ep,\mu)$, where
$\mu\ge 2$.

Using (5.37) and the definitions of $\eta$ and $\delta$ of (2.31), (5.22) we
obtain that:
$$\eqalign{
&{\rm if}\ \ 0<\tau\le\tau_1(\alpha,\mu,\mu_0,\sigma,N),
\ \ E\ge E_1(\alpha,\mu,\mu_0,\sigma,N),\ \
{\rm then\ conditions}\cr
&(2.18), (5.22)\ \ {\rm and}\ \ (5.31)\ \
{\rm are\ fulfilled},\cr}\eqno(5.37)$$
where $\tau_1$ and $E_1$ are appropriate constants such that
$\tau_1\in ]0,1[$ is sufficiently small and $E_1\ge 1$ is sufficiently great,
$0<\alpha<1$, $2\le\mu_0\le\mu$, $0<\sigma<1$, $d=3$.

As a corollary of Propositions 1,2 and property (5.37), we obtain the
following result.

\vskip 2 mm
{\bf Corollary 1.}
{\it Let assumptions} (2.4) {\it be fulfilled, where} $d=3$, $\mu\ge 2$,
{\it and} $\|\hat v\|_{\alpha,\mu}\le N$ {\it for some} $N>0$.
{\it Let}
$$0<\tau\le\tau_1(\alpha,\mu,\mu_0,\sigma,N),
\ \ E\ge E_1(\alpha,\mu,\mu_0,\sigma,N),$$
where $2\le\mu_0\le\mu$, $0<\sigma<1$. {\it Then}
$H_{\pm}(\lambda,p,E)$,
$\lambda\in {\cal T}$, $p\in {\cal B}_{2\tau\sqrt{E}}\b {\cal L}_{\nu}$,
{\it determines} ({\it via} (5.11), (5.34))
$H(\lambda,p,E)$, $\lambda\in\C\b ({\cal T}\cup 0)$,
$p\in {\cal B}_{2\tau\sqrt{E}}\b {\cal L}_{\nu}$, {\it up to}
$O(E^{-(\mu-\mu_0)/2})$ {\it in the norm} $|||\cdot|||_{E,\tau,\mu_0}$
 {\it as} $E\to +\infty$  ({\it due to estimate} (5.33), {\it where}
, {\it for example}, $r=\max\,(r_1,r_2)$).

\vskip 2 mm
{\bf 6. Finding
 $H_{\pm}$ on $Re\,\Omega_E$ and   $H^0_{E,\tau}$ on
$(\C\b ({\cal T}\cup 0))\times ({\cal B}_{2\tau\sqrt{E}}\b {\cal L}_{\nu})$
from $f$ on  ${\cal M}_E$}

Under assumptions (2.4), (2.18), due to (2.27), (2.28) we have that
$$\eqalignno{
&\|f\|_{C({\cal M}_E),\mu}\le g_1N,&(6.1)\cr
&\|f\|_{C^{\alpha}({\cal M}_E),\mu}\le g_2N,&(6.2)\cr
&g_1=(1-\eta)^{-1},\ \ g_2=c_2(\mu)(1-\eta)^{-1},&(6.3)\cr}$$
where $\eta$ is given by (2.31).

To find
 $H_{\pm}$ on $Re\,\Omega_E$ and   $H^0_{E,\tau}$ on
$(\C\b ({\cal T}\cup 0))\times ({\cal B}_{2\tau\sqrt{E}}\b {\cal L}_{\nu})$
from $f$ on  ${\cal M}_E$  (for $d=3$) we proceed from
(2.20), (5.1) and (5.9a), (5.11). Due to (2.20) we have that
$$h_{\gamma}(k,\cdot)=f(k,\cdot)+B_{\gamma}(k)h_{\gamma}(k,\cdot),\ \
\gamma\in\S^{d-1},\ \ k\in\S^{d-1}_{\sqrt{E}},\eqno(6.4)$$
where the operator $B_{\gamma}(k)$ is defined by (2.38).

Let us decompose $h_{\gamma}(k,l)$ as
$$\eqalignno{
&h_{\gamma}(k,l)=h^{(n)}_{\gamma}(k,l)+t^{(n)}_{\gamma}(k,l),&(6.5)\cr
&h^{(n)}_{\gamma}(k,\cdot)=\sum_{j=0}^n(B_{\gamma}(k))^jf(k,\cdot),&(6.6)\cr
&t^{(n)}_{\gamma}(k,\cdot)=(B_{\gamma}(k))^{n+1}f(k,\cdot)
+B_{\gamma}(k)t^{(n)}_{\gamma}(k,\cdot),&(6.7)\cr}$$
where $\gamma\in\S^{d-1},\ \ k,l\in\S^{d-1}_{\sqrt{E}}$, $n\in\N\cup 0$.

Further, using (2.8), (5.1), (5.11), (6.5) let us decompose
$H_{\pm}$ and $H^0_{E,\tau}$ as:
$$\eqalignno{
&H_{\pm}(k,p)=H^{(n)}_{\pm}(k,p)+T^{(n)}_{\pm}(k,p),&(6.8)\cr
&H^{(n)}_{\pm}(k,p)=h^{(n)}_{\gamma^{\pm}(k,p)}(k,k-p),\ \
T^{(n)}_{\pm}(k,p)=t^{(n)}_{\gamma^{\pm}(k,p)}(k,k-p),&(6.9)\cr}$$
where $n\in\N\cup 0$, $(k,p)\in Re\,\Omega^1_E$, $|p|\ne 0$, $E>0$;
$$H^0_{E,\tau}(\lambda,p)=
H^{0,n}_{E,\tau}(\lambda,p)+T^{0,n}_{E,\tau}(\lambda,p),\eqno(6.10)$$
where $n\in\N\cup 0$,
$\lambda\in\C\b ({\cal T}\cup 0)$, $p\in {\cal B}_{2\tau\sqrt{E}}\b
{\cal L}_{\nu}$, $E>0$, $\tau\in ]0,1[$, and
$H^{0,n}_{E,\tau}$, $T^{0,n}_{E,\tau}$ are defined by (5.11) with
$H_{\pm}(\zeta,p,E)$ replaced by  $H^{(n)}_{\pm}(k(\zeta,p,E),p)$ and
$T^{(n)}_{\pm}(k(\zeta,p,E),p)$, respectively, where $k(\lambda,p,E)$ is
defined in (3.8).

\vskip 2 mm
{\bf Lemma 9.}
{\it Let} $d=3$. {\it Let} $f$ {\it satisfy} (6.1), (6.2) {\it and}
$$\eqalign{
&\delta_1=c_3(0,\mu,\sigma,3)E^{-\sigma/2}g_1N<1,\cr
&\delta_2=c_3(\alpha,\mu,\sigma,3)E^{-\sigma/2}g_2N,\cr}\eqno(6.11)$$
{\it for some} $\alpha\in ]0,1[$, $\mu>1$, $\sigma\in ]0,\min\,(1,\mu-1)[$,
$E\ge 1$
{\it and some} $g_1$, $g_2$, $N\in ]0,+\infty [$, $g_1<g_2$.
{\it Then} (6.4), (6.7) {\it for fixed} $\gamma$, $k$ {\it and} $n$
 {\it are uniquely solvable for} $h_{\gamma}(k,\cdot),
t^{(n)}_{\gamma}(k,\cdot)\in
C^{\alpha}(\S^2_{\sqrt{E}})$ ({\it by the method of successive
approximations) and the following estimates hold:}
$$\eqalignno{
&|h_{\gamma}(k,l)|\le {g_1N\over
(1-\delta_1)(1+|k-l|^2)^{\mu/2}},&(6.12)\cr
&|h_{\gamma}(k,l)-h_{\gamma}(k,l^{\prime})|\le
{(1+c_2^{\prime\prime}(\mu)\delta_2)
g_2N|l-l^{\prime}|^{\alpha}\over
(1-\delta_1)(1+|k-l|^2)^{\mu/2}},&(6.13)\cr}$$
{\it where} $\gamma\in\S^2$, $k,l,l^{\prime}\in\S^2_{\sqrt{E}}$,
$|l-l^{\prime}|\le 1$;
$$\eqalign{
&|h_{\gamma}(k,l)-h_{\gamma^{\prime}}(k^{\prime},l)|\le
{g_2N|k-k^{\prime}|^{\alpha}\over
(1-\delta_1)(1+|k-l|^2)^{\mu/2}}+\cr
&{c_9(\beta,\mu)(g_1N)^2
|\gamma-\gamma^{\prime}|^{\beta}\over
(1-\delta_1)^2(1+|k-l|^2)^{\mu/2}},\cr}\eqno(6.14)$$
{\it where} $\gamma,\gamma^{\prime}\in\S^2$, $k,k^{\prime},l\in\S^2_{\sqrt{E}}
$, $\gamma k=\gamma^{\prime}k^{\prime}=0$,
$|\gamma-\gamma^{\prime}|\le 1$, $|k-k^{\prime}|\le 1$, $0<\beta<1/2$;
$$\eqalign{
&|H_{\pm}(k(\lambda,p,E),p)-H_{\pm}(k(\lambda^{\prime},p,E),p)|\le \cr
&{c_{10}(\mu)(1+\delta_2)
g_2N(E-p^2/4)^{\alpha/2}
|\lambda-\lambda^{\prime}|^{\alpha}\over (1-\delta_1)
(1+p^2)^{\mu/2}}+
{c_9(\beta,\mu))(g_1N)^2
|\lambda-\lambda^{\prime}|^{\beta}\over (1-\delta_1)^2
(1+p^2)^{\mu/2}},\cr}\eqno(6.15)$$
{\it where}  $\lambda,\lambda^{\prime}\in {\cal T}$,
$|\lambda-\lambda^{\prime}|\le (E-p^2/4)^{-1/2}$,
$p\in {\cal B}_{2\sqrt{E}}\b {\cal L}_{\nu}$, $k(\lambda,p,E)$
{\it is defined in} (3.8), $0<\beta<1/2$, {\it and}
$H_{\pm}(k,p)$ {\it are defined by} (5.1);
$$\eqalign{
&|H^0_{E,\tau}(\lambda,p)|\le\biggl(
{c_{11}g_1N(1+\ln\,E)\over {1-\delta_1}}+\cr
&{c_{12}(\alpha)c_{10}(\mu)(1+\delta_2)
g_2N\over {1-\delta_1}}+
{c_{12}(\beta)c_9(\beta,\mu)(g_1N)^2\over (1-\delta_1)^2E^{\beta/2}}
\biggr){1\over (1+p^2)^{\mu/2}},\cr}\eqno(6.16)$$
{\it where} $\lambda\in\C\b ({\cal T}\cup 0)$,
$p\in {\cal B}_{2\sqrt{E}}\b {\cal L}_{\nu}$, $\tau\in ]0,1[$,
{\it and} $H^0_{E,\tau}$ {\it is defined by} (5.11);

$$\eqalignno{
&|t^{(n)}_{\gamma}(k,l)|\le {\delta_1^{n+1}g_1N\over
(1-\delta_1)(1+|k-l|^2)^{\mu/2}},&(6.17)\cr
&|t^{(n)}_{\gamma}(k,l)-t^{(n)}_{\gamma}(k,l^{\prime})|\le
{c_2^{\prime\prime}(\mu)\delta_2\delta_1^ng_1N|l-l^{\prime}|^{\alpha}\over
(1-\delta_1)(1+|k-l|^2)^{\mu/2}},&(6.18)\cr}$$
{\it where} $n\in\N\cup 0$,
$\gamma\in\S^2$, $k,l,l^{\prime}\in\S^2_{\sqrt{E}}$, $|l-l^{\prime}|\le
1$;
$$\eqalign{
&|t^{(n)}_{\gamma}(k,l)-t^{(n)}_{\gamma^{\prime}}(k^{\prime},l)|\le
{\delta_1^{n+1}g_2N|k-k^{\prime}|^{\alpha}\over
(1-\delta_1)(1+|k-l|^2)^{\mu/2}}+\cr
&{c_9(\beta,\mu)\delta_1^n(1+n(1-\delta_1))(g_1N)^2
|\gamma-\gamma^{\prime}|^{\beta}\over
(1-\delta_1)^2(1+|k-l|^2)^{\mu/2}},\cr}\eqno(6.19)$$
{\it where} $n\in\N\cup 0$,
$\gamma,\gamma^{\prime}\in\S^2$, $k,k^{\prime},l\in\S^2_{\sqrt{E}}$,
$\gamma k=\gamma^{\prime} k^{\prime}=0$, $|\gamma-\gamma^{\prime}|\le 1$,
$|k-k^{\prime}|\le1$, $0<\beta<1/2$;
$$\eqalign{
&|T^{(n)}_{\pm}(k(\lambda,p,E),p)-T^{(n)}_{\pm}(k(\lambda^{\prime},p,E),p)|
\le\cr
&{(\delta_1+c_2(\mu)\delta_2)
\delta_1^ng_2N(E-p^2/4)^{\alpha/2}
|\lambda-\lambda^{\prime}|^{\alpha}\over (1-\delta_1)
(1+p^2)^{\mu/2}}+\cr
&{c_9(\beta,\mu))\delta_1^n(1+n(1-\delta_1))(g_1N)^2
|\lambda-\lambda^{\prime}|^{\beta}\over (1-\delta_1)^2
(1+p^2)^{\mu/2}},\cr}\eqno(6.20)$$
{\it where} $n\in\N\cup 0$,
$\lambda,\lambda^{\prime}\in {\cal T}$,
$|\lambda-\lambda^{\prime}|\le (E-p^2/4)^{-1/2}$,
$p\in {\cal B}_{2\sqrt{E}}\b {\cal L}_{\nu}$, $k(\lambda,p,E)$
{\it is defined in} (3.8), $0<\beta<1/2$, {\it and}
$T^{(n)}_{\pm}(k,p)$ {\it are defined in} (6.9);
$$\eqalign{
&|T^{0,n}_{E,\tau}(\lambda,p)|\le\biggl(
{c_{11}\delta_1g_1N(1+\ln\,E)\over {1-\delta_1}}+\cr
&{c_{12}(\alpha)(\delta_1+c_2(\mu)\delta_2)
g_2N\over {1-\delta_1}}+
{c_{12}(\beta)c_9(\beta,\mu)(1+n(1-\delta_1))
(g_1N)^2\over (1-\delta_1)^2E^{\beta/2}}
\biggr)\times\cr
&{\delta_1^n\over (1+p^2)^{\mu/2}},\cr}\eqno(6.21)$$
{\it where} $n\in\N\cup 0$,
$\lambda\in\C\b ({\cal T}\cup 0)$,
$p\in {\cal B}_{2\sqrt{E}}\b {\cal L}_{\nu}$, $\tau\in ]0,1[$,
{\it and} $T^{0,n}_{E,\tau}$ {\it is defined in} (6.10).

Note that in Lemma 9 we do not suppose that $f$ is the scattering
amplitude for some potential. Lemma 9 is proved in Section 11.

\vskip 2 mm
{\bf Lemma 10.}
{\it Let the assumptions of Lemma 9 be fulfilled. Let estimates}
(6.1), (6.2) {\it be also fulfilled for} $\tilde f$ ({\it in place of} $f$).
{\it Let} ${\tilde h}_{\gamma}$, ${\tilde H}_{\pm}$, ${\tilde H}^0_{E,\tau}$
{\it correspond to} $\tilde f$ {\it as well as} $h_{\gamma}$, $H_{\pm}$,
$H^0_{E,\tau}$ {\it correspond to} $f$. {\it Then}:
$$|h_{\gamma}(k,l)-{\tilde h}_{\gamma}(k,l)|\le
{\|f-\tilde f\|_{C({\cal M}_E),\mu}\over
(1-\delta_1)^2(1+|k-l|^2)^{\mu/2}}\, \ \ \gamma\in\S^2,\ \
k,l\in\S^2_{\sqrt{E}};\eqno(6.22)$$
$$\eqalign{
&|(H_{\pm}-{\tilde H}_{\pm})(k(\lambda,p,E),p)-
(H_{\pm}-{\tilde H}_{\pm})(k(\lambda^{\prime},p,E),p)|\le\cr
&2\max\,(1,c_{10}(\mu)(1+\delta_2)(1-\delta_1)g_2N+
c_9(\beta,\mu)(g_1N)^2E^{-\beta/2})
\times\cr
&{E^{\ep\beta/2}|\lambda-\lambda^{\prime}|^{\ep\beta}
(\|f-\tilde f\|_{C({\cal M}_E),\mu})^{1-\ep}\over (1-\delta_1)^2
(1+p^2)^{\mu/2}},\cr}
\eqno(6.23)$$
{\it where} $\lambda,\lambda^{\prime}\in {\cal T}$,
$|\lambda-\lambda^{\prime}|\le (E-p^2/4)^{-1/2}$,
$p\in {\cal B}_{2\sqrt{E}}\b {\cal L}_{\nu}$, $0<\beta< min\,(\alpha,1/2)$,
$0\le\ep\le 1$;
$$\eqalign{
&|H^0_{E,\tau}(\lambda,p)-{\tilde H}^0_{E,\tau}(\lambda,p)|\le
\biggl(c_{11}(1+\ln\,E)
(\|f-\tilde f\|_{C({\cal M}_E),\mu})^{\ep}+\cr
&2c_{12}(\ep\beta)
\max\,(1,c_{10}(\mu)(1+\delta_2)(1-\delta_1)g_2N+
c_9(\beta,\mu)(g_1N)^2E^{-\beta/2})\biggr)
\times\cr
&{(\|f-\tilde f\|_{C({\cal M}_E),\mu})^{1-\ep}\over (1-\delta_1)^2
(1+p^2)^{\mu/2}},
\cr}\eqno(6.24)$$
{\it where} $\lambda\in\C\b ({\cal T}\cup 0)$,
$p\in {\cal B}_{2\sqrt{E}}\b {\cal L}_{\nu}$, $0<\tau< 1$,
$0<\beta<\min\,(\alpha,1/2)$, $0<\ep\le 1$.

Note that in Lemma 10 we do not suppose that $f$  and $\tilde f$
are the scattering
amplitudes for some potential. Lemma 10 is proved in Section 11.

Let
$$\eqalign{
&u(s,s_1,s_2)=1\ \ {\rm for}\ \ s\in [0,s_1],\cr
&u(s,s_1,s_2)={{s_2-s}\over {s_2-s_1}}\ \ {\rm for}\ \ s\in [s_1,s_2],\cr
&u(s,s_1,s_2)=0\ \ {\rm for}\ \ s\in [s_2,+\infty [,\cr}\eqno(6.25)$$
where $0<s_1<s_2$.

\vskip 2 mm
{\bf Lemma 11.}
{\it Let} $f$ {\it satisfy} (6.1), (6.2) {\it for some} $E>0$,
$\mu>0$, $\alpha\in ]0,1[$ {\it and} $g_1,g_2,N\in ]0, +\infty [$.
{\it Let}
$${\tilde f}(k,l)=f(k,l)u(|k-l|,2\tau_0\sqrt{E},2\tau\sqrt{E}),\ \
(k,l)\in {\cal M}_E,\eqno(6.26)$$
{\it where} $u$ {\it is defined by} (6.25), $0<\tau_0<\tau<1$. {\it Then}:
$$\eqalignno{
&\|\tilde f\|_{C({\cal M}_E),\mu}\le g_1N,&(6.27)\cr
&\|\tilde f\|_{C^{\alpha}({\cal M}_E),\mu}\le
\biggl(g_2+{g_1\over 2(\tau-\tau_0)\sqrt{E}}\biggr)N,&(6.28)\cr
&\|f-\tilde f\|_{C({\cal M}_E),\mu_0}\le {g_1N\over
(1+4\tau_0^2E)^{(\mu-\mu_0)/2}}\ \ {\it for}\ \ \mu_0\in [0,\mu].&(6.29)\cr}
$$

Lemma 11 is proved in Section 11.

\vskip 2 mm
{\bf 7. Approximate finding $\hat v$ on
${\cal B}_{2\tau\sqrt{E}}\b {\cal L}_{\nu}$ from ${\tilde H}_{E,\tau}$}
{\bf on
$(\C\b ({\cal T}\cup 0))\times ({\cal B}_{2\tau\sqrt{E}}\b {\cal L}_{\nu})$}

\vskip 2 mm
Consider, first, $H_{E,\tau}$ defined in Proposition 1. Under assumptions
(2.4), (2.18), formulas (2.26), (3.8), (3.10), (5.8) imply that
$$\eqalign{
&H_{E,\tau}(\lambda,p)\to {\hat v}(p)\ \ {\rm as}\ \ \lambda\to 0,\cr
&H_{E,\tau}(\lambda,p)\to {\hat v}(p)\ \ {\rm as}\ \ \lambda\to\infty,\cr}
\eqno(7.1)$$
where $p\in  {\cal B}_{2\tau\sqrt{E}}\b {\cal L}_{\nu}$, $\tau\in ]0,1[$.

Consider now ${\tilde H}_{E,\tau}$ defined in Proposition 2.
Under the assumptions of Proposition 2, the following formulas hold:
$$\eqalignno{
&{\tilde H}_{E,\tau}(\lambda,p)\to {\hat v}_+(p,E,\tau)\ \ {\rm as}\ \
\lambda\to 0,&(7.2a)\cr
&{\tilde H}_{E,\tau}(\lambda,p)\to {\hat v}_-(p,E,\tau)\ \ {\rm as}\ \
\lambda\to\infty,&(7.2b)\cr}$$
where
$$\eqalign{
&{\hat v}_+(p,E,\tau)={1\over 2\pi i}\int\limits_{\cal T}H_+(\zeta,p,E)
{d\zeta\over \zeta}-\cr
&{1\over \pi}\int\!\!\!\int\limits_{{\cal D}_+}
({\tilde H}_{E,\tau},{\tilde H}_{E,\tau})_{E,\tau}(\zeta,p)
{d Re\,\zeta d Im\,\zeta\over \zeta},\cr}\eqno(7.3a)$$
$$\eqalign{
&{\hat v}_-(p,E,\tau)={1\over 2\pi i}\int\limits_{\cal T}H_-(\zeta,p,E)
{d\zeta\over \zeta}+\cr
&{1\over \pi}\int\!\!\!\int\limits_{{\cal D}_-}
({\tilde H}_{E,\tau},{\tilde H}_{E,\tau})_{E,\tau}(\zeta,p)
{d Re\,\zeta d Im\,\zeta\over \zeta},\cr}\eqno(7.3b)$$
where  $p\in  {\cal B}_{2\tau\sqrt{E}}\b {\cal L}_{\nu}$, $\tau\in ]0,1[$,
$H_{\pm}$ are defined by (5.9a),
$({\tilde H}_{E,\tau},{\tilde H}_{E,\tau})_{E,\tau}$ is defined by means of
(5.13). Formulas (7.2), (7.3) follow from the definition of
${\tilde H}_{E,\tau}$ as the solution of (5.28), where
$u^0=H^0_{E,\tau}$, $|||{\tilde H}_{E,\tau}|||_{E,\tau,\mu}\le r$
(see Proposition 2), formulas (5.11), (5.12) and Lemma 3.
Formulas (5.16), (7.1), (7.2) imply that
$$\|\hat v-{\hat v}_{\pm}(\cdot,E,\tau)\|_{E,\tau,\mu}\le
|||H_{E,\tau}-{\tilde H}_{E,\tau}|||_{E,\tau,\mu},\eqno(7.4)$$
where
$$\|w\|_{E,\tau,\mu}=\sup_{\scriptstyle
p\in  {\cal B}_{2\tau\sqrt{E}}\b {\cal L}_{\nu}}(1+|p|)^{\mu}|w(p)|,
\eqno(7.5)$$
$E\ge 1$, $\tau\in ]0,1[$, $\mu<0$, $w$ is a test function.
Under the assumptions of Proposition 2 (or under the assumptions of
Corollary 1), formulas (5.33), (7.4) imply that $\hat v$  on
${\cal B}_{2\tau\sqrt{E}}\b {\cal L}_{\nu}$ can be approximately determined
from ${\tilde H}_{E,\tau}$ on
$(\C\b ({\cal T}\cup 0))\times
{\cal B}_{2\tau\sqrt{E}}\b {\cal L}_{\nu}$ as ${\hat v}_{\pm}(\cdot,E,\tau)$
of (7.2), (7.3) and
$$\|\hat v- {\hat v}_{\pm}(\cdot,E,\tau)\|_{E,\tau,\mu_0}=
O(E^{-(\mu-\mu_0)/2})\ \ {\rm as}\ \ E\to +\infty.\eqno(7.6)$$

\vskip 2 mm
{\bf 8. Approximate finding $\hat v$ on ${\cal B}_{2\tau\sqrt{E}}\b
{\cal L}_{\nu}$ from $f$ on ${\cal M}_E$ for $d=3$}

\vskip 2 mm
In this section we summarize our method (developed in Sections 3,4,5,6,7)
for approximate finding   $\hat v$ on  ${\cal B}_{2\tau\sqrt{E}}$ from $f$
on ${\cal M}_E$.

Consider $\tau_1(\alpha,\mu,\mu_0,\sigma,N)$ and
$E_1(\alpha,\mu,\mu_0,\sigma,N)$ of (5.37), where $0<\alpha<1$,
$2\le\mu_0\le\mu$, $0<\sigma<1$, $N>0$. Let $g_1,g_2$ be some fixed numbers
such that
$$\eqalign{
&g_1\ge \bigl(1-E^{-\sigma/2}c_1(\alpha,\mu,\sigma,3)N\bigr)^{-1},\cr
&g_2\ge c_2(\mu)(1-E^{-\sigma/2}c_1(\alpha,\mu,\sigma,3)N)^{-1},\ \ g_1<g_2,
\cr
&{\rm for}
\ \ E\ge E_1(\alpha,\mu,\mu_0,\sigma,N).\cr}\eqno(8.1)$$
Consider $E_2(\mu,\sigma,N,g_1)\ge 1$ such that
$$\delta_1<1\ \ {\rm for}\ \ E\ge E_2(\mu,\sigma,N,g_1),\eqno(8.2)$$
where $\delta_1$ is defined in (6.11),  $2\le\mu$,
$0<\sigma<1$, $N>0$.

\vskip 2 mm
{\bf Theorem 1.}
{\it Let} $\hat v$ {\it satisfy} (2.4), {\it where} $d=3$, $\mu\ge 2$
{\it and} $\|\hat v\|_{\alpha,\mu}\le N$ {\it for some} $N>0$. {\it Let}
$$0<\tau<\tau_1(\alpha,\mu,\mu_0,\sigma,N),\ \
E\ge \max\,(E_1(\alpha,\mu,\mu_0,\sigma,N),\ E_2(\mu,\sigma,N,g_1)),
\eqno(8.3)$$
{\it where} $2\le\mu_0\le\mu$, $0<\sigma<1$ {\it and}  $g_1$ {\it satisfies}
(8.1). {\it Let} $f$ {\it be the scattering amplitude for equation} (1.1)
{\it and} ${\hat v}_{\pm}(p,E,\tau)$ {\it be determined from} $f$ {\it on}
${\cal M}_E$ {\it via the following (stable) reconstruction procedure:}
$$\eqalign{
&f\ \ {\it on}\ \ {\cal M}_E\buildrel (2.20) \over \rightarrow h_{\gamma}(k,l)
,\ (k,l)\in {\cal M}_E,\ \gamma\in\S^2,\ \gamma k=0\cr
&\buildrel
(2.8),(5.1),(5.9a) \over \rightarrow
H_{\pm}(\lambda,p,E),\ \lambda\in {\cal T},\ p\in {\cal B}_{2\tau\sqrt{E}}\b
 {\cal L}_{\nu},\cr
&\buildrel (5.11),(5.25b) \over \rightarrow H^0_{E,\tau}
(\lambda,p),\
\lambda\in\C\b ({\cal T}\cup 0),\
p\in {\cal B}_{2\tau\sqrt{E}}\b {\cal L}_{\nu},\cr
&\buildrel (5.34) \over \rightarrow {\tilde H}_{E,\tau}(\lambda,p),\
\lambda\in\C\b ({\cal T}\cup 0),\
p\in {\cal B}_{2\tau\sqrt{E}}\b {\cal L}_{\nu},\buildrel (7.2),(7.3) \over
\rightarrow {\hat v}_{\pm}(p,E,\tau),\  p\in {\cal B}_{2\tau\sqrt{E}},\cr}
\eqno(8.4)$$
{\it where integral equations} (2.20), (5.34) {\it are solved by the method
of successive approximations}. {\it Then}
$$\|{\hat v}-{\hat v}_{\pm}(\cdot,E,\tau)\|_{E,\tau,\mu_0}\le
{3c_5c_4(\mu_0,\tau,E)2^{\mu}N^2\over (1-2c_5c_4(\mu_0,\tau,E)r)(1-\eta)
(1+2\tau\sqrt{E})^{\mu-\mu_0}},\eqno(8.5)$$
{\it where} $\|\cdot\|_{E,\tau,\mu_0}$ {\it is defined by} (7.5),
$\eta=E^{-\sigma/2}c_1(\alpha,\mu,\sigma,3)N$, $r=\max\,(r_1,r_2)$,
{\it where} $r_1,r_2$ {\it are defined by} (5.32), {\it and thus}
$$\|{\hat v}-{\hat v}_{\pm}(\cdot,E,\tau)\|_{E,\tau,\mu_0}=
O(E^{-(\mu-\mu_0)/2})\ \ {\it as}\ \ E\to +\infty \eqno(8.6)$$
{\it at fixed} $\alpha,\mu,\mu_0,\tau,N$.

Theorem 1 follows from estimates (6.1)-(6.3) and Lemma 9 of Section 6,
Lemmas 5,7,8, Proposition 2 and property (5.37) of Section 5, formulas
(7.2)-(7.4) of Section 7 and the definition of $E_2(\alpha,\mu,\sigma,N,g_2)$.

Note that for finding $H^0_{E,\tau}$ from $H_{\pm}$ in (8.4) we mention also
estimate (5.25b), in addition to formula (5.11). This means that: (1) we
determine $H^0_{E,\tau}$ from $H_{\pm}$ by (5.11); (2) we redefine
$H^0_{E,\tau}$ by the formulas:
$$\eqalign{
&H^0_{E,\tau}(\lambda,p)\ \rightarrow\  H^0_{E,\tau}(\lambda,p)\ \
 {\rm if}\cr
&|H^0_{E,\tau}(\lambda,p)|\le
\biggl(2^{\mu/2}N+ {c_7(\alpha,\mu,\sigma,\beta)N^2\over (1-\eta)^2
(1-\delta)E^{\beta/2}}\biggr){1\over (1+|p|)^{\mu}},\cr}\eqno(8.7a)$$
$$\eqalign{
&H^0_{E,\tau}(\lambda,p)\rightarrow\
\biggl(2^{\mu/2}N+ {c_7(\alpha,\mu,\sigma,\beta)N^2\over (1-\eta)^2
(1-\delta)E^{\beta/2}}\biggr){1\over (1+|p|)^{\mu}}
{H^0_{E,\tau}(\lambda,p)\over |H^0_{E,\tau}(\lambda,p)|}\ \ {\rm if}\cr
&|H^0_{E,\tau}(\lambda,p)|>
\biggl(2^{\mu/2}N+ {c_7(\alpha,\mu,\sigma,\beta)N^2\over (1-\eta)^2
(1-\delta)E^{\beta/2}}\biggr){1\over (1+|p|)^{\mu}},\cr}\eqno(8.7b)$$
where $\eta$ is defined by (2.31), $d=3$, $\delta$ is defined by (5.22),
$0<\beta<\min\,(\alpha,\sigma,1/2)$.
If, under the assumptions of Theorem 1, the scattering amplitude $f$ is
given with no errors and all calculations based on (2.20), (2.8), (5.1),
(5.9a), (5.11) of (8.4) are fulfilled with no errors, then estimate (5.25b)
is fulfilled automatically. However, if some of these errors are present,
then estimate (5.25b) taken into account according to (8.7) permits to
improve the stability of the reconstruction procedure (8.4).

\vskip 2 mm
{\bf Theorem 2.}
{\it Let} $g_1$, $g_2$ satisfy (8.1).
{\it Let} $\hat v$, $N$, $\tau$, $E$, $\alpha$, $\mu$, $\mu_0$, $\sigma$
{\it satisfy the assumptions of Theorem 1. Let} $f$ {\it be the
scattering amplitude for equation} (1.1) {\it and, as in Theorem 1,}
${\hat v}_{\pm}(\cdot,E,\tau)$ {\it be determined from} $f$ {\it on}
${\cal M}_E$ {\it via} (8.4). {\it Let} $\tilde f$ {\it be an
approximation to} $f$, {\it where estimates} (6.1), (6.2) {\it are valid
also for}
 $\tilde f$ {\it in place of} $f$.
{\it Let} ${\tilde {\hat v}}_{\pm}
(\cdot,E,\tau)$ {\it be determined from} $\tilde f$ {\it on} ${\cal M}_E$
{\it via} (8.4) ({\it including} (8.7)) {\it with} $\tilde f$,
${\tilde h}_{\gamma}$, ${\tilde H}_{\pm}$, ${\tilde H}^0_{E,\tau}$,
${\tilde {\tilde H}}_{E,\tau}$, ${\tilde {\hat v}}(\cdot,E,\tau)$ {\it in
place of} $f$,
$h_{\gamma}$, $H_{\pm}$, $H^0_{E,\tau}$,
${\tilde H}_{E,\tau}$, ${\hat v}(\cdot,E,\tau)$, {\it respectively. Then}
$$\eqalign{
&\|{\hat v}_{\pm}(\cdot,E,\tau)-
{\tilde {\hat v}}_{\pm}(\cdot,E,\tau)\|_{E,\tau,\mu_0}\le
\bigl(c_{11}(1+\ln\,E)
(\|f-\tilde f\|_{C({\cal M}_E),\mu})^{\ep}+\cr
&2c_{12}(\ep\beta)
\max\,(1,c_{10}(\mu)(1+\delta_2)(1-\delta_1)g_2N+
c_9(\beta,\mu)(g_1N)^2E^{-\beta/2})\bigr)
\times\cr
&{(\|f-\tilde f\|_{C({\cal M}_E),\mu})^{1-\ep}\over (1-\delta_1)^2
(1-2c_5c_4(\mu_0,\tau,E)r)},\ \ r=\max\,(r_1,r_2),\cr}\eqno(8.8)$$
{\it where} $0<\ep\le 1$, $0<\beta<\min\,(\alpha,1/2)$, $\delta_1$,
$\delta_2$ {\it are defined in} (6.11), $r_1$, $r_2$ {\it are defined by}
(5.32).

Theorem 2 follows from Lemma 10 of Section 6, Lemma 8 of Section 5 and the
following formula
$$\|{\hat v}_{\pm}(\cdot,E,\tau)-
  {\tilde {\hat v}}_{\pm}(\cdot,E,\tau)\|_{E,\tau,\mu_0}\le
|||{\tilde H}_{E,\tau}-{\tilde {\tilde H}}_{E,\tau}|||_{E,\tau,\mu_0}.
\eqno(8.9)$$

Consider ${\cal M}_{E,\tau}$ defined by (1.14). Consider the function
$\tilde f$  defined by (6.25) on ${\cal M}_E$
, where $f$ is the
scattering amplitude for equation (1.1). Note that $\tilde f\equiv 0$ on
${\cal M}_E\b {\cal M}_{E,\tau}$ and $\tilde f$ is completely determined
from $f$ on ${\cal M}_{E,\tau}$. As a corollary of Theorems 1 and 2 of
this section, estimates (6.1)-(6.3) and Lemma 11 of Section 6 we obtain
the following results.

\vskip 2 mm
{\bf Corollary 2.}
{\it Let} $\hat v$, $N$, $\tau$, $E$, $\alpha$, $\mu$, $\mu_0$, $\sigma$
{\it satisfy the assumptions of Theorem 1.}
 {\it Let} $f$ {\it be the scattering amplitude for
equation} (1.1) {\it and} $\tilde f$ {\it on} ${\cal M}_E$ {\it be
defined in terms of} $f$ {\it on} ${\cal M}_{E,\tau}$ {\it by} (6.25). 
{\it Let,
further}, ${\tilde {\hat v}}(\cdot,E,\tau)$ {\it be reconstructed from}
$\tilde f$ {\it on} ${\cal M}_E$ {\it via} (8.4) ({\it including} (8.7))
{\it as in Theorem 2. Then}
$$\|{\hat v}-
{\tilde {\hat v}}(\cdot,E,\tau)\|_{E,\tau,\mu_0}=
O\biggl({1\over E^{(1-\ep)(\mu-\mu_0)/2}}\biggr)\ \ {\it as}\ \
E\to +\infty \eqno(8.10)$$
{\it for fixed} $\ep\in ]0,1[$ {\it and} $\alpha$, $\mu$, $\mu_0$,
$\tau$, $\tau_0$, $N$.

Let us remind that if $v$ satisfies (1.2), then $\hat v$ satisfies (2.3).
Using this remark and Theorem 1 we obtain the following result.

\vskip 2 mm
{\bf Corollary 3.}
{\it Let} $v$ satisfy (1.2), {\it where} $d=3$, $n>3$. {\it Then:}

(1) $\hat v$ {\it satisfies the assumptions of Theorem 1 for} $\mu=n$ {\it
and some} $\alpha\in ]0,1[$ {\it and} $N>0$;

(2) {\it the scattering amplitude} $f$ {\it on} ${\cal M}_E$ {\it determines}
$v$ {\it on} $\R^3$ {\it up to} $O(E^{-(n-3-\ep)/2})$ {\it in the
uniform norm as} $E\to +\infty$ {\it for any fixed arbitrary small}
$\ep>0$ {\it via} (8.4) ({\it where} $\tau$ {\it and} $E$ {\it satisfy}
(8.3) {\it with} $\alpha$, $\mu$ {\it and} $N$ {\it of the item} (1),
$\mu_0=3+\ep$ {\it and some} $\sigma\in ]0,1[$) {\it and the formula}
$$v(x)=v_{appr}^{\pm}(x,E,\tau)+v_{err}^{\pm}(x,E,\tau),\eqno(8.11)$$
{\it where}
$$\eqalign{
&v_{appr}^{\pm}(x,E,\tau)=\int\limits_{{\cal B}_{2\tau\sqrt{E}}}
e^{-ipx}{\hat v}_{\pm}(p,E,\tau)dp,\cr
&v_{err}^{\pm}(x,E,\tau)=\int\limits_{{\cal B}_{2\tau\sqrt{E}}}
e^{-ipx}({\hat v}(p)-{\hat v}_{\pm}(p,E,\tau))dp
+\int\limits_{\R^3\b {\cal B}_{2\tau\sqrt{E}}}{\hat v}(p)dp.\cr}\eqno(8.12)$$

\vskip 2 mm
{\bf 9. Proof of Lemma 2}

First of all, for $d=3$ we write equation (2.22) as
$${\pa\over \pa\bar k_j}H(k,p)=-{\pi\over 2}
\int\limits_{\{\xi\in\R^3:\ \xi^2+2k\xi=0\}}\xi_jH(k,-\xi)H(k+\xi,p+\xi)
{ds\over |Im\,k|\,|Re\,k|} \eqno(9.1)$$
for $j=1,2,3$, $k\in\C^3\b\R^3$, $k^2=E$, $p\in\R^3$, where $ds$ is
arc-length measure on the circle $\{\xi\in\R^3:\ \xi^2+2k\xi=0\}$.

Taking into account (3.10), (3.8) we have that
$$\eqalign{
&{\pa\over \pa\bar\lambda}H(k(\lambda,p,E),p)=\sum_{j=1}^3
{\pa {\bar k}_j\over \pa\bar\lambda}{\pa H(k,p)\over \pa {\bar k}_j}=\cr
&\sum_{j=1}^3\biggl({\pa {\bar\kappa}_1\over \pa\bar\lambda}\theta_j+
{\pa {\bar\kappa}_2\over \pa\bar\lambda}\omega_j\biggr)
{\pa H(k,p)\over \pa {\bar k}_j},\ \ \lambda\in\C\b 0,\
p\in {\cal B}_{2\sqrt{E}}\b {\cal L}_{\nu},\cr}\eqno(9.2)$$
where $k=k(\lambda,p,E)$, $\kappa_1=\kappa_1(\lambda,p,E)$,
$\kappa_2=\kappa_2(\lambda,p,E)$, $\theta=\theta(p)$, $\omega=\omega(p)$,
see (3.8), (3.4).

Formulas (9.1), (9.2) imply that
$$\eqalign{
&{\pa\over \pa\bar\lambda}H(k(\lambda,p,E),p)=-{\pi\over 2}
\int\limits_{\{\xi\in\R^3:\ \xi^2+2k\xi=0\}}
\biggl({\pa {\bar\kappa}_1\over \pa\bar\lambda}\theta\xi+
{\pa {\bar\kappa}_2\over \pa\bar\lambda}\omega\xi\biggr)\times\cr
&H(k,-\xi)H(k+\xi,p+\xi){ds\over |Im\,k|\,|Re\,k|},\ \lambda\in\C\b 0,\
p\in {\cal B}_{2\tau\sqrt{E}}\b {\cal L}_{\nu},\cr}\eqno(9.3)$$
where $k=k(\lambda,p,E)$, $\kappa_1=\kappa_1(\lambda,p,E)$,
$\kappa_2=\kappa_2(\lambda,p,E)$, $\theta=\theta(p)$, $\omega=\omega(p)$,
$\tau\in ]0,1]$.

Let the circle  $\{\xi\in\R^3:\ \xi^2+2k\xi=0\}$ (where $k=k(\lambda,p,E)$)
be parametrized by $\v\in ] -\pi,\pi [$ according to (4.2). In the
parametrization (4.2) the following formula holds:
$$ds=|Re\,k|d\v,\eqno(9.4)$$
where $ds$ is arc-length measure on $\{\xi\in\R^3:\ \xi^2+2k\xi=0\}$.

Further,
$$\eqalign{
&k^{\perp}\buildrel (4.3),(3.8) \over = |Im\,k|^{-1}
(Im\,\kappa_1\,\theta+Im\,\kappa_2\,\omega)\times
(Re\,\kappa_1\,\theta+Re\,\kappa_2\,\omega+p/2)=\cr
&|Im\,k|^{-1}
(Im\,\kappa_1\,\theta\times Re\,\kappa_1\,\theta +
Im\,\kappa_1\,\theta\times Re\,\kappa_2\,\omega + Im\,\kappa_1\,\theta\times
p/2+\cr
&Im\,\kappa_2\,\omega\times Re\,\kappa_1\,\theta +
Im\,\kappa_2\,\omega\times Re\,\kappa_2\,\omega + Im\,\kappa_2\,\omega\times
p/2).\cr}\eqno(9.5)$$
Formula (9.5) and the formulas
$$\theta\times\omega\buildrel (3.4),(3.5a) \over =\hat p,\ \
\theta\times\hat p\buildrel (3.4),(3.5a) \over = -\omega,\ \
\omega\times\hat p\buildrel (3.4),(3.5a) \over =\theta,\ \
\theta\times\theta=0,\ \ \omega\times\omega=0,\eqno(9.6)$$
where $\hat p=p/|p|$, imply that
$$k^{\perp}=|Im\,k|^{-1}
(Im\,\kappa_1\,Re\,\kappa_2-Im\,\kappa_2\,Re\,\kappa_1)\hat p +
Im\,\kappa_2\,{|p|\over 2}\theta-Im\,\kappa_1\,{|p|\over 2}\omega).\eqno(9.7)
$$
Formulas (3.4), (3.8), (4.2), (9.7) imply that
$$\eqalign{
&\theta\xi=Re\,\kappa_1(\cos\v-1)+(2|Im\,k|)^{-1}Im\,\kappa_2 |p|\sin\v,\cr
&\omega\xi=Re\,\kappa_2(\cos\v-1)-(2|Im\,k|)^{-1}Im\,\kappa_1 |p|\sin\v.\cr}
\eqno(9.8)$$
Due to (9.8) we have that
$$\eqalign{
&{\pa {\bar\kappa}_1\over \pa\bar\lambda}\theta\xi+
{\pa {\bar\kappa}_2\over \pa\bar\lambda}\omega\xi=
\biggl(
{\pa {\bar\kappa}_1\over \pa\bar\lambda}Re\,\kappa_1+
{\pa {\bar\kappa}_2\over \pa\bar\lambda}Re\,\kappa_2\biggr)
(\cos\v-1)+\cr
&{|p|\over 2|Im\,k|}
\biggl(
{\pa {\bar\kappa}_1\over \pa\bar\lambda}Im\,\kappa_2-
{\pa {\bar\kappa}_2\over \pa\bar\lambda}Im\,\kappa_1\biggr)\sin\v.\cr}
\eqno(9.9)$$
The definition of $\kappa_1$, $\kappa_2$ (see (3.8)) implies that
$${\pa {\bar\kappa}_1\over \pa\bar\lambda}=
\bigl(1-{1\over {\bar\lambda}^2}\bigr){(E-p^2/4)^{1/2}\over 2},\ \
{\pa {\bar\kappa}_2\over \pa\bar\lambda}=
\bigl(1+{1\over {\bar\lambda}^2}\bigr){i(E-p^2/4)^{1/2}\over 2},\eqno(9.10)$$
$$\eqalign{
&Re\,\kappa_1=\bigl(\lambda+\bar\lambda+{1\over \lambda}+{1\over \bar\lambda}
\bigr){(E-p^2/4)^{1/2}\over 4},\ \
Im\,\kappa_1=\bigl(\lambda-\bar\lambda+{1\over \lambda}-{1\over \bar\lambda}
\bigr){(E-p^2/4)^{1/2}\over 4i},\cr
&Re\,\kappa_2=\bigl({1\over \lambda}-{1\over \bar\lambda}-
\lambda+\bar\lambda
\bigr){i(E-p^2/4)^{1/2}\over 4},\ \
Im\,\kappa_2=\bigl({1\over \lambda}+{1\over \bar\lambda}-
\lambda-\bar\lambda
\bigr){(E-p^2/4)^{1/2}\over 4},\cr}\eqno(9.11)$$
where $\lambda\in\C\b 0$, $p\in {\cal B}_{2\sqrt{E}}\b {\cal L}_{\nu}$.
Due to (9.10), (9.11) we have that
$$\eqalign{
&{\pa {\bar\kappa}_1\over \pa\bar\lambda}Re\,\kappa_1+
{\pa {\bar\kappa}_2\over \pa\bar\lambda}Re\,\kappa_2=\cr
&{{E-p^2/4}\over 8}
\biggl(\bigl(1-{1\over {\bar\lambda}^2}\bigr)
\bigl(\lambda+\bar\lambda+{1\over \lambda}+{1\over \bar\lambda}\bigr)-
\bigl(1+{1\over {\bar\lambda}^2}\bigr)
\bigl({1\over \lambda}-{1\over \bar\lambda}-\lambda+\bar\lambda\bigr)\biggr)=
\cr
&{{E-p^2/4}\over 8}
\bigl(\lambda+\bar\lambda+{1\over \lambda}+{1\over \bar\lambda}-
{\lambda\over {\bar\lambda}^2}-{1\over \bar\lambda}-
{1\over {\bar\lambda}^2\lambda}-{1\over {\bar\lambda}^3}-{1\over \lambda}+
{1\over \bar\lambda}+\cr
&\lambda-\bar\lambda-{1\over {\bar\lambda}^2\lambda}+
{1\over {\bar\lambda}^3}+{\lambda\over {\bar\lambda}^2}-
{1\over \bar\lambda}\bigr)=\cr
&{{E-p^2/4}\over 4}\bigl(\lambda-{1\over {\bar\lambda}^2\lambda}\bigr)=
{{E-p^2/4}\over 4}{{|\lambda|^4-1}\over |\lambda|^2\bar\lambda},\cr}
\eqno(9.12)$$
$$\eqalign{
&{\pa {\bar\kappa}_1\over \pa\bar\lambda}Im\,\kappa_2-
{\pa {\bar\kappa}_2\over \pa\bar\lambda}Im\,\kappa_1=\cr
&{{E-p^2/4}\over 8}
\biggl(\bigl(1-{1\over {\bar\lambda}^2}\bigr)
\bigl({1\over \lambda}+{1\over \bar\lambda}
-\lambda-\bar\lambda\bigr)-
\bigl(1+{1\over {\bar\lambda}^2}\bigr)
\bigl(\lambda-\bar\lambda+
{1\over \lambda}-{1\over \bar\lambda}\bigr)\biggr)=
\cr
&{{E-p^2/4}\over 8}
\bigl({1\over \lambda}+{1\over \bar\lambda}-
\lambda-\bar\lambda-{1\over {\bar\lambda}^2\lambda}-
{1\over {\bar\lambda}^3}+{\lambda\over {\bar\lambda}^2}
+{1\over \bar\lambda}-\cr
&\lambda+\bar\lambda-{1\over \lambda}+{1\over \bar\lambda}-
{\lambda\over {\bar\lambda}^2}+{1\over \bar\lambda}-
{1\over {\bar\lambda}^2\lambda}+
{1\over {\bar\lambda}^3}\bigr)=\cr
&{{E-p^2/4}\over 4}\bigl(-\lambda+{2\over \bar\lambda}
-{1\over {\bar\lambda}^2\lambda}\bigr)=
-{{E-p^2/4}\over 4}{(|\lambda|^2-1)^2\over |\lambda|^2\bar\lambda}.\cr}
\eqno(9.13)$$
The $\bar\pa$- equation (4.1) follows from (9.3), (9.4), (9.9), (9.12),
(9.13), (3.9).

Lemma 2 is proved.

\vskip 2 mm
{\bf 10. Proof of Lemma 3}

Let us show, first, that
$$\{U_1,U_2\}(\cdot,\cdot,E)\in L^{\infty}_{local}
((\C\b ({\cal T}\cup 0))\times ({\cal B}_{2\tau\sqrt{E}}\b
{\cal L}_{\nu})),\ \tau\in ]0,1[.\eqno(10.1)$$
Property (10.1) follows from formula (4.6), the properties
$$
U_1(k,-\xi(k,\v))\in L^{\infty}((\Sigma_E\b Re\,\Sigma_E)\times [0,2\pi])\ \
{\rm (as\ a\ function\ of}\ \ k,\v),\eqno(10.2)$$
$$\eqalign{
&U_2(k+\xi(k,\v),p+\xi(k,\v))\in L^{\infty}((\Omega_E\b Re\,\Omega_E)
\times [0,2\pi])\cr
&{\rm (as\ a\ function\ of}\ \ k,p,\v),\cr}\eqno(10.3)$$
where
$$\eqalignno{
&\Sigma_E=\{k\in\C^3:\ k^2=E\},\ \ Re\,\Sigma_E=\{k\in\R^3:\ k^2=E\},&(10.4)
\cr
&\xi(k,\v)=Re\,k(\cos\v-1)+k^{\perp}\sin\v,\ \ k^{\perp}=
{Im\,k\times Re\,k\over |Im\,k|},&(10.5)\cr}$$
and Lemma 1. In turn, (10.2) follows from
$U_1\in L^{\infty}(\Omega_E\b Re\,\Omega_E)$, definition (2.11) ($d=3$) and
the fact that $p=-\xi(k,\v)$, $\v\in [0,2\pi]$, is a parametrization of the
set $\{p\in\R^3:\ p^2=2kp\}$, $k\in\Sigma_E\b Re\,\Sigma_E$. To prove (10.3)
consider
$$\eqalign{
&\Omega_E^{\prime}=\{k\in\C^3,\ l\in\C^3:\ k^2=l^2=E,\ Im\,k=Im\,l\},\cr
&Re\,\Omega_E^{\prime}=\{k\in\R^3,\ l\in\C^3:\ k^2=l^2=E\}.\cr}\eqno(10.6)$$
Note that
$$\eqalign{
&\Omega_E^{\prime}\b Re\,\Omega_E^{\prime}\approx \Omega_E\b Re\,\Omega_E,\cr
&(k,l)\in\Omega_E^{\prime}\b Re\,\Omega_E^{\prime}\Rightarrow
(k,k-l)\in \Omega_E\b Re\,\Omega_E,\cr
&(k,p)\in\Omega_E\b Re\,\Omega_E\Rightarrow
(k,k-p)\in \Omega_E^{\prime}\b Re\,\Omega_E^{\prime}.\cr}\eqno(10.7)$$
Consider
$$u_2(k,l)=U_2(k,k-l),\ \ (k,l)\in \Omega_E^{\prime}\b Re\,\Omega_E^{\prime}.
\eqno(10.8)$$

The property $U_2\in L^{\infty}(\Omega_E\b Re\,\Omega_E)$  is equivalent to
the property

\noindent
$u_2\in L^{\infty}(\Omega_E^{\prime}\b Re\,\Omega_E^{\prime})$.
Property (10.3) is equivalent to the property
$$u_2(k+\xi(k,\v),l)\in L^{\infty}((\Omega_E^{\prime}\b Re\,\Omega_E^{\prime})
\times [0,2\pi])\ \ {\rm (as\ a\ function\ of}\ \ k,l,\v).\eqno(10.9)$$
Property (10.9) follows from the property
$$\eqalign{
&u_2(\zeta(l,\psi,\v)+i Im\,l,l)\in L^{\infty}((\Sigma_E\b Re\,\Sigma_E)
\times [0,2\pi]\times [0,2\pi]),\cr
&{\rm (as\ a\ function\ of}\ \ l,\psi,\v),\cr}\eqno(10.10)$$
where
$$\zeta(l,\psi,\v)=Re\,l\,\cos(\v-\psi)+l^{\perp}\sin(\v-\psi),\ \ l^{\perp}=
{Im\,l\times Re\,l\over |Im\,l|}.\eqno(10.11)$$
Note that $k=\zeta(l,\psi,\v)$, $\v\in [0,2\pi]$, at fixed $\psi\in [0,2\pi]$
is a parametrization of the set
$S_l=\{k\in\C^3:\ k^2=l^2,\ Im\,k=Im\,l\},\ l\in\Sigma_E\b Re\,\Sigma_E$.
In turn, (10.10) follows from
$u_2\in L^{\infty}(\Omega_E^{\prime}\b Re\,\Omega_E^{\prime})$, definition
(10.6) and the aforementioned fact concerning the parametrization of $S_l$.
Thus, properties (10.10), (10.9), (10.3) are proved. This completes the
proof of (10.1).

Let us prove now (4.11).

We have that
$$\{U_1,U_2\}(\lambda,p,E)=\{U_1,U_2\}_1(\lambda,p,E)+
\{U_1,U_2\}_2(\lambda,p,E),\eqno(10.12)$$
where
$$\{U_1,U_2\}_1(\lambda,p,E)=-{\pi\over 4}(E-p^2/4)^{1/2}
{sgn\,(|\lambda|^2-1)(|\lambda|^2+1)\over \bar\lambda |\lambda|}
\{U_1,U_2\}_3(\lambda,p,E),\eqno(10.13a)$$
$$\eqalign{
&\{U_1,U_2\}_3(\lambda,p,E)=\int_{-\pi}^{\pi}(\cos\v-1)\times\cr
&U_1(k(\lambda,p,E),-\xi(\lambda,p,E,\v))
U_2(k(\lambda,p,E)+\xi(\lambda,p,E,\v),p+\xi(\lambda,p,E,\v))d\v,\cr}
\eqno(10.13b)$$
$$\{U_1,U_2\}_2(\lambda,p,E)={\pi\over 4}{|p|\over \bar\lambda}
\{U_1,U_2\}_4(\lambda,p,E),\eqno(10.14a)$$
$$\eqalign{
&\{U_1,U_2\}_4(\lambda,p,E)=\int_{-\pi}^{\pi}\sin\v\times\cr
&U_1(k(\lambda,p,E),-\xi(\lambda,p,E,\v))
U_2(k(\lambda,p,E)+\xi(\lambda,p,E,\v),p+\xi(\lambda,p,E,\v))d\v,\cr}
\eqno(10.14b)$$
$\lambda\in\C\b ({\cal T}\cup 0)$, $p\in {\cal B}_{2\tau\sqrt{E}}\b
{\cal L}_{\nu}$, $\tau\in ]0,1[$.

Formulas (4.2), (4.3) imply that
$$|\xi|^2=|Re\,k|^2((\cos\v-1)^2+(\sin\v)^2)=4|Re\,k|^2(\sin(\v/2))^2,
\eqno(10.15)$$
where $\xi=\xi(\lambda,p,E,\v)$, $k=k(\lambda,p,E)$.

The relation $p^2=2k(\lambda,p,E)p$, $p\in\R^3$, implies that
$$p=-Re\,k(\lambda,p,E)(\cos\psi-1)-k^{\perp}(\lambda,p,E)\sin\psi
\eqno(10.16)$$
for some $\psi=\psi(\lambda,p,E)\in [-\pi,\pi]$, where
$k^{\perp}(\lambda,p,E)$ is defined by (4.3). Formulas (4.2), (4.3), (10.16)
imply that
$$\eqalign{
&|p+\xi|^2=|Re\,k|^2((\cos\v-\cos\psi)^2+(\sin\v-\sin\psi)^2)=
4|Re\,k|^2\bigl(\sin {{\v-\psi}\over 2}\bigr)^2,\cr
&|p|^2=4|Re\,k|^2\bigl(\sin {\psi\over 2}\bigr)^2,\cr}\eqno(10.17)$$
where $\xi=\xi(\lambda,p,E,\v)$, $k=k(\lambda,p,E)$, $\psi=\psi(\lambda,p,E)$.

Using the assumptions of Lemma 3 and formulas (10.13b), (10.14b), (10.15),
(10.17) we obtain that
$$\eqalignno{
&|\{U_1,U_2\}_3(\lambda,p,E)|\le\int\limits_{-\pi}^{\pi}
{(1-\cos\v)|||U_1|||_{E,\mu}|||U_2|||_{E,\mu}d\v\over
(1+2r|\sin(\v/2)|)^{\mu} (1+2r|\sin\bigl({{\v-\psi}\over 2}\bigr)|)^{\mu}},
&(10.18)\cr
&|\{U_1,U_2\}_4(\lambda,p,E)|\le\int\limits_{-\pi}^{\pi}
{|\sin\v|\, |||U_1|||_{E,\mu}|||U_2|||_{E,\mu}d\v\over
(1+2r|\sin(\v/2)|)^{\mu} (1+2r|\sin\bigl({{\v-\psi}\over 2}\bigr)|)^{\mu}},
&(10.19)\cr}$$
where $r=Re\,k(\lambda,p,E)$, $\psi=\psi(\lambda,p,E)$.

Consider
$$\eqalignno{
&A(r,\psi,\alpha,\beta)=\int\limits_{-\pi}^{\pi}
{(1-\cos\v)d\v\over
(1+2r|\sin(\v/2)|)^{\alpha} (1+2r|\sin\bigl({{\v-\psi}
\over 2}\bigr)|)^{\beta}},
&(10.20a)\cr
&B(r,\psi,\alpha,\beta)=\int\limits_{-\pi}^{\pi}
{|\sin\v|\,d\v\over
(1+2r|\sin(\v/2)|)^{\alpha} (1+2r|\sin\bigl({{\v-\psi}
\over 2}\bigr)|)^{\beta}},
&(10.20b)\cr}$$
where $r\ge 0$, $\psi\in [-\pi,\pi]$, $\alpha\ge 2$, $\beta\ge 2$. Due to
(10.18)-(10.20) we have that
$$\eqalignno{
&|\{U_1,U_2\}_3(\lambda,p,E)|\le |||U_1|||_{E,\mu}|||U_2|||_{E,\mu}
A(r,\psi,\mu,\mu),&(10.21a)\cr
&|\{U_1,U_2\}_4(\lambda,p,E)|\le |||U_1|||_{E,\mu}|||U_2|||_{E,\mu}
B(r,\psi,\mu,\mu),&(10.21b)\cr}$$
where $r=Re\,k(\lambda,p,E)$, $\psi=\psi(\lambda,p,E)$.

\vskip 2 mm
{\bf Lemma 12.}
{\it Let} $r\ge 0$, $\psi\in [-\pi,\pi]$, $\rho=2r|\sin(\psi/2)|$,
$\alpha\ge 2$, $\beta\ge 2$. {\it Then}
$$\eqalignno{
&A(r,\psi,\alpha,\beta)\le\sum_{j=1}^4A_j(r,\psi,\alpha,\beta),&(10.22)\cr
&A_1(r,\psi,\alpha,\beta)\le \min\,\biggl({\rho^3\over 6r^3},{\rho\over
r^3}\biggr)\,{1\over (1+\rho/2)^{\beta}},&(10.23)\cr
&A_2(r,\psi,\alpha,\beta)\le  {\rho^3\over r^3}
\,{1\over (1+\rho/2)^{\alpha+1}},&(10.24)\cr
&A_3(r,\psi,\alpha,\beta)\le  {4\rho^3\over r^3}
\,{1\over (1+\rho)^{\alpha}(1+\rho/2)},&(10.25)\cr
&A_4(r,\psi,\alpha,\beta)\le \biggl({3\over {1+r^2}}+
{2\pi\over (1+\sqrt{2}r)^{\alpha}}
\biggr)\,{1\over (1+\rho/2)^{\beta}},&(10.26)\cr
&B(r,\psi,\alpha,\beta)\le\sum_{j=1}^4B_j(r,\psi,\alpha,\beta),&(10.27)\cr
&B_1(r,\psi,\alpha,\beta)\le \min\,\biggl({\rho^2\over 2r^2},
{\sqrt{2}\rho\over
r^2}\biggr)\,{1\over (1+\rho/2)^{\beta}},&(10.28)\cr
&B_2(r,\psi,\alpha,\beta)\le  {2\rho^2\over r^2}
\,{1\over (1+\rho/2)^{\alpha+1}},&(10.29)\cr
&B_3(r,\psi,\alpha,\beta)\le  {4\rho^2\over r^2}
\,{1\over (1+\rho)^{\alpha}(1+\rho/2)},&(10.30)\cr
&B_4(r,\psi,\alpha,\beta)\le \biggl({5\over {1+r}}+
{3\over (1+\sqrt{2}r)^{\alpha}}
\biggr)\,{1\over (1+\rho/2)^{\beta}}.&(10.31)\cr}$$

\vskip 2 mm
{\it Proof of Lemma 12.}
Note that
$$A(r,\psi,\alpha,\beta)=A(r,-\psi,\alpha,\beta),\ \
B(r,\psi,\alpha,\beta)=B(r,-\psi,\alpha,\beta).\eqno(10.32)$$
Therefore, it is sufficient to prove Lemma 12 for $\psi\in [0,\pi]$.
Let
$$\eqalign{
&W_1(r,\psi,\alpha,\beta,\v)=
{{1-\cos\v}\over
(1+2r|\sin(\v/2)|)^{\alpha} (1+2r|\sin\bigl({{\v-\psi}
\over 2}\bigr)|)^{\beta}},\cr
&W_2(r,\psi,\alpha,\beta,\v)=
{\sin\v\over
(1+2r|\sin(\v/2)|)^{\alpha} (1+2r|\sin\bigl({{\v-\psi}
\over 2}\bigr)|)^{\beta}}.\cr}\eqno(10.33)$$
For $\psi\in [0,\pi]$ we have that:
$$A(r,\psi,\alpha,\beta)\le 2\int\limits_0^{\pi}W_1(r,\psi,\alpha,\beta,\v)
d\v=\sum_{j=1}^4 A_j(r,\psi,\alpha,\beta),\eqno(10.34)$$
where
$$\eqalignno{
&A_1(r,\psi,\alpha,\beta)=2\int\limits_0^{\psi/2}W_1(r,\psi,\alpha,\beta,\v)
d\v,&(10.35)\cr
&A_2(r,\psi,\alpha,\beta)=2\int\limits_{\psi/2}^{\psi}
W_1(r,\psi,\alpha,\beta,\v)d\v,&(10.36)\cr
&A_3(r,\psi,\alpha,\beta)=2\int\limits_{\psi}^{\min\,(3\psi/2,\pi)}
W_1(r,\psi,\alpha,\beta,\v)d\v,&(10.37)\cr
&A_4(r,\psi,\alpha,\beta)=2\int\limits_{\min\,(3\psi/2,\pi)}^{\pi}
W_1(r,\psi,\alpha,\beta,\v)d\v;&(10.38)\cr
&B(r,\psi,\alpha,\beta)\le 2\int\limits_0^{\pi}W_2(r,\psi,\alpha,\beta,\v)
d\v=\sum_{j=1}^4 B_j(r,\psi,\alpha,\beta),&(10.39)\cr}$$
where
$$\eqalignno{
&B_1(r,\psi,\alpha,\beta)=2\int\limits_0^{\psi/2}W_2(r,\psi,\alpha,\beta,\v)
d\v,&(10.40)\cr
&B_2(r,\psi,\alpha,\beta)=2\int\limits_{\psi/2}^{\psi}
W_2(r,\psi,\alpha,\beta,\v)d\v,&(10.41)\cr
&B_3(r,\psi,\alpha,\beta)=2\int\limits_{\psi}^{\min\,(3\psi/2,\pi)}
W_2(r,\psi,\alpha,\beta,\v)d\v,&(10.42)\cr
&B_4(r,\psi,\alpha,\beta)=2\int\limits_{\min\,(3\psi/2,\pi)}^{\pi}
W_2(r,\psi,\alpha,\beta,\v)d\v.&(10.43)\cr}$$

To prove Lemma 12 it remains to prove estimates (10.23)-(10.26),
(10.28)-(10.31) for $A_1,A_2,A_3,A_4,B_1,B_2,B_3,B_4$ defined by
(10.35)-(10.38), (10.40)-(10.43) for $\psi\in [0,\pi]$. Note that in these
proofs given below we largely use the following formulas
$$\eqalign{
&\rho=2r\sin(\psi/2)=4r\sin(\psi/4)\,\cos(\psi/4),\cr
&\rho/2\le 2r\sin(\psi/4),\ \ \sin(\psi/4)\le {\rho\over 2\sqrt{2}r},\cr}
\eqno(10.44)$$
where $\psi\in [0,\pi]$.

\vskip 2 mm
{\it Proof of estimate} (10.23) {\it for} $A_1$ {\it of} (10.35).
We have that
$$\eqalign{
&A_1=\int\limits_0^{\psi/2}{4(\sin(\v/2))^2d\v\over
(1+2r\sin(\v/2))^{\alpha}(1+2r|\sin\bigl({{\v-\psi}\over 2}\bigr)|)^{\beta}}
\le\cr
&{1\over (1+2r\sin(\psi/4))^{\beta}}\int\limits_0^{\psi/2}
{4(\sin(\v/2))^2d\v\over (1+2r\sin(\v/2))^{\alpha}},\cr}\eqno(10.45)$$
$${1\over (1+2r\sin(\psi/4))^{\beta}}\buildrel (10.44) \over \le
{1\over (1+\rho/2)^{\beta}},\eqno(10.46)$$
$$\eqalign{
&\int\limits_0^{\psi/2}
{4(\sin(\v/2))^2d\v\over (1+2r\sin(\v/2))^{\alpha}}\le
\int\limits_0^{\psi/2}4(\sin(\v/2))^2d\v\le 8\sqrt{2}
\int\limits_0^{\psi/2}(\sin(\v/2))^2d\sin(\v/2)\le\cr
&{8\sqrt{2}\over 3}(\sin(\psi/4))^3\buildrel (10.44) \over \le
{8\sqrt{2}\over 3}\biggl({\rho\sqrt{2}\over 4r}\biggr)^3={1\over 6}
\biggl({\rho\over r}\biggr)^3,\cr}\eqno(10.47a)$$
$$\eqalign{
&\int\limits_0^{\psi/2}
{4(\sin(\v/2))^2d\v\over (1+2r\sin(\v/2))^{\alpha}}\le
\int\limits_0^{\psi/2}{4d\v\over 4r^2(1+2r\sin(\v/2))^{\alpha-2}}\le\cr
&{\psi\over 2}{1\over r^2}\le {\pi\over 2}\sin(\psi/2){1\over r^2}=
{\pi\over 4}{\rho\over r^3},\cr}\eqno(10.47b)$$
where $\psi\in [0,\pi]$, $\alpha\ge 2$, $\beta\ge 2$. Estimate (10.23)
for $A_1$ of (10.35) follows from (10.45)-(10.47).

\vskip 2 mm
{\it Proof of estimates} (10.24) {\it for} $A_2$ {\it of} (10.36).
We have that
$$\eqalign{
&A_2=\int\limits_{\psi/2}^{\psi}{4(\sin(\v/2))^2d\v\over
(1+2r\sin(\v/2))^{\alpha}(1+2r|\sin\bigl({{\v-\psi}\over 2}\bigr)|)^{\beta}}
\le\cr
&{4(\sin(\psi/2))^2\over (1+2r\sin(\psi/4))^{\alpha}}\int\limits_0^{\psi/2}
{d\v\over (1+2r\sin(\v/2))^{\beta}},\cr}\eqno(10.48)$$
$${4(\sin(\psi/2))^2\over (1+2r\sin(\psi/4))^{\alpha}}
\buildrel (10.44) \over \le {(\rho/r)^2\over (1+\rho/2)^{\alpha}},\eqno(10.49)
$$
$$\eqalign{
&\int\limits_0^{\psi/2}{d\v\over(1+2r\sin(\v/2))^{\beta}}=
\int\limits_0^{\psi/2}{(2/\cos(\v/2))d\sin(\v/2)\over
(1+2r\sin(\v/2))^{\beta}}\le\cr
&2\sqrt{2}\int\limits_0^{\sin(\psi/4)}{dt\over (1+2rt)^{\beta}}=
{\sqrt{2}\over r(\beta-1)}\biggl(1-{1\over (1+2r\sin(\psi/4))^{\beta-1}}
\biggr)\le\cr
&{\sqrt{2}2r\sin(\psi/4)\over r(1+2r\sin(\psi/4))}\buildrel (10.44) \over
\le {\rho\over r(1+\rho/2)},\cr}\eqno(10.50)$$
where $\psi\in [0,\pi]$, $\alpha\ge 2$, $\beta\ge 2$. Estimate (10.24)
for $A_2$ of (10.36) follows from (10.48)-(10.50).

\vskip 2 mm
{\it Proof of estimate} (10.25) {\it for} $A_3$ {\it of} (10.37).
We have that
$$\eqalign{
&A_3=\int\limits_{\psi}^{\min\,(\pi,3\psi/2)}
{16\bigl(\sin(\v/4)\cos(\v/4))^2d\v\over
(1+2r\sin(\v/2))^{\alpha}(1+2r\sin({{\v-\psi}\over 2})\bigr)^{\beta}}\le\cr
&{16(\sin(3\psi/8)^2\over (1+2r\sin(\psi/2))^{\alpha}}
\int\limits_{\psi}^{\min\,(\pi,3\psi/2)}
{d\v\over (1+2r\sin({{\v-\psi}\over 2})\bigr)^{\beta}}\le\cr
&{16(\sin(\psi/2)^2\over (1+2r\sin(\psi/2))^{\alpha}}
\int\limits_0^{\psi/2}
{d\v\over (1+2r\sin(\v/2))^{\beta}}\le\cr
&\buildrel (10.44),(10.50) \over \le {4\rho^3\over
r^3(1+\rho)^{\alpha}(1+\rho/2)},\cr}\eqno(10.51)$$
where $\psi\in [0,\pi]$, $\alpha\ge 2$, $\beta\ge 2$. Estimate (10.25)
for $A_3$ of (10.37) is proved.

\vskip 2 mm
{\it Proof of estimate} (10.26) {\it for} $A_4$ {\it of} (10.38).
We have that
$$\eqalign{
&A_4=\int\limits_{\min\,(\pi,3\psi/2)}^{\pi}
{4(\sin(\v/2))^2d\v\over
(1+2r\sin(\v/2))^{\alpha}(1+2r\sin\bigl({{\v-\psi}\over 2}\bigr))^{\beta}}\le
\cr
&{1\over (1+2r\sin(\psi/4))^{\beta}}
\int\limits_0^{\pi}{4(\sin(\v/2))^2d\v\over (1+2r\sin(\v/2))^{\alpha}}\le\cr
&{1\over (1+\rho/2)^{\beta}}\biggl(\int\limits_0^{\pi/2}+
\int\limits_{\pi/2}^{\pi}\biggr)
{4(\sin(\v/2))^2d\v\over (1+2r\sin(\v/2))^{\alpha}},\cr}\eqno(10.52)$$
$$\eqalignno{
&\int\limits_0^{\pi/2}
{4(\sin(\v/2))^2d\v\over (1+2r\sin(\v/2))^{\alpha}}\buildrel (10.47a) \over
\le {8\sqrt{2}\over 3}(\sin(\pi/4))^3={4\over 3},&(10.53a)\cr
&\int\limits_0^{\pi/2}
{4(\sin(\v/2))^2d\v\over (1+2r\sin(\v/2))^{\alpha}}\buildrel (10.47b) \over
\le {\pi\over 2r^2},&(10.53b)\cr
&\int\limits_{\pi/2}^{\pi}
{4(\sin(\v/2))^2d\v\over (1+2r\sin(\v/2))^{\alpha}}
\le {2\pi\over (1+2r\sin(\pi/4))^{\alpha}},&(10.54)\cr}$$
where $\psi\in [0,\pi]$, $\alpha\ge 2$, $\beta\ge 2$. Formulas
(10.52)-(10.54) imply that
$$A_4\le {4\over 3}\min\,\bigl(1,{5\over 4r^2}\bigr)+
{2\pi\over (1+\sqrt{2}r)^{\alpha}}.\eqno(10.55)$$
Estimate (10.26) for $A_4$ of (10.38) follows from (10.55) and
the inequality
$${{1+c}\over {1+s}}\ge \min\,\bigl(1,{c\over s}\bigr),\ c\ge 0,\ s\ge 0.
\eqno(10.56)$$

\vskip 2 mm
{\it Proof of estimate} (10.28) {\it for} $B_1$ {\it of} (10.40).
We have that
$$\eqalign{
&B_1=\int\limits_0^{\psi/2}
{2\sin(\v)d\v\over
(1+2r\sin(\v/2))^{\alpha}(1+2r|\sin\bigl({{\v-\psi}\over 2}\bigr)|)^{\beta}}
\le\cr
&{1\over (1+2r\sin(\psi/4))^{\beta}}
\int\limits_0^{\psi/2}{2\sin(\v)d\v\over (1+2r\sin(\v/2))^{\alpha}}\le\cr
&{1\over (1+\rho/2)^{\beta}}\int\limits_0^{\psi/2}
{4\sin(\v/2)\cos(\v/2)d\v\over (1+2r\sin(\v/2))^{\alpha}}=
{1\over (1+\rho/2)^{\beta}}\int\limits_0^{\sin(\psi/4)}
{8tdt\over (1+2rt)^{\alpha}},\cr}\eqno(10.57)$$
$$\eqalignno{
&\int\limits_0^{\sin(\psi/4)}{8tdt\over (1+2rt)^{\alpha}}\le
\int\limits_0^{\sin(\psi/4)}8tdt=4(\sin(\psi/4))^2\buildrel (10.44) \over \le
{\rho^2\over 2r^2},&(10.58a)\cr
&\int\limits_0^{\sin(\psi/4)}{8tdt\over (1+2rt)^{\alpha}}\le
{4\sin(\psi/4)\over r}\buildrel (10.44) \over \le {\sqrt{2}\rho\over r^2},
&(10.58b)\cr}$$
where $\psi\in [0,2\pi]$, $\alpha\ge 2$, $\beta\ge 2$. Estimate (10.28)
for $B_1$ of (10.40) follows from (10.59), (10.60).

\vskip 2 mm
{\it Proof of estimate} (10.29) {\it for} $B_2$ {\it of} (10.41).
We have that
$$\eqalign{
&B_2=\int\limits_{\psi/2}^{\psi}{4\sin(\v/2)\,\cos(\v/2)d\v\over
(1+2r\sin(\v/2))^{\alpha}(1+2r|\sin\bigl({{\v-\psi}\over 2}\bigr)|)^{\beta}}
\le\cr
&{4\sin(\psi/2)\over (1+2r\sin(\psi/4))^{\alpha}}\int\limits_0^{\psi/2}
{d\v\over (1+2r\sin(\v/2))^{\beta}}\cr
&\buildrel (10.44),(10.50) \over \le {2\rho^2\over r^2(1+\rho/2)^{\alpha+1}}
,\cr}\eqno(10.59)$$
where $\psi\in [0,\pi]$, $\alpha\ge 2$, $\beta\ge 2$. Estimate (10.29)
for $B_2$ of (10.41) is proved.

\vskip 2 mm
{\it Proof of estimate} (10.30) {\it for} $B_3$ {\it of} (10.42).
We have that
$$\eqalign{
&B_3=\int\limits_{\psi}^{\min\,(\pi,3\psi/2)}
{8\sin(\v/4)\,\cos(\v/4)\,\cos(\v/2)d\v\over
(1+2r\sin(\v/2))^{\alpha}(1+2r\sin\bigl({{\v-\psi}\over 2}\bigr))^{\beta}}
\le\cr
&{8\sin(3\psi/8)\over (1+2r\sin(\psi/2))^{\alpha}}
\int\limits_{\psi}^{\min\,(\pi,3\psi/2)}
{d\v\over (1+2r\sin\bigl({{\v-\psi}\over 2}\bigr))^{\beta}}\le\cr
&{8\sin(\psi/2)\over (1+2r\sin(\psi/2))^{\alpha}}
\int\limits_0^{\psi/2}{d\v\over  (1+2r\sin(\v/2))^{\beta}}\cr
&\buildrel (10.44),(10.50) \over \le {4\rho^2\over r^2(1+\rho)^{\alpha}
(1+\rho/2)},\cr}\eqno(10.60)$$
where $\psi\in [0,\pi]$, $\alpha\ge 2$, $\beta\ge 2$. Estimate (10.30)
for $B_3$ of (10.42) is proved.

\vskip 2 mm
{\it Proof of estimate} (10.31) {\it for} $B_4$ {\it of} (10.43).
We have that
$$\eqalign{
&B_4=\int\limits_{\min\,(\pi,3\psi/2)}^{\pi}
{2\sin(\v)d\v\over
(1+2r\sin(\v/2))^{\alpha}(1+2r\sin\bigl({{\v-\psi}\over 2}\bigr))^{\beta}}\le
\cr
&{1\over (1+2r\sin(\psi/4))^{\beta}}
\int\limits_0^{\pi}{4\sin(\v/2)\cos(\v/2)
d\v\over (1+2r\sin(\v/2))^{\alpha}}\le\cr
&{1\over (1+\rho/2)^{\beta}}\biggl(\int\limits_0^{\pi/2}+
\int\limits_{\pi/2}^{\pi}\biggr)
{8\sin(\v/2)d\sin(\v/2)
\over (1+2r\sin(\v/2))^{\alpha}}=\cr
&{1\over (1+\rho/2)^{\beta}}\biggl(\int\limits_0^{\sin(\pi/4)}+
\int\limits_{\sin(\pi/4)}^{\sin(\pi/2)}\biggr)
{8tdt\over (1+2rt)^{\alpha}},\cr}\eqno(10.61)$$
$$\eqalignno{
&\int\limits_0^{\sin(\pi/4)}{8tdt\over (1+2rt)^{\alpha}}\buildrel (10.58)
\over \le 2\min\,\bigl(1,{\sqrt{2}\over r}\bigr)\buildrel (10.56) \over \le
{2(1+\sqrt{2})\over {1+r}},&(10.62)\cr
&\int\limits_{\sin(\pi/4)}^{\sin(\pi/2)} {8tdt\over (1+2rt)^{\alpha}}\le
{4(2-\sqrt{2})\over (1+\sqrt{2}r)^{\alpha}},&(10.63)\cr}$$
where $\psi\in [0,\pi]$, $\alpha\ge 2$, $\beta\ge 2$. Estimate (10.31)
for $B_4$ of (10.43) follows from (10.61)-(10.63).

Lemma 12 is proved.

\vskip 2 mm
{\bf Lemma 13.}
{\it Let}
$$\eqalign{
&r=r(\lambda,\rho,E)=((E-\rho^2/4)(|\lambda|+|\lambda|^{-1})^2+\rho^2)^{1/2}/
2,\cr
&|\sin(\psi/2)|=\rho/(2r),\cr}\eqno(10.64)$$
{\it where} $\lambda\in\C\b 0$, $\rho\in [0,2\tau\sqrt{E}]$, $E>0$,
$\tau\in ]0,1[$, $\psi\in [-\pi,\pi]$. {\it Let}
$$z=(1-\tau^2)/(4\tau^2).\eqno(10.65)$$
{\it Then:}
$$\eqalignno{
&{(E-\rho^2/4)^{1/2}(|\lambda|^2+1)\over |\lambda|^2}A_1\le
{4|\lambda|\over (|\lambda|^2+1)^2z(1+\rho/2)^{\beta}},&(10.66)\cr
&{(E-\rho^2/4)^{1/2}(|\lambda|^2+1)\over |\lambda|^2}A_2\le
{16|\lambda|\over (|\lambda|^2+1)^2z(1+\rho/2)^{\alpha}},&(10.67)\cr
&{(E-\rho^2/4)^{1/2}(|\lambda|^2+1)\over |\lambda|^2}A_3\le
{64|\lambda|\over (|\lambda|^2+1)^2z(1+\rho)^{\alpha}},&(10.68)\cr
&{(E-\rho^2/4)^{1/2}(|\lambda|^2+1)\over |\lambda|^2}A_4\le
{4\sqrt{2}(3+\pi)\over (|\lambda|^2+1)\sqrt{E}\min\,(1,2\sqrt{z})
(1+\rho/2)^{\beta}},&(10.69)\cr
&{\rho B_1\over |\lambda|}\le {4\sqrt{2}|\lambda|\over
(|\lambda|^2+1)^2z(1+\rho/2)^{\beta}},&(10.70)\cr
&{\rho B_2\over |\lambda|}\le {16|\lambda|\over
(|\lambda|^2+1)^2z(1+\rho/2)^{\alpha}},&(10.71)\cr
&{\rho B_3\over |\lambda|}\le {32|\lambda|\over
(|\lambda|^2+1)^2z(1+\rho)^{\alpha}},&(10.72)\cr
&{\rho B_4\over |\lambda|}\le {15\over
(|\lambda|^2+1)\sqrt{z}(1+\rho/2)^{\beta}},&(10.73)\cr}$$
{\it where} $A_j=A_j(r,|\psi|,\alpha,\beta)$,
$B_j=B_j(r,|\psi|,\alpha,\beta)$ {\it are the same that in Lemma 12},
$j=1,2,3,4$, $\alpha\ge 2$, $\beta\ge 2$.

\vskip 2 mm
{\it Proof of Lemma 13.}
The property $\rho\in [0,2\tau\sqrt{E}]$ and formula (10.65) imply that
$$E-\rho^2/4\ge z\rho^2.\eqno(10.74)$$
Further, proceeding from (10.23)-(10.26), (10.28)-(10.31), (10.64) and
(10.74) we prove separately each of estimates (10.67)-(10.74).

\vskip 2 mm
{\it Proof of} (10.66).
Note that
$$\eqalign{
&{(E-\rho^2/4)^{1/2}\rho^3\over 6r^3}\buildrel (10.64) \over =
{8(E-\rho^2/4)^{1/2}\rho^3\over
6((E-\rho^2/4)(|\lambda|+|\lambda|^{-1})^2+\rho^2)^{3/2}}\le\cr
&{4\rho^3\over 3(|\lambda|+|\lambda|^{-1})
((E-\rho^2/4)(|\lambda|+|\lambda|^{-1})^2+\rho^2)}\buildrel (10.74) \over \le
{4\rho\over 3z(|\lambda|+|\lambda|^{-1})^3},\cr}\eqno(10.75)$$
$$\eqalign{
&{(E-\rho^2/4)^{1/2}\rho\over r^3}\buildrel (10.64) \over =
{8(E-\rho^2/4)^{1/2}\rho\over
((E-\rho^2/4)(|\lambda|+|\lambda|^{-1})^2+\rho^2)^{3/2}}\le\cr
&{8\rho\over (|\lambda|+|\lambda|^{-1})
((E-\rho^2/4)(|\lambda|+|\lambda|^{-1})^2+\rho^2)}\buildrel (10.74) \over \le
{8\over \rho z(|\lambda|+|\lambda|^{-1})^3},\cr}\eqno(10.76)$$
$$\eqalign{
&(E-\rho^2/4)^{1/2}\min\,\bigl({\rho^3\over 6r^3},{\rho\over r^3}\bigr)
\buildrel (10.75)-(10.76) \over \le\cr
&{8\min\,(\rho/6,1/\rho)\over z(|\lambda|+|\lambda|^{-1})^3}\le
{8\over \sqrt{6}(|\lambda|+|\lambda|^{-1})^3}.\cr}\eqno(10.77)$$
Estimate (10.66) follows from (10.23), (10.77).

\vskip 2 mm
{\it Proof of} (10.67).
Estimate (10.67) follows from (10.24) and (10.75).

\vskip 2 mm
{\it Proof of} (10.68).
Estimate (10.68) follows from (10.25) and (10.75).

\vskip 2 mm
{\it Proof of} (10.69).
Note that
$$\eqalign{
&(E-\rho^2/4)^{1/2}\biggl({3\over {1+r^2}}+{2\pi\over (1+\sqrt{2}r)^{\alpha}}
\biggr)\le {(E-\rho^2/4)^{1/2}(3+\pi)\over r^2}\buildrel (10.64) \over \le\cr
&{4(3+\pi)\over (|\lambda|+|\lambda|^{-1})
((E-\rho^2/4)(|\lambda|+|\lambda|^{-1})^2+\rho^2)^{1/2}}\le
{4(3+\pi)\over
(E-\rho^2/4)^{1/2}(|\lambda|+|\lambda|^{-1})^2}.\cr}\eqno(10.78)$$
In addition,
$$\eqalign{
&\rho^2\le 2E\Rightarrow E-\rho^2/4\ge E/2;\cr
&\rho^2\ge 2E,\ E-\rho^2/2\ge z\rho^2\ \ ({\rm see}\ \ (10.74))
\Rightarrow E-\rho^2/2\ge 2zE.\cr}\eqno(10.79)$$
Thus,
$$E-\rho^2/2\ge \min\,(E/2,2zE)\ \ {\rm for}\ \ \rho\in [0,2\tau\sqrt{E}].
\eqno(10.80)$$
Estimate (10.69) follows from (10.26), (10.78), (10.80).

\vskip 2 mm
{\it Proof of} (10.70).
Note that
$$\eqalignno{
&{\rho^3\over 2r^2}\buildrel (10.64) \over = {2\rho^3\over
(E-\rho^2/4)(|\lambda|+|\lambda|^{-1})^2+\rho^2}\buildrel (10.74) \over \le
{2\rho\over z(|\lambda|+|\lambda|^{-1})^2},&(10.81)\cr
&{\sqrt{2}\rho^2\over r^2}\buildrel (10.64) \over = {4\sqrt{2}\rho^2\over
(E-\rho^2/4)(|\lambda|+|\lambda|^{-1})^2+\rho^2}\buildrel (10.74) \over \le
{4\sqrt{2}\over z(|\lambda|+|\lambda|^{-1})^2}.&(10.82)\cr}$$
Estimate (10.70) follows from (10.28), (10.81), (10.82).

\vskip 2 mm
{\it Proof of} (10.71).
Estimate (10.71) follows from (10.29), (10.82).

\vskip 2 mm
{\it Proof of} (10.72).
Estimate (10.72) follows from (10.30), (10.82).

\vskip 2 mm
{\it Proof of} (10.73).
Note that
$$\eqalign{
&\rho\biggl({3\over {1+r}}+{3\over (1+\sqrt{2}r)^{\alpha}}
\biggr)\le {(5+3/\sqrt{2})\rho\over r}\buildrel (10.64) \over =\cr
&{(10+3\sqrt{2})\rho\over
((E-\rho^2/4)(|\lambda|+|\lambda|^{-1})^2+\rho^2)^{1/2}}\le
{15\over
\sqrt{z}(|\lambda|+|\lambda|^{-1})}.\cr}\eqno(10.83)$$
Estimate (10.73) follows from (10.31), (10.83).

Lemma 13 is proved.

Estimate (4.11) follows from (10.12)-(10.14), (10.21), (10.22) and Lemma 13.

Property (4.10) follows from (10.1), (4.11).

Lemma 3 is proved.

\vskip 4 mm
{\bf 11. Proofs of Lemmas 5, 9, 10, 11}

\vskip 2 mm
{\it Proof of Lemma 5.}
Estimate (5.23) follows from (2.8), (2.29a), (5.1), (5.9a). To prove
estimate (5.24) we proceed from estimates (2.27), (2.29) and equation (2.20).
Due to (2.8), (2.29), we have, in particular, that
$$\eqalignno{
&|h_{\gamma}(k,l)|\le {N\over (1-\eta)(1+|k-l|^2)^{\mu/2}},&(11.1a)\cr
&|h_{\gamma}(k,l)-h_{\gamma}(k,l^{\prime})-{\hat v}(k-l)+
{\hat v}(k-l^{\prime})|\le
{c_2^{\prime\prime}(\mu)
\eta N|l-l^{\prime}|^{\alpha}\over (1-\eta)(1+|k-l|^2)^{\mu/2}},&(11.1b)\cr}
$$
where $k,l,l^{\prime}\in\S^2_{\sqrt{E}}$, $\gamma\in\S^2$,
$|l-l^{\prime}|\le 1$ (and $\eta$ is given by (2.31)).

Consider formulas (6.5)-(6.7) for $n=0$. Note that
$$h_{\gamma}^{(0)}(k,l)=f(k,l),\ \ h_{\gamma}(k,l)=f(k,l)+
t_{\gamma}^{(0)}(k,l),\eqno(11.2)$$
where $\gamma\in\S^2$, $k,l\in\S^2_{\sqrt{E}}$.
The following estimate holds:
$$|t_{\gamma}^{(0)}(k,l)-t_{\gamma^{\prime}}^{(0)}(k^{\prime},l)|\le
{N^2(c_2(\mu)c_3(0,\mu,\sigma,3)E^{-\sigma/2}|k-k^{\prime}|^{\alpha}+
c_{13}(\beta,\mu)|\gamma-\gamma^{\prime}|^{\beta})\over
(1-\eta)^2(1-\delta)(1+|k-l|^2)^{\mu/2}},\eqno(11.3)$$
where $k,k^{\prime},l\in\S^2_{\sqrt{E}}$, $\gamma,\gamma^{\prime}\in\S^2$,
$\gamma k=\gamma^{\prime} k^{\prime}=0$, $|\gamma-\gamma^{\prime}|\le 1$,
$|k-k^{\prime}|\le 1$, $0<\beta\le 1/2$ (and $\eta$, $\delta$ are given by
(2.31), (5.22)).

\vskip 2 mm
{\it Proof of} (11.3).
Using (6.7) for $n=0$ we obtain that
$$\eqalign{
&t_{\gamma}^{(0)}(k,\cdot)-t_{\gamma^{\prime}}^{(0)}(k^{\prime},\cdot)=
B_{\gamma}(k)f(k,\cdot)-B_{\gamma^{\prime}}(k^{\prime})f(k^{\prime},\cdot)+
B_{\gamma}(k)t_{\gamma}^{(0)}(k,\cdot)-\cr
&B_{\gamma^{\prime}}(k^{\prime})
t_{\gamma^{\prime}}^{(0)}(k^{\prime},\cdot),\cr}\eqno(11.4)$$
$$\eqalign{
&t_{\gamma}^{(0)}(k,\cdot)-t_{\gamma^{\prime}}^{(0)}(k^{\prime},\cdot)=
B_{\gamma}(k)(f(k,\cdot)-f(k^{\prime},\cdot))+
(B_{\gamma}(k)-B_{\gamma^{\prime}}(k^{\prime}))\times\cr
&(f(k^{\prime},\cdot)+
t_{\gamma^{\prime}}^{(0)}(k^{\prime},\cdot))+
B_{\gamma}(k)
(t_{\gamma}^{(0)}(k,\cdot)-t_{\gamma^{\prime}}^{(0)}(k^{\prime},\cdot)),\cr}
\eqno(11.5)$$
where $k,k^{\prime}\in\S^2_{\sqrt{E}}$, $\gamma,\gamma^{\prime}\in\S^2$.
We consider (11.5) as a linear integral equation for

\noindent
$t_{\gamma}^{(0)}(k,\cdot)-t_{\gamma^{\prime}}^{(0)}(k^{\prime},\cdot)$.
In addition, the following estimates hold:
$$\eqalignno{
&|B_{\gamma}(k)(f(k,\cdot)-f(k^{\prime},\cdot))(l)|
\le
{\delta c_2(\mu)N^2|k-k^{\prime}|^{\alpha}\over
(1-\eta)^2(1+|k-l|^2)^{\mu/2}},&(11.6)\cr
&|(B_{\gamma}(k)-B_{\gamma^{\prime}}(k^{\prime}))(f(k^{\prime},\cdot)+
t_{\gamma^{\prime}}^{(0)}(k^{\prime},\cdot))(l)|\le
{c_2^{\prime}(\mu)
c_{13}(\beta,\mu)N^2|\gamma-\gamma^{\prime}|^{\beta}\over
(1-\eta)^2(1+|k-l|^2)^{\mu/2}},&(11.7)\cr}$$
where $k,k^{\prime},l\in\S^2_{\sqrt{E}}$, $\gamma,\gamma^{\prime}\in\S^2$,
$\gamma k=\gamma^{\prime} k^{\prime}=0$, $|\gamma-\gamma^{\prime}|\le 1$,
$|k-k^{\prime}|\le 1$, $0<\beta< 1/2$, $E\ge 1$.
Estimate (11.6) follows from (2.39) (for $\beta=0$), (2.27a), (2.28a),
(5.22). Estimate (11.7) follows from (11.1a), (11.2) and the following lemma:

\vskip 2 mm
{\bf Lemma 14.}
{\it Let} $B_{\gamma}(k)$ {\it and} $\Lambda_k$ {\it be defined by} (2.38),
(2.40). {\it Then:}
$$\eqalign{
&\|\Lambda_k^{\mu}(B_{\gamma}(k)-B_{\gamma^{\prime}}(k^{\prime}))
\Lambda_k^{-\mu}u\|_{C(\S^2_{\sqrt{E}})}\le
c_{13}(\beta,\mu)E^{\beta-1/2}\|f\|_{C({\cal M}_E),\mu}\times\cr
&\|u\|_{C(\S^2_{\sqrt{E}})} |\gamma-\gamma^{\prime}|^{\beta},\cr}\eqno(11.8)$$
where $k,k^{\prime}\in\S^2_{\sqrt{E}}$, $\gamma,\gamma^{\prime}\in\S^2$,
$\gamma k=\gamma^{\prime} k^{\prime}=0$, $|\gamma-\gamma^{\prime}|\le 1$,
$\mu\ge 2$, $0<\beta\le 1/2$, $E\ge 1$.

\vskip 2 mm
{\it Proof of Lemma} 14.
From (2.38) it follows that
$$(B_{\gamma}(k)-B_{\gamma^{\prime}}(k^{\prime}))U(l)={\pi i\over \sqrt{E}}
\int\limits_{\S^2_{\sqrt{E}}}U(m)(\chi(m\gamma)-\chi(m\gamma^{\prime}))f(m,l)
dm,\eqno(11.9)$$
$$\eqalign{
&|(B_{\gamma}(k)-B_{\gamma^{\prime}}(k^{\prime}))\Lambda_k^{-\mu}u(l)|\le
{\pi\over \sqrt{E}}\|f\|_{C({\cal M}_E),\mu}
\|u\|_{C(\S^2_{\sqrt{E}})}\times\cr
&\int\limits_{\S^2_{\sqrt{E}}}{|\chi(m\gamma)-\chi(m\gamma^{\prime})|dm\over
(1+|k-m|^2)^{\mu/2} (1+|m-l|^2)^{\mu/2}},\cr}\eqno(11.10)$$
where $k,k^{\prime}\in\S^2_{\sqrt{E}}$, $\gamma,\gamma^{\prime}\in\S^2$,
$\gamma k=\gamma^{\prime} k^{\prime}=0$.
Consider
$$\eqalign{
&{\cal D}_{k,l}=\{m\in\S^2_{\sqrt{E}}:\ |m-k|<|m-l|\},\cr
&{\cal D}_{l,k}=\{m\in\S^2_{\sqrt{E}}:\ |m-k|>|m-l|\},\ \
k,l\in\S^2_{\sqrt{E}}.\cr}\eqno(11.11)$$
Note that
$$|m-l|>|k-l|/2\ \ {\rm for}\ \ m\in {\cal D}_{k,l},\ \
|m-k|>|k-l|/2\ \ {\rm for}\ \ m\in {\cal D}_{l,k}.\eqno(11.12)$$
We have that
$$\eqalign{
&{1\over \sqrt{E}}
\int\limits_{\S^2_{\sqrt{E}}}{|\chi(m\gamma)-\chi(m\gamma^{\prime})|dm\over
(1+|k-m|^2)^{\mu/2} (1+|m-l|^2)^{\mu/2}}\buildrel (11.12) \over \le
{1\over \sqrt{E}(1+|k-l|^2/4)^{\mu/2}}\times\cr
&\biggl(\int\limits_{{\cal D}_{k,l}}
{|\chi(m\gamma)-\chi(m\gamma^{\prime})|dm\over (1+|k-m|^2)^{\mu/2}}+
\int\limits_{{\cal D}_{l,k}}
{|\chi(m\gamma)-\chi(m\gamma^{\prime})|dm\over (1+|m-l|^2)^{\mu/2}}\biggr)\le
\cr
&{1\over \sqrt{E}(1+|k-l|^2/4)^{\mu/2}}
\biggl(\int\limits_{\S^2_{\sqrt{E}}}
{|\chi(m\gamma)-\chi(m\gamma^{\prime})|dm\over (1+|k-m|^2)^{\mu/2}}+
\int\limits_{\S^2_{\sqrt{E}}}
{|\chi(m\gamma)-\chi(m\gamma^{\prime})|dm\over (1+|m-l|^2)^{\mu/2}}\biggr)\le
\cr
&{\sqrt{E}\over (1+|k-l|^2/4)^{\mu/2}}
\biggl(\int\limits_{\S^2_{\sqrt{E}}}
{|\chi(n\gamma)-\chi(n\gamma^{\prime})|dn\over (1+E|n-\hat k|^2)^{\mu/2}}+
\int\limits_{\S^2_{\sqrt{E}}}
{|\chi(n\gamma)-\chi(n\gamma^{\prime})|dn\over (1+E|n-\hat l|^2)^{\mu/2}}
\biggr),\cr}\eqno(11.13)$$
where $k,l\in\S^2_{\sqrt{E}}$, $\hat k=k/\sqrt{E}$, $\hat l=l/\sqrt{E}$.
Further, by the H\"older inequality we have that
$$\eqalign{
&\int\limits_{\S^2}
{|\chi(n\gamma)-\chi(n\gamma^{\prime})|dn\over (1+E|n-\omega|^2)^{\mu/2}}\le
\biggl(\int\limits_{\S^2}|\chi(n\gamma)-\chi(n\gamma^{\prime})|^pdn
\biggr)^{1/p}\times\cr
&\biggl(\int\limits_{\S^2}{dn\over (1+E|n-\omega|^2)^{p\mu/(2(p-1))}}
\biggr)^{(p-1)/p},\ \ \gamma,\gamma^{\prime},\omega\in\S^2,\ p\in ]1,+\infty
[.\cr}\eqno(11.14)$$
In addition,
$$\int\limits_{\S^2}{dn\over (1+E|n-\omega|^2)^{p/(p-1)}}\le
{c_{14}(p)\over E},\ \ p\in ]1,+\infty [,\eqno(11.15)$$
$$\eqalign{
&\int\limits_{\S^2}
|\chi(n\gamma)-\chi(n\gamma^{\prime})|^pdn=
\int\limits_{\S^2}
|\chi(n\gamma)-\chi(n\gamma^{\prime})|dn=\cr
&2\int\limits_0^{\pi}\sin(\psi)\int\limits_0^{\v(\gamma,\gamma^{\prime})}
d\v d\psi=4\v(\gamma,\gamma^{\prime})\le {4\pi\over 3}|\gamma-
\gamma^{\prime}|,\cr}\eqno(11.16)$$
where $|\gamma-\gamma^{\prime}|\le 1$ and $\v(\gamma,\gamma^{\prime})$ is
(absolute value of) the angle between $\gamma$ and $\gamma^{\prime}$.
Note that in (11.16) we have used the following: (1) the property that
$\chi(n\gamma)-\chi(n\gamma^{\prime})$ takes values in the set
$\{-1,0,1\}$; (2) Euclidean basis $e_1$, $e_2$, $e_3$, where
$e_1=\gamma$, $e_3={\gamma\times\gamma^{\prime}\over
|\gamma\times\gamma^{\prime}|}$, $e_2=-e_1\times e_3$, and the standard
spherical coordinates related with this basis; (3) the formulas
$|\gamma-\gamma^{\prime}|/2=\sin(\v(\gamma,\gamma^{\prime})/2)$,
$1/2=\sin(\pi/6)$, $(\pi/3)\sin(\v/2)\ge\v/2$ for $\v\in [0,\pi/3]$.

Estimate (11.8) follows from (11.10), (11.13)-(11.16), where $p=1/\beta$.
Lemma 14 is proved. Thus, estimate (11.7) is also proved.

Estimate (11.3) follows from (11.5)-(11.7) and (2.39) (for $\beta=0$),
(5.22). This completes the proof of (11.3).

Using (11.2) we obtain that
$$\eqalign{
&h_{\gamma}(k,l)-h_{\gamma^{\prime}}(k^{\prime},l^{\prime})=
{\tilde h}_{\gamma}(k,l)-{\tilde h}_{\gamma^{\prime}}(k^{\prime},l^{\prime})=
\cr
&{\tilde h}_{\gamma}(k,l)-{\tilde h}_{\gamma^{\prime}}(k^{\prime},l)+
{\tilde h}_{\gamma^{\prime}}(k^{\prime},l)-{\tilde h}_{\gamma^{\prime}}
(k^{\prime},l^{\prime})=\cr
&{\tilde f}(k,l)-{\tilde f}(k^{\prime},l)+
t_{\gamma}^{(0)}(k,l)-t_{\gamma^{\prime}}^{(0)}(k^{\prime},l)+\cr
&{\tilde h}_{\gamma^{\prime}}(k^{\prime},l)-{\tilde h}_{\gamma^{\prime}}
(k^{\prime},l^{\prime})\ \ {\rm for}\ \ k-l=k^{\prime}-l^{\prime},\cr}
\eqno(11.17)$$
where
$${\tilde h}_{\gamma}(k,l)=h_{\gamma}(k,l)-{\hat v}(k-l),\ \
{\tilde f}(k,l)=f(k,l)-{\hat v}(k-l),\eqno(11.18)$$
$k,l,k^{\prime},l^{\prime}\in\S^2_{\sqrt{E}}$, $\gamma,\gamma^{\prime}\in\S^2$
.

Estimate (5.24) follows from (11.17), (11.18), (2.28b), (11.1b), (11.3),
(2.31), (2.8), (5.1), (5.9a) and the formulas
$$\eqalign{
&|k(\lambda,p,E)-k(\lambda^{\prime},p,E)|=
|l(\lambda,p,E)-l(\lambda^{\prime},p,E)|=(E-p^2/4)^{1/2}
|\lambda-\lambda^{\prime}|,\cr
&|\gamma^{\pm}(k(\lambda,p,E),p)-\gamma^{\pm}(k(\lambda^{\prime},p,E),p)|\le
|\lambda-\lambda^{\prime}|,\cr}\eqno(11.19)$$
$$\Delta^s\ge\Delta^t\ \ {\rm for}\ \ 0<s\le t,\ 0\le\Delta\le 1,\
(E-p^2/4)^{1/2}\le E^{1/2},\eqno(11.20)$$
where $k(\lambda,p,E)$ is given by (3.8),  $l(\lambda,p,E)= k(\lambda,p,E)-p$,
  $\gamma^{\pm}(k,p)$ are given in (5.1), $\lambda,\lambda^{\prime}\in
{\cal T}$, $p\in {\cal B}_{2\tau\sqrt{E}}\b {\cal L}_{\nu}$, $\tau\in
]0,1[$. This completes the proof of (5.24).

To obtain (5.25b) we  proceed from (5.11), (5.23), (5.24). First, using
(5.23), (5.24) we obtain that: $H^0_{E,\tau}(\cdot,p)$ defined by (5.11a) is
a bounded holomorphic function on ${\cal D}_+$ and is extended continuously
on ${\cal D}_+\cup {\cal T}$;  $H^0_{E,\tau}(\cdot,p)$
defined by (5.11b) is
a bounded holomorphic function on ${\cal D}_-$ and is extended continuously
on ${\cal D}_-\cup {\cal T}$. Therefore, due to the maximum principle for
holomorphic functions to prove (5.25b) it is sufficient to prove that
$$|H^0_{E,\tau,\pm}(\lambda,p)|\le 2^{\mu/2}N+
{c_7(\alpha,\mu,\sigma,\beta)N^2\over
(1-\eta)^2(1-\delta)E^{\beta/2}},\ \ \lambda\in {\cal T},\ \
p\in {\cal B}_{2\tau\sqrt{E}}\b {\cal L}_{\nu},\eqno(11.21)$$
where
$$H^0_{E,\tau,\pm}(\lambda,p)=H^0_{E,\tau}(\lambda(1\mp 0),p),\ \
\lambda\in {\cal T},\ \
p\in {\cal B}_{2\tau\sqrt{E}}\b {\cal L}_{\nu}.\eqno(11.22)$$
Proceeding from (11.22), (5.11) we obtain that
$$\eqalign{
&H^0_{E,\tau,+}(\lambda,p)={\hat v}(p)+{1\over 2\pi i}\int\limits_{\cal T}
(H_+(\zeta,p,E)-{\hat v}(p)){d\zeta\over {\zeta-\lambda(1-0)}}=\cr
&{\hat v}(p)+{1\over 2}\bigl(H_+(\lambda,p,E)-{\hat v}(p)\bigr)+
{1\over 2\pi i}p.v.\int\limits_{\cal T}
(H_+(\zeta,p,E)-{\hat v}(p)){d\zeta\over {\zeta-\lambda}},\cr
&H^0_{E,\tau,-}(\lambda,p)={\hat v}(p)-{1\over 2\pi i}\int\limits_{\cal T}
(H_-(\zeta,p,E)-{\hat v}(p)){\lambda d\zeta\over \zeta(\zeta-\lambda(1+0))}=
\cr
&{\hat v}(p)+{1\over 2}\bigl(H_-(\lambda,p,E)-{\hat v}(p)\bigr)-
{1\over 2\pi i}p.v.\int\limits_{\cal T}
(H_-(\zeta,p,E)-{\hat v}(p)){\lambda d\zeta\over \zeta(\zeta-\lambda)},\cr
}\eqno(11.23)$$
where
$\lambda\in {\cal T}$, $p\in {\cal B}_{2\tau\sqrt{E}}\b {\cal L}_{\nu}$.

Estimate (11.21) follows from formulas (11.23), where we use the
decomposition
$$\eqalign{
&p.v.\int_{\cal T}=p.v.\int_{{\cal T}_{\lambda,E}}+
\int_{{\cal T}\b {\cal T}_{\lambda,E}},\cr
&{\cal T}_{\lambda,E}=\{\zeta\in {\cal T}:\ |\zeta-\lambda|\le E^{-1/2}\},\cr}
\eqno(11.24a)$$
and from the estimation of $p.v.\int\limits_{{\cal T}_{\lambda,E}}$ using
(5.23), (5.24) and the formulas
$$\eqalignno{
&\bigg|p.v.\int\limits_{{\cal T}_{\lambda,E}}{d\zeta\over {\zeta-\lambda}}
\bigg|\le {c_{15}\over E^{1/2}},\ \
\bigg|p.v.\int\limits_{{\cal T}_{\lambda,E}}{d\zeta\over \zeta
(\zeta-\lambda)}\bigg|\le {c_{15}\over E^{1/2}},\ \ E\ge 1,&(11.24b)\cr
&\int\limits_{{\cal T}_{\lambda,E}}{|\zeta-\lambda|^{\beta}
|d\zeta|\over
|\zeta-\lambda|}\le {c_{16}\over \beta E^{\beta/2}},\ \ E\ge 1,\ \
0<\beta\le 1,&(11.24c)\cr}$$
and of $\int\limits_{{\cal T}\b {\cal T}_{\lambda,E}}$ using (5.23),
(2.31) and the formula
$$\int\limits_{{\cal T}\b {\cal T}_{\lambda,E}}{|d\zeta|\over
|\zeta-\lambda|}\le c_{17}\ln\,(1+E),\ \ E\ge 1.\eqno(11.24d)$$

 This completes the proof of (5.25b).

Lemma 5 is proved.

\vskip 2 mm
{\it Proof of Lemma 9}.
Estimate (6.12) follows from  (6.1), (6.4) and (2.39) for $\beta=0$.
Estimate (6.13) follows from (6.2), (6.12), (6.4), (2.39) for $\beta=\alpha$
(and the property $g_2>g_1$).

Further, using (6.4) we obtain that
$$
h_{\gamma}(k,\cdot)-h_{\gamma^{\prime}}(k^{\prime},\cdot)=
f(k,\cdot)-f(k^{\prime},\cdot)+
B_{\gamma}(k)h_{\gamma}(k,\cdot)-
B_{\gamma^{\prime}}(k^{\prime})
h_{\gamma^{\prime}}(k^{\prime},\cdot),\eqno(11.25)$$
$$\eqalign{
&h_{\gamma}(k,\cdot)-h_{\gamma^{\prime}}(k^{\prime},\cdot)=
f(k,\cdot)-f(k^{\prime},\cdot)+
(B_{\gamma}(k)-B_{\gamma^{\prime}}(k^{\prime}))
h_{\gamma^{\prime}}(k^{\prime},\cdot)+\cr
&B_{\gamma}(k)
(h_{\gamma}(k,\cdot)-h_{\gamma^{\prime}}(k^{\prime},\cdot)),\cr}
\eqno(11.26)$$
where $k,k^{\prime}\in\S^2_{\sqrt{E}}$, $\gamma,\gamma^{\prime}\in\S^2$.
We consider (11.26) as a linear integral equation for
$h_{\gamma}(k,\cdot)-h_{\gamma^{\prime}}(k^{\prime},\cdot)$. In addition:
$$\eqalignno{
&|f(k,l)-f(k^{\prime},l)|\buildrel (6.2) \over \le
{g_2N|k-k^{\prime}|^{\alpha}\over (1+|k-l|^2)^{\mu/2}},&(11.27)\cr
&|(B_{\gamma}(k)-B_{\gamma^{\prime}}(k^{\prime}))
h_{\gamma^{\prime}}(k^{\prime},\cdot)(l)|\buildrel (6.12),(11.8) \over \le
{c_2^{\prime}(\mu)
c_{13}(\beta,\mu)(g_1N)^2|\gamma-\gamma^{\prime}|^{\beta}\over
(1-\delta_1)(1+|k-l|^2)^{\mu/2}},&(11.28)\cr}$$
where $k,k^{\prime},l\in\S^2_{\sqrt{E}}$, $\gamma,\gamma^{\prime}\in\S^2$,
$\gamma k=\gamma^{\prime} k^{\prime}=0$, $|\gamma-\gamma^{\prime}|\le 1$,
$|k-k^{\prime}|\le 1$, $0<\beta\le 1/2$. Estimate (6.14) follows from
(11.26)-(11.28) and (2.39) for $\beta=0$.

Further, we have that
$$h_{\gamma}(k,l)-h_{\gamma^{\prime}}(k^{\prime},l^{\prime})=
(h_{\gamma}(k,l)-h_{\gamma^{\prime}}(k^{\prime},l))+
(h_{\gamma^{\prime}}(k^{\prime},l)-
h_{\gamma^{\prime}}(k^{\prime},l^{\prime})),\eqno(11.29)$$
where $k,k^{\prime},l,l^{\prime}\in\S^2_{\sqrt{E}}$,
$\gamma,\gamma^{\prime}\in\S^2$. Estimate (6.15) follows from (11.29),
(6.13), (6.14), (2.8), (5.1), (5.9a) and (11.19).

To obtain (6.16) we proceed from (5.11), (2.8), (5.1), (5.9), (6.12), (6.15).
In a similar way with the proof of (5.25b), to prove (6.16) it is
sufficient to prove (6.16) for $H^0_{E,\tau}$ replaced by its limits
$$\eqalign{
&H^0_{E,\tau,+}(\lambda,p)=H^0_{E,\tau}(\lambda(1-0),p)=
{1\over 2\pi i}\int\limits_{\cal T}
H_+(\zeta,p,E){d\zeta\over {\zeta-\lambda(1-0)}}=\cr
&{1\over 2}H_+(\lambda,p,E)+
{1\over 2\pi i}p.v.\int\limits_{\cal T}
H_+(\zeta,p,E){d\zeta\over {\zeta-\lambda}},\cr
&H^0_{E,\tau,-}(\lambda,p)=H^0_{E,\tau}(\lambda(1+0),p)=
-{1\over 2\pi i}\int\limits_{\cal T}
H_-(\zeta,p,E){\lambda d\zeta\over \zeta(\zeta-\lambda(1+0))}=\cr
&{1\over 2}H_-(\lambda,p,E)+
{1\over 2\pi i}p.v.\int\limits_{\cal T}
H_-(\zeta,p,E){\lambda d\zeta\over \zeta(\zeta-\lambda)},\cr
}\eqno(11.30)$$
where
$\lambda\in {\cal T}$, $p\in {\cal B}_{2\tau\sqrt{E}}\b {\cal L}_{\nu}$.
Estimate (6.16) (with $H^0_{E,\tau}$ replaced by $H^0_{E,\tau,\pm}$)
follows from (11.30), where we use the decomposition
(11.24), and from the estimation of
 $p.v.\int\limits_{{\cal T}_{\lambda,E}}$ using (6.12)
(with (2.8), (5.1), (5.9)), (6.15), (11.24b), (11.24c)
 and of $\int\limits_{{\cal T}\b {\cal T}_{\lambda,E}}$ using (6.12)
(with (2.8), (5.1), (5.9)), (11.24d).

Estimate (6.17) follows from (6.1), (6.7) and (2.39) for $\beta=0$.
Estimate (6.18) follows from (6.1), (6.7), (6.17) and (2.39) for
$\beta=\alpha$.

Further, using (6.7) we obtain that
$$\eqalign{
&t_{\gamma}^{(n)}(k,\cdot)-t_{\gamma^{\prime}}^{(n)}(k^{\prime},\cdot)=
(B_{\gamma}(k))^{n+1}f(k,\cdot)-(B_{\gamma^{\prime}}(k^{\prime}))^{n+1}
f(k^{\prime},\cdot)+\cr
&B_{\gamma}(k)t_{\gamma}^{(n)}(k,\cdot)-
B_{\gamma^{\prime}}(k^{\prime})
t_{\gamma^{\prime}}^{(n)}(k^{\prime},\cdot),\cr}\eqno(11.31)$$
$$\eqalign{
&t_{\gamma}^{(n)}(k,\cdot)-t_{\gamma^{\prime}}^{(n)}(k^{\prime},\cdot)=
((B_{\gamma}(k))^{n+1}-(B_{\gamma^{\prime}}(k^{\prime}))^{n+1})
f(k^{\prime},\cdot)+\cr
&(B_{\gamma}(k))^{n+1}(f(k,\cdot)-f(k^{\prime},\cdot))+
(B_{\gamma}(k)-B_{\gamma}^{\prime}(k^{\prime}))
t_{\gamma^{\prime}}^{(n)}(k^{\prime},\cdot))+\cr
&B_{\gamma}(k)
(t_{\gamma}^{(n)}(k,\cdot)-t_{\gamma^{\prime}}^{(n)}(k^{\prime},\cdot)),\cr}
\eqno(11.32)$$
where $k,k^{\prime}\in\S^2_{\sqrt{E}}$, $\gamma,\gamma^{\prime}\in\S^2$,
$n\in\N\cup 0$. We consider (11.32) as a linear integral equation for
$t_{\gamma}^{(n)}(k,\cdot)-t_{\gamma^{\prime}}^{(n)}(k^{\prime},\cdot)$.
In addition, the following estimates hold:
$$\eqalignno{
&|(B_{\gamma}(k))^{n+1}(f(k,\cdot)-f(k^{\prime},\cdot))(l)|\le
{\delta_1^{n+1}g_2N|k-k^{\prime}|^{\alpha}\over
(1+|k-l|^2)^{\mu/2}},&(11.33)\cr
&|(B_{\gamma}(k)-B_{\gamma^{\prime}}(k^{\prime}))
t_{\gamma^{\prime}}^{(n)}(k^{\prime},\cdot)(l)|\le
{c_{13}(\beta,\mu)\delta_1^{n+1}(g_1N)^2|\gamma-\gamma^{\prime}|^{\beta}\over
(1-\delta_1)(1+|k-l|^2)^{\mu/2}},&(11.34)\cr
&|((B_{\gamma}(k))^{n+1}-(B_{\gamma^{\prime}}(k^{\prime}))^{n+1})
f(k^{\prime},\cdot)(l)|\le
{(n+1)
c_{13}(\beta,\mu)\delta_1^n(g_1N)^2\over
(1+|k-l|^2)^{\mu/2}},&(11.35)\cr}$$
where $n\in\N\cup 0$,
$k,k^{\prime},l\in\S^2_{\sqrt{E}}$, $\gamma,\gamma^{\prime}\in\S^2$,
$\gamma k=\gamma^{\prime} k^{\prime}=0$, $|\gamma-\gamma^{\prime}|\le 1$,
$|k-k^{\prime}|\le 1$, $0<\beta\le 1/2$. Estimate (11.33) follows from
(6.2) and (2.39) for $\beta=0$.  Estimate (11.34) follows from
(6.17), (11.8), (6.1).

To obtain (11.35) we use, in particular, the formulas
$$\eqalignno{
&A_1^{n+1}-A_2^{n+1}=(A_1-A_2)A_1^n+A_2(A_1^n-A_2^n),&(11.36)\cr
&\|A_1^{n+1}-A_2^{n+1}\|\le \|A_1-A_2\|\|A_1\|^n+\|A_2\|\|A_1^n-A_2^n\|,
&(11.37)\cr
&\|A_1^{n+1}-A_2^{n+1}\|\le (n+1)\|A_1-A_2\|\,(\max\,(\|A_1\|,\|A_2\|))^n,
&(11.38)\cr}$$
where $A_1$, $A_2$ are bounded linear operators in $C(\S^2_{\sqrt{E}})$,
$\|\cdot\|$ denotes the norm of operators in  $C(\S^2_{\sqrt{E}})$,
$n\in\N\cup 0$. Note that (11.38) follows from (11.37) by the induction
method. Estimate (11.35) follows from (6.1), (6.2), (11.8), (11.38) for
$A_1=\Lambda_k^{\mu}B_{\gamma}(k)\Lambda_k^{-\mu}$,
$A_2=\Lambda_k^{\mu}B_{\gamma^{\prime}}(k^{\prime})\Lambda_k^{-\mu}$,
(2.38) and (2.39) for $\beta=0$.

Estimate (6.19) follows from (11.32)-(11.35) and (2.39) for $\beta=0$.
Further, we have that
$$t_{\gamma}^{(n)}(k,l)-t_{\gamma^{\prime}}^{(n)}(k^{\prime},l^{\prime})=
(t_{\gamma}^{(n)}(k,l)-t_{\gamma^{\prime}}^{(n)}(k^{\prime},l))+
(t_{\gamma^{\prime}}^{(n)}(k^{\prime},l)
-t_{\gamma^{\prime}}^{(n)}(k^{\prime},l^{\prime})),\eqno(11.39)$$
where $n\in\N\cup 0$,
$k,k^{\prime},l,l^{\prime}\in\S^2_{\sqrt{E}}$,
$\gamma,\gamma^{\prime}\in\S^2$. Estimate (6.20) follows from (11.39),
(6.18), (6.19), (6.9), (11.19), (2.28c), (2.28d).

To obtain (6.21) we proceed from (6.9), (6.17), (6.20) and  the definition
of $T^{0,n}_{E,\tau}$. In a similar way with the proofs of (5.25b) and
(6.16), to prove (6.21) it is sufficient to prove (6.21) for
$T^{0,n}_{E,\tau}$   replaced by its limits
$$\eqalign{
&T^{0,n}_{E,\tau,+}(\lambda,p)=T^{0,n}_{E,\tau}(\lambda(1-0),p)=
{1\over 2\pi i}\int\limits_{\cal T}
T_+^{(n)}(k(\zeta,p,E),p){d\zeta\over {\zeta-\lambda(1-0)}},\cr
&T^{0,n}_{E,\tau,-}(\lambda,p)=T^{0,n}_{E,\tau}(\lambda(1+0),p)=
{1\over 2\pi i}\int\limits_{\cal T}
T_-^{(n)}(k(\zeta,p,E),p){\lambda d\zeta\over \zeta(\zeta-\lambda(1+0))},\cr
}\eqno(11.40)$$
where $\lambda\in {\cal T}$, $p\in {\cal B}_{2\tau\sqrt{E}}\b {\cal L}_{\nu}$.
Further, the proof of (6.21) (with $T^{0,n}_{E,\tau}$ replaced by
$T^{0,n}_{E,\tau,\pm}$) is completely similar to the proof of (6.16) (with
$H^0_{E,\tau}$ replaced by $H^0_{E,\tau,\pm}$).

Lemma 9 is proved.

{\it Proof of Lemma 10.}
Using (6.4) we obtain that
$$
h_{\gamma}(k,\cdot)-{\tilde h}_{\gamma}(k,\cdot)=
f(k,\cdot)-{\tilde f}(k,\cdot)+
B_{\gamma}(k)h_{\gamma}(k,\cdot)-
{\tilde B}_{\gamma}(k){\tilde h}_{\gamma}(k,\cdot)
,\eqno(11.41)$$
$$\eqalign{
&h_{\gamma}(k,\cdot)-{\tilde h}_{\gamma}(k,\cdot)=
f(k,\cdot)-{\tilde f}(k,\cdot)+
(B_{\gamma}(k)-{\tilde B}_{\gamma}(k))
{\tilde h}_{\gamma}(k,\cdot)+\cr
&B_{\gamma}(k)
(h_{\gamma}(k,\cdot)-{\tilde h}_{\gamma}(k,\cdot)),\cr}
\eqno(11.42)$$
where $k\in\S^2_{\sqrt{E}}$, $\gamma\in\S^2$, ${\tilde B}_{\gamma}(k)$ is
defined by (2.38) with $f$ replaced by $\tilde f$. We consider (11.42)
as a linear integral equation for
$h_{\gamma}(k,\cdot)-{\tilde h}_{\gamma}(k,\cdot)$. Using estimate (6.12)
with  $h_{\gamma}(k,l)$ replaced by  ${\tilde h}_{\gamma}(k,l)$  and
estimate (2.39) (for $\beta=0$) with $f$ replaced by $f-\tilde f$, we
obtain that
$$\eqalign{
&|(B_{\gamma}(k)-{\tilde B}_{\gamma}(k)){\tilde h}_{\gamma}(k,\cdot)(l)|
\le {c_3(0,\mu,\sigma,3)g_1N
\|f-\tilde f\|_{C({\cal M}_E),\mu}\over E^{\sigma/2}
(1-\delta_1)(1+|k-l|^2)^{\mu/2}}\cr
&\buildrel (6.11) \over
\le {\delta_1
\|f-\tilde f\|_{C({\cal M}_E),\mu}\over
(1-\delta_1)(1+|k-l|^2)^{\mu/2}},\cr}\eqno(11.43)$$
where $\gamma\in\S^2$, $k,l\in\S^2_{\sqrt{E}}$. Estimate (6.22) follows
from (11.42), (11.43) and (2.39) for $\beta=0$.

To prove (6.23) we proceed from (6.15), (6.22) and the definitions of
$H_{\pm}$ and  ${\tilde H}_{\pm}$. Due to (6.22) and the definitions of
$H_{\pm}$ and  ${\tilde H}_{\pm}$, we have that
$$\eqalign{
&|(H_{\pm}-{\tilde H}_{\pm})(k(\lambda,p,E),p)-
(H_{\pm}-{\tilde H}_{\pm})(k(\lambda^{\prime},p,E),p)|\le\cr
&{2\|f-\tilde f\|_{C({\cal M}_E),\mu}\over (1-\delta_1)^2
(1+p^2)^{\mu/2}},\cr}\eqno(11.44)$$
where $\lambda,\lambda^{\prime}\in {\cal T}$,
$p\in {\cal B}_{2\sqrt{E}}\b {\cal L}_{\nu}$. Using (6.15) we obtain
that
$$\eqalign{
&|(H_{\pm}-{\tilde H}_{\pm})(k(\lambda,p,E),p)-
(H_{\pm}-{\tilde H}_{\pm})(k(\lambda^{\prime},p,E),p)|=\cr
&|H_{\pm}(k(\lambda,p,E),p)-H_{\pm}(k(\lambda^{\prime},p,E),p)-
({\tilde H}_{\pm}(k(\lambda,p,E),p)-
{\tilde H}_{\pm}(k(\lambda^{\prime},p,E),p))|\cr
&\le 2\biggl({c_{10}(\mu)(1+\delta_2)
g_2N\over {1-\delta_1}}+
{c_9(\beta,\mu)
(g_1N)^2\over (1-\delta_1)^2E^{\beta/2}}
\biggr)
{E^{\beta/2}|\lambda-\lambda^{\prime}|^{\beta}\over
(1+p^2)^{\mu/2}},\cr}\eqno(11.45)$$
where $\lambda,\lambda^{\prime}\in {\cal T}$,
$p\in {\cal B}_{2\sqrt{E}}\b {\cal L}_{\nu}$, $|\lambda-\lambda^{\prime}|\le
(E-p^2/4)^{-1/2}$, $0<\beta<\min\,(\alpha,1/2)$.
In addition:
$$\eqalign{
&\|f-\tilde f\|_{C({\cal M}_E),\mu}=
\bigl(\|f-\tilde f\|_{C({\cal M}_E),\mu}\bigr)^{1-\ep}
\bigl(\|f-\tilde f\|_{C({\cal M}_E),\mu}\bigr)^{\ep}\le\cr
&\bigl(\|f-\tilde f\|_{C({\cal M}_E),\mu}\bigr)^{1-\ep}
\bigl(E^{\beta/2}|\lambda-\lambda^{\prime}|^{\beta}\bigr)^{\ep}\ \ {\rm for}\
\  \|f-\tilde f\|_{C({\cal M}_E),\mu}\le
E^{\beta/2}|\lambda-\lambda^{\prime}|^{\beta},\cr}\eqno(11.46)$$
$$\eqalign{
&E^{\beta/2}|\lambda-\lambda^{\prime}|^{\beta}=
\bigl(E^{\beta/2}|\lambda-\lambda^{\prime}|^{\beta}\bigr)^{1-\ep}
\bigl(E^{\beta/2}|\lambda-\lambda^{\prime}|^{\beta}\bigr)^{\ep}\le\cr
&\bigl(\|f-\tilde f\|_{C({\cal M}_E),\mu}\bigr)^{1-\ep}
\bigl(E^{\beta/2}|\lambda-\lambda^{\prime}|^{\beta}\bigr)^{\ep}\ \ {\rm for}\
\  \|f-\tilde f\|_{C({\cal M}_E),\mu}\ge
E^{\beta/2}|\lambda-\lambda^{\prime}|^{\beta},\cr}\eqno(11.47)$$
where $0\le\ep\le 1$. Estimate (6.23) follows from (11.44)-(11.47).

To obtain (6.24) we proceed from (6.22), (6.23) and the definitions of
$H_{\pm}$, ${\tilde H}_{\pm}$, $H^0_{E,\tau}$, ${\tilde H}^0_{E,\tau}$.
This proof is completely similar to the proof of (6.16).

Lemma 10 is proved.

\vskip 2 mm
{\it Proof of Lemma 11.}
Estimate (6.27) follows from (2.36) (for $\alpha=0$), (6.1) and the
property
$$|u(|k-l|,2\tau_0\sqrt{E},2\tau\sqrt{E})|\le 1,\ \ (k,l)\in {\cal M}_E.
\eqno(11.48)$$
Estimate (6.28) follows from (2.36), (6.1), (6.2), (6.27), the formula
$$\eqalign{
&{\tilde f}(k,l)-{\tilde f}(k^{\prime},l^{\prime})=
(u(|k-l|,2\tau_0\sqrt{E},2\tau\sqrt{E})-
u(|k^{\prime}-l^{\prime}|,2\tau_0\sqrt{E},2\tau\sqrt{E})) f(k,l)+\cr
&u(|k^{\prime}-l^{\prime}|,2\tau_0\sqrt{E},2\tau\sqrt{E})
(f(k,l)-f(k^{\prime},l^{\prime})),\ \
(k,l)\in {\cal M}_E,\ \ (k^{\prime},l^{\prime})\in {\cal M}_E,\cr}
\eqno(11.49)$$
property (11.48) and the inequalities
$$\eqalign{
&|u(|k-l|,2\tau_0\sqrt{E},2\tau\sqrt{E})-
u(|k^{\prime}-l^{\prime}|,2\tau_0\sqrt{E},2\tau\sqrt{E})|\le\cr
&{|\,|k-l|-|k^{\prime}-l^{\prime}|\,|\over 2(\tau-\tau_0)\sqrt{E}}\le
{{|k-k^{\prime}|+|l-l^{\prime}|}\over 2(\tau-\tau_0)\sqrt{E}},\ \
(k,l)\in {\cal M}_E,\ \ (k^{\prime},l^{\prime})\in {\cal M}_E.\cr}
\eqno(11.50)$$
Estimate (6.29) follows from (2.36) (for $\alpha=0$), (6.1) and the
formulas
$$(1+|k-l|^2)^{\mu_0/2}|f(k,l)-{\tilde f}(k,l)|\le
{\|f\|_{C({\cal M}_E),\mu}(1-u(|k-l|,2\tau_0\sqrt{E},2\tau\sqrt{E}))\over
(1+|k-l|^2)^{(\mu-\mu_0)/2}},\eqno(11.51)$$
$$\eqalign{
&1-u(|k-l|,2\tau_0\sqrt{E},2\tau\sqrt{E})=0\ \ {\rm for}\ \
|k-l|\le 2\tau_0\sqrt{E},\cr
&|1-u(|k-l|,2\tau_0\sqrt{E},2\tau\sqrt{E})|\le 1,\cr}\eqno(11.52)$$
where  $(k,l)\in {\cal M}_E$.

Lemma 11 is proved.

\vskip 4 mm

{\bf 12. Proof of Lemmas 7 and 8}

{\it Proof of Lemma 7.}
For
$$\eqalign{
&U^0,U,U_1,U_2\in
 L^{\infty}((\C\b ({\cal T}\cup 0))\times
({\cal B}_{2\tau\sqrt{E}}\b {\cal L}_{\nu})),\cr
&|||U^0|||_{E,\tau,\mu}\le r/2,\ \ |||U|||_{E,\tau,\mu}\le r,\ \
|||U_1|||_{E,\tau,\mu}\le r,\ \ |||U_2|||_{E,\tau,\mu}\le r,\cr}\eqno(12.1)$$
using Lemma 4 and the assumptions of Lemma 7 we obtain that
$$\eqalign{
&M_{E,\tau,U^0}(U)\in
 L^{\infty}((\C\b ({\cal T}\cup 0))\times
({\cal B}_{2\tau\sqrt{E}}\b {\cal L}_{\nu})),\cr
&|||M_{E,\tau,U^0}(U)|||_{E,\tau,\mu}\le
|||U^0|||_{E,\tau,\mu}+|||M_{E,\tau}(U)|||_{E,\tau,\mu}\le\cr
&r/2 + c_5c_4(\mu,\tau,E)r^2<r,\cr}\eqno(12.2)$$
$$\eqalign{
&|||M_{E,\tau,U^0}(U_1)-M_{E,\tau,U^0}(U_2)|||_{E,\tau,\mu}\le
2c_5c_4(\mu,\tau,E)r\,|||U_1-U_2|||_{E,\tau,\mu},\cr
&2c_5c_4(\mu,\tau,E)r<1,\cr}\eqno(12.3)$$
where
$$M_{E,\tau,U^0}(U)=U^0+M_{E,\tau}(U).\eqno(12.4)$$
Due to (12.1)-(12.4), $M_{E,\tau,U^0}$ is a contraction map of the ball
$U\in L^{\infty}((\C\b ({\cal T}\cup 0))\times
({\cal B}_{2\tau\sqrt{E}}\b {\cal L}_{\nu}))$,
$|||U|||_{E,\tau,\mu}\le r$. Using now the lemma about contraction maps
and using the formulas
$$\eqalignno{
&|||U-M^n_{E,\tau,U^0}(0)|||_{E,\tau,\mu}\le\sum_{j=n}^{\infty}
|||M^{j+1}_{E,\tau,U^0}(0)-M^j_{E,\tau,U^0}(0)|||_{E,\tau,\mu},&(12.5)\cr
&|||M_{E,\tau,U^0}(0)-0|||_{E,\tau,\mu}=|||U^0|||_{E,\tau,\mu}\le r/2,&(12.6)
\cr}$$
$$\eqalign{
&|||M^{j+1}_{E,\tau,U^0}(0)-M^j_{E,\tau,U^0}(0)|||_{E,\tau,\mu}
\buildrel (12.3) \over \le
2c_5c_4(E,\tau,\mu)r\times\cr
&|||M^j_{E,\tau,U^0}(0)-M^{j-1}_{E,\tau,U^0}(0)|||_{E,\tau,\mu},\ \
j=1,2,3...,\cr}\eqno(12.7)$$
where $M^0_{E,\tau,U^0}(0)=0$, we obtain Lemma 7.

\vskip 2 mm
{\it Proof of Lemma 8.}
We have that
$$\eqalignno{
&U-\tilde U=U^0-{\tilde U}^0+M_{E,\tau}(U)-M_{E,\tau}(\tilde U),&(12.8a)\cr
&M_{E,\tau}(U)-M_{E,\tau}(\tilde U)=M_{E,\tau}(U-\tilde U,U)+
M_{E,\tau}(\tilde U,U-\tilde U),&(12.8b)\cr}$$
where
$$\eqalign{
&M_{E,\tau}(U_1,U_2)(\lambda,p)=M^+_{E,\tau}(U_1,U_2)(\lambda,p)=\cr
&-{1\over \pi}\int\!\!\!\int\limits_{{\cal D}_+}(U_1,U_2)_{E,\tau}(\zeta,p)
{dRe\,\zeta dIm\,\zeta\over {\zeta-\lambda}},\ \ \lambda\in {\cal D}_+\b 0,\ \
 p\in {\cal B}_{2\tau\sqrt{E}}\b {\cal L}_{\nu},\cr}\eqno(12.9a)$$
$$\eqalign{
&M_{E,\tau}(U_1,U_2)(\lambda,p)=M^-_{E,\tau}(U_1,U_2)(\lambda,p)=\cr
&-{1\over \pi}\int\!\!\!\int\limits_{{\cal D}_-}(U_1,U_2)_{E,\tau}(\zeta,p)
{\lambda dRe\,\zeta dIm\,\zeta\over \zeta(\zeta-\lambda)},\ \
\lambda\in {\cal D}_-,\ \
 p\in {\cal B}_{2\tau\sqrt{E}}\b {\cal L}_{\nu},\cr}\eqno(12.9b)$$
where $(U_1,U_2)(\zeta,p)$ is defined by (5.13).

In view of (12.8b) we can consider (12.8a) as a linear integral equation for
"unknown" $U-\tilde U$ with given  $U^0-{\tilde U}^0$, $U$, $\tilde U$.

As well as (5.19), using also Lemma 7 we obtain that
$$|||M_{E,\tau}(U-\tilde U,U)-M_{E,\tau}(\tilde U,U-\tilde U)|||_{E,\tau,\mu}
\le 2c_5c_4(\mu,\tau,E)r|||U-\tilde U|||_{E,\tau,\mu}.\eqno(12.10)$$
Using (12.8b), (12.10) and solving (12.8a) by the method of successive
approximations we obtain (5.29). Lemma 8 is proved.

\vskip 8 mm
{\bf References}
\vskip 2 mm
\item{[ ABF]} M.J.Ablowitz, D.Bar Yaacov and A.S.Fokas, {\it On the inverse
scattering transform for the Kadomtsev-Petviashvili equation},
Stud. Appl. Math. {\bf 69} (1983), 135-143.
\item{[ ABR]} N.V.Alexeenko, V.A.Burov, O.D.Rumyantseva,
{\it Solution of three-dimensional acoustical inverse scattering problem
based on Novikov-Henkin algorithm}, Acoustical Journal, to appear
(in Russian).
\item{[  BC]} R.Beals and R.R.Coifman, {\it Multidimensional inverse
scattering and nonlinear partial differential equations}, Proc. Symp. Pure
Math. {\bf 43} (1985), 45-70.
\item{[BBMR]} A.V.Bogatyrev, V.A.Burov, S.A.Morozov, O.D.Rumyantseva and
E.G.Sukhov, {\it Numerical realization of algorithm for exact solution of
two-dimensional monochromatic inverse problem of acoustical scattering},
Acoustical Imaging {\bf 25} (2000), 65-70 (Kluwer Academic/Plenum Publishers,
New York).
\item{[ BMR]} V.A.Burov, S.A.Morozov and O.D.Rumyantseva, {\it Reconstruction
of fine-scale structure of acoustical scatterer on large-scale contrast
background},
Acoustical Imaging {\bf 26} (2002), 231-238 (Kluwer Academic/Plenum
Publishers, New York).
\item{[  Ch]} Y.Chen, {\it Inverse scattering via Heisenberg's
uncertainty principle}, Inverse Problems {\bf 13} (1997), 253-282.
\item{[ ER1]} G.Eskin and J.Ralston, {\it The inverse back-scattering problem
in three dimensions}, Commun. Math. Phys. {\bf 124} (1989), 169-215.
\item{[ ER2]} G.Eskin and J.Ralston, {\it Inverse back-scattering
in two dimensions}, Commun. Math. Phys. {\bf 138} (1991), 451-486.
\item{[  F1]} L.D.Faddeev, {\it Mathematical aspects of the three-body problem
in the quantum scattering theory}, Trudy MIAN {\bf 69} (1963).
\item{[  F2]} L.D.Faddeev, {\it Growing solutions of the Schr\"odinger equation}
, Dokl. Akad. Nauk SSSR {\bf 165} (1965), 514-517 (in Russian); English
Transl.: Sov. Phys. Dokl. {\bf 10} (1966), 1033-1035.
\item{[  F3]} L.D.Faddeev, {\it Inverse problem of quantum scattering theory II}
, Itogi Nauki i Tekhniki, Sovr. Prob. Math. {\bf 3} (1974), 93-180 (in
Russian); English transl.: J.Sov. Math. {\bf 5} (1976), 334-396.
\item{[  FM]} L.D.Faddeev and S.P.Merkuriev, {\it Quantum Scattering Theory
for Multi-particle Systems}, Nauka, Moscow, 1985 (in Russian); English
transl.: Math. Phys. Appl. Math. {\bf 11} (1993), Kluwer Academic
Publishers Group, Dordrecht.
\item{[ Gel]} I.M.Gel'fand, {\it Some problems of functional analysis and
algebra}, Proceedings of the International Congress of Mathematicians
Held at Amsterdam (1954), 253-276.
\item{[  GM]} P.G.Grinevich and S.V.Manakov, {\it The inverse scattering
problem for the two-dimen-

\noindent
\item{      }  sional Schr\"odinger operator, the} $\bar\partial$-
{\it method and non-linear equations}, Funkt. Anal. i Pril. {\bf 20(2)}
(1986), 14-24 (in Russian); English transl.: Funct. Anal. and Appl. {\bf 20}
(1986), 94-103.
\item{[  GN]} P.G.Grinevich and R.G.Novikov, {\it Analogues of multisoliton
potentials for the two-dimensional Schr\"odinger equations and a nonlocal
Riemann problem}, Dokl. Akad. Nauk SSSR {\bf 286} (1986), 19-22 (in
Russian); English transl.: Sov. Math. Dokl. {\bf 33} (1986), 9-12.
\item{[ Gro]} M.Gromov, {\it Possible trends in mathematics in the
coming decades}, Mathematics Unlimited - 2001 and Beyond (B.Engquist and
W.Schmid, eds) Springer, Berlin, 2001, 525-527.
\item{[  HN]} G.M.Henkin and R.G.Novikov, {\it The} $\bar\partial$- {\it
equation in the multidimensional inverse scattering problem}, Uspekhi Mat.
Nauk {\bf 42(3)} (1987), 93-152 (in Russian); English transl.: Russ. Math.
Surv. {\bf 42(3)} (1987), 109-180.
\item{[   M]} S.V.Manakov, {The inverse scattering transform for the time
dependent Schr\"odinger equation and Kadomtsev-Petviashvili equation},
Physica D {\bf  3(1,2)} (1981), 420-427.
\item{[Mand]} N.Mandache, {\it Exponential instability in an inverse problem
for the Schr\"odinger equation}, Inverse Problems {\bf 17} (2001),
1435-1444.
\item{[ No1]} R.G.Novikov, {\it Construction of a two-dimensional Schr\"odinger
operator with a given scattering amplitude at fixed energy}, Teoret. Mat.
Fiz. {\bf 66} (1986), 234-240.
\item{[ No2]} R.G.Novikov, {\it Reconstruction of a two-dimensional
Schr\"odinger
operator from the scattering amplitude at fixed energy}, Funkt. Anal. i Pril.
{\bf 20(3)} (1986), 90-91 (in Russian); English transl.: Funct. Anal. and
Appl. {\bf 20} (1986), 246-248.
\item{[ No3]} R.G.Novikov, {\it Multidimensional inverse spectral problem for
the equation} $-\Delta\psi+(v(x)-E u(x))\psi=0$, Funkt. Anal. i Pril.
{\bf 22(4)} (1988), 11-22 (in Russian); English transl.: Funct. Anal. and
Appl. {\bf 22} (1988), 263-272.
\item{[ No4]} R.G.Novikov, {\it The inverse scattering problem on a fixed
energy level for the two-dimensional Schr\"odinger operator}, J.Funct. Anal.
{\bf 103} (1992), 409-463.
\item{[ No5]} R.G.Novikov, {\it The inverse scattering problem at fixed
energy  for the three-dimensional Schr\"odinger equation
with an exponentially decreasing potential}, Commun. Math. Phys.
{\bf 161} (1994), 569-595.
\item{[ No6]} R.G.Novikov, {\it Rapidly converging approximation in inverse
quantum scattering in dimension 2}, Physics Letters A {\bf 238} (1998),
73-78.
\item{[ No7]} R.G.Novikov, {\it Approximate inverse quantum scattering at
fixed energy in dimension 2}, Proceedings of the Steklov Mathematical
Institute {\bf 225} (1999), 285-302.

\end